\documentclass[%
 reprint,
superscriptaddress,
floatfix,
 amsmath,amssymb,
 aps, prx
]{revtex4-2}

\usepackage[utf8]{inputenc}
\usepackage{amsmath,amssymb}
\usepackage{wrapfig}
\usepackage{graphicx}
\usepackage{bm}
\usepackage[colorlinks=true,linkcolor=blue,citecolor=blue,urlcolor=blue]{hyperref}
\usepackage{makecell}
\usepackage{siunitx}
\usepackage{cases}

\newcommand{\citeasnoun}[1]{Ref.~\cite{#1}}

\newcommand{\Figref}[1]{Figure~\ref{fig:#1}}
\newcommand{\figref}[1]{Fig.~\ref{fig:#1}}
\renewcommand{\eqref}[1]{Eq.~(\ref{eq:#1})}

\newcommand{\eqreftwo}[2]{Eqs.~(\ref{eq:#1},\ref{eq:#2})}
\newcommand{\eqrefrange}[2]{Eqs.~(\ref{eq:#1})--(\ref{eq:#2})}

\renewcommand{\Im}{\operatorname{Im}}
\newcommand{\vect}[1]{\boldsymbol{\mathbf{#1}}}

\newcommand{\secref}[1]{Sec.~\ref{sec:#1}}

\newcommand*{\Pscat}{P_{\rm scat}}
\newcommand*{\Pabs}{P_{\rm abs}}
\newcommand*{\Pext}{P_{\rm ext}}
\newcommand*{\psiscat}{\psi_{\rm scat}}
\newcommand*{\psiinc}{\psi_{\rm inc}}
\newcommand*{\cinc}{c_{\rm inc}}
\newcommand*{\sscat}{\sigma_{\rm scat}}

\newcommand*{\ImGO}{\Im \Gamma_0}

\newcommand*{\Ev}{\vect{E}}
\newcommand*{\Hv}{\vect{H}}
\newcommand*{\Pv}{\vect{P}}
\newcommand*{\Mv}{\vect{M}}
\newcommand*{\vv}{\vect{v}}

\newcommand*{\SM}{SM}
\newcommand*{\BB}{\mathbb{B}}
\newcommand*{\Gbb}{\mathbb{G}}
\newcommand*{\Vbb}{\mathbb{V}}

\newcommand{\hl}{}

\usepackage[english]{babel}
\makeatletter
\def\bbl@set@language#1{%
  \edef\languagename{%
    \ifnum\escapechar=\expandafter`\string#1\@empty
    \else\string#1\@empty\fi}%
  \@ifundefined{babel@language@alias@\languagename}{}{%
    \edef\languagename{\@nameuse{babel@language@alias@\languagename}}%
  }%
  \select@language{\languagename}%
  \expandafter\ifx\csname date\languagename\endcsname\relax\else
    \if@filesw
      \protected@write\@auxout{}{\string\select@language{\languagename}}%
      \bbl@for\bbl@tempa\BabelContentsFiles{%
        \addtocontents{\bbl@tempa}{\xstring\select@language{\languagename}}}%
      \bbl@usehooks{write}{}%
    \fi
  \fi}
\newcommand{\DeclareLanguageAlias}[2]{%
  \global\@namedef{babel@language@alias@#1}{#2}%
}
\makeatother

\DeclareLanguageAlias{en}{english}
\DeclareLanguageAlias{EN}{english}

\begin{document}

\preprint{APS/123-QED}

\title{Maximal single-frequency electromagnetic response}

\author{Zeyu Kuang}
\affiliation{Department of Applied Physics and Energy Sciences Institute, Yale University, New Haven, Connecticut 06511, USA}
\author{Lang Zhang}
\affiliation{Department of Applied Physics and Energy Sciences Institute, Yale University, New Haven, Connecticut 06511, USA}
\author{Owen D. Miller}
\affiliation{Department of Applied Physics and Energy Sciences Institute, Yale University, New Haven, Connecticut 06511, USA}

\date{\today}
\begin{abstract}
    \hl{Modern nanophotonic and meta-optical devices utilize a tremendous number of structural degrees of freedom to enhance light--matter interactions. A fundamental question is how large such enhancements can be.} We develop an analytical framework to derive upper bounds to single-frequency electromagnetic response, across near- and far-field regimes, for any materials, naturally incorporating the tandem effects of material- and radiation-induced losses. Our framework relies on a power-conservation law for the polarization fields induced in any scatterer. It unifies previous theories on optical scattering bounds and reveals new insight for optimal nanophotonic design, with applications including far-field scattering, near-field local-density-of-states engineering, \hl{optimal wavefront shaping}, and the design of perfect absorbers. Our bounds predict strikingly large minimal thicknesses for arbitrarily patterned perfect absorbers, ranging from 50--\SI{100}{nm} for typical materials at visible wavelengths to $\mu$m-scale thicknesses for polar dielectrics at infrared wavelengths. We \hl{use inverse design to} discover metasurface structures approaching the \hl{minimum-thickness} perfect-absorber bounds.
\end{abstract}

\pacs{Valid PACS appear here}
\maketitle

\section{Introduction}
Electromagnetic scattering at a single frequency is constrained by two loss mechanisms: material dissipation (absorption) and radiative coupling (scattering). There has been substantial research probing the limits of light--matter interactions subject to constraint of either mechanism~\cite{Yaghjian1996,hamam_coupled-mode_2007,kwon2009optimal, ruan2011design, liberal2014least, liberal2014upper,miller_fundamental_2016,miller2017limits,yang2017low,pendry1999radiative, thongrattanasiri_complete_2012,hugonin_fundamental_2015,miller2015shape, rahimzadegan2017fundamental,liu2018optimal,yang2018maximal,michon2019limits, nordebo2019optimal, shim2019fundamental, ivanenko2019optical,dias_fundamental_2019}, yet no general theory simultaneously accounting for both. In this Article, we develop a framework for upper bounds to electromagnetic response, across near- and far-field regimes, for any materials, naturally incorporating the tandem effects of material- and radiation-induced losses. Our framework relies on a power-conservation law for the polarization currents induced in any medium via a volume-integral version of the optical theorem~\cite{jackson1999classical, newton1976optical, bohren2008absorption, carney2004generalized}. An illustrative example is that of plane-wave scattering, where our bounds unify two previously separate approaches: radiative-coupling constraints leading to maximum cross-sections proportional to the square wavelength~\cite{Yaghjian1996,hamam_coupled-mode_2007,kwon2009optimal, ruan2011design, liberal2014least, liberal2014upper}, $\max \sigma\sim \lambda^2$, and material-dissipation constraints leading to cross-section bounds inversely proportional to material loss~\cite{miller_fundamental_2016,miller2017limits,yang2017low}, $\max \sigma \sim |\chi|^2 / \Im \chi$. Our framework contains more than a dozen previous results~\cite{Yaghjian1996,hamam_coupled-mode_2007,kwon2009optimal, ruan2011design, liberal2014least,miller_fundamental_2016,miller2017limits,yang2017low, thongrattanasiri_complete_2012,hugonin_fundamental_2015, rahimzadegan2017fundamental,liu2018optimal,yang2018maximal,michon2019limits} as asymptotic limits, it regularizes unphysical divergences in these results, and it reveals new insight for optimal nanophotonic design, with applications including far-field scattering, near-field local-density-of-states engineering, and the design of perfect absorbers. The ramifications of our bounds for perfect absorbers are striking: we prove that independent of the geometric patterning, the minimum thickness of perfect or near-perfect absorbers comprising conventional materials is typically on the order of 50--$\SI{100}{nm}$ at visible wavelengths, and closer to 1 $\mu$m at infrared wavelengths where polar-dielectric materials are resonant. These values are larger than the material skin depths, and roughly 100$\times$ larger than \hl{those} suggested by previous material-loss bounds~\cite{miller_fundamental_2016}. We use inverse design to discover ultrathin absorber designs closely approaching the bounds. We show that these bounds can further be utilized for the ``reverse'' problem of identifying optimal illumination fields, a critical element of the burgeoning field of wavefront shaping~\cite{Popoff2014,Vellekoop2015,Horstmeyer2015,Jang2018}. The framework developed here has immediate applicability to \emph{any} linear and quadratic response functions in electromagnetic scattering problems, including those that arise in near-field radiative heat transfer (NFRHT)~\cite{Polder1971,Otey2014,rodriguez_fluctuating-surface-current_2012}, optical force/torque~\cite{Mazilu2011,rahimzadegan2017fundamental,Lee2017,Liu2019,Horodynski2019}, \hl{high-NA metalenses~\cite{lalanne_design_1999,lalanne_metalenses_2017-1,chung_high-na_2020}}, and more general nanophotonic mode coupling~\cite{Miller2019}.

For many years, there was a single ``channel bound'' approach underlying the understanding of bounds to single-frequency electromagnetic response~\cite{Yaghjian1996,hamam_coupled-mode_2007,kwon2009optimal, ruan2011design, liberal2014least, liberal2014upper,pendry1999radiative, thongrattanasiri_complete_2012, hugonin_fundamental_2015, rahimzadegan2017fundamental,liu2018optimal, ivanenko2019optical}. The approach identifies ``channels'' (typically infinite in number) that carry power towards and away from the scattering body~\cite{newton2013scattering,mahaux1969shell,jalas2013and,rotter2017light}, use intuition or asymptotic arguments to restrict the scattering process to a finite number of channels, and then apply energy-conservation within those channels to arrive at maximal power-exchange quantities. The canonical example is in bounds for scattering cross-sections, i.e., the total scattered power divided by the intensity of an incoming plane wave. It has long been known that the maximal cross-section of a subwavelength electric-dipole antenna~\cite{Stutzman2012}, or even a single two-level atomic transition~\cite{Loudon2000}, is proportional to the square wavelength; for scattering cross-sections, the bound is $\sscat \leq 3\lambda^2/2\pi$. These bounds are consequences of properties of the incident waves (\emph{not} the scatterers): though plane waves carry infinite total power, they carry a finite amount of power in each vector-spherical-wave basis function, and $3\lambda^2/2\pi$ scattering corresponds simply to scattering all of the power in the electric-dipole channel. Related arguments can be used to bound NFRHT rates, which are constrained by restricting near-field coupling to only finite-wavenumber evanescent waves~\cite{pendry1999radiative}, absorption rates in ultrathin films, which are constrained by symmetry to have nonzero coupling to up/down plane-wave channels~\cite{thongrattanasiri_complete_2012}, and maximal antenna directivity~\cite{liberal2014upper}. All such channel bounds are consequences of radiative-coupling constraints, with optimal power-flow dynamics corresponding to ideal coupling to every channel that interacts with the scattering system. The drawbacks of channel bounds are two-fold: (1) they do not account for absorptive losses in the scatterers, and (2) except in the simplest (e.g. dipolar) systems, it is typically impossible to predict \emph{a priori} how many channels may actually contribute in optimal scattering processes. Without any such restrictions, the bounds diverge.

In recent years, an alternative approach has been developed: material-absorption bounds~\cite{miller_fundamental_2016, shim2019fundamental, miller2015shape, miller2017limits, michon2019limits, liu2018optimal, yang2018maximal,yang2017low, ivanenko2019optical,dias_fundamental_2019,nordebo2019optimal} that rectify the two drawbacks of the channel approaches. These bounds identify upper limits to responses including cross-sections~\cite{miller_fundamental_2016}, local density of states~\cite{shim2019fundamental}, NFRHT~\cite{miller2015shape}, and 2D-material response~\cite{miller2017limits} that are determined by the lossiness of the material comprising the scattering body. The independence from channels provides generality and convenience, but with the key drawback that they do not account for necessary radiative damping. Very recently, for the special case of incoherent thermal or zero-point-field excitations, radiative and absorptive losses are separately identified using the $\mathbb{T}$ operator, yielding upper bounds for incoherent response functions~\cite{molesky2019t,molesky2019fundamental,venkataram2019fundamental}.

In this work, we identify a \emph{single} constraint that incorporates the cooperative effects of absorptive and radiative losses at any level of coherence. The constraint is the volume-integral formulation of the optical theorem (\secref{GF}), which is an energy-conservation constraint that imposes the condition that absorption plus scattered power equal extinction, for any incident field. Channel bounds distill in essence to loosening this constraint to an inequality that scattered power is bounded above by extinction. Material-absorption bounds distill to loosening the optical-theorem constraint to an inequality that absorbed power is bounded above by extinction. Our key innovation is the recognition that one can retain the entire constraint, and enforce the requirement that the sum of absorption and scattered power equal extinction. We describe the use of Lagrangian duality to solve the resulting optimization problems, ultimately yielding very general bounds to arbitrary response functions. For the important case of plane-wave scattering (\secref{plane-wave}), we derive explicit bound expressions and also identify an important application: perfect absorbers. We show that our framework enables predictions of the minimal scatterer thicknesses at which perfect or near-perfect absorption may be possible, thicknesses much larger than any previous framework predicted. Our bounds explicitly account for the precise form of the incident waves; for a given material and designable region, then, we can treat the illumination-field degrees of freedom as the variables and identify the optimal incoming-wave excitation (\secref{IF}). As one example, we show that in certain parameter regimes the extinction of an unpatterned sphere under the optimal illumination field exceeds the upper bound under plane-wave excitation, which means that, as long as the incident field is a plane wave, there is no patterning of any kind that can reach the same power-response level of the optimal illumination. In the final section (\secref{Disc}), we discuss the simplicity with which our framework can be applied to numerous other scenarios, and discuss remaining open problems.
\begin{figure} [t!]
	\includegraphics[width=0.45\textwidth]{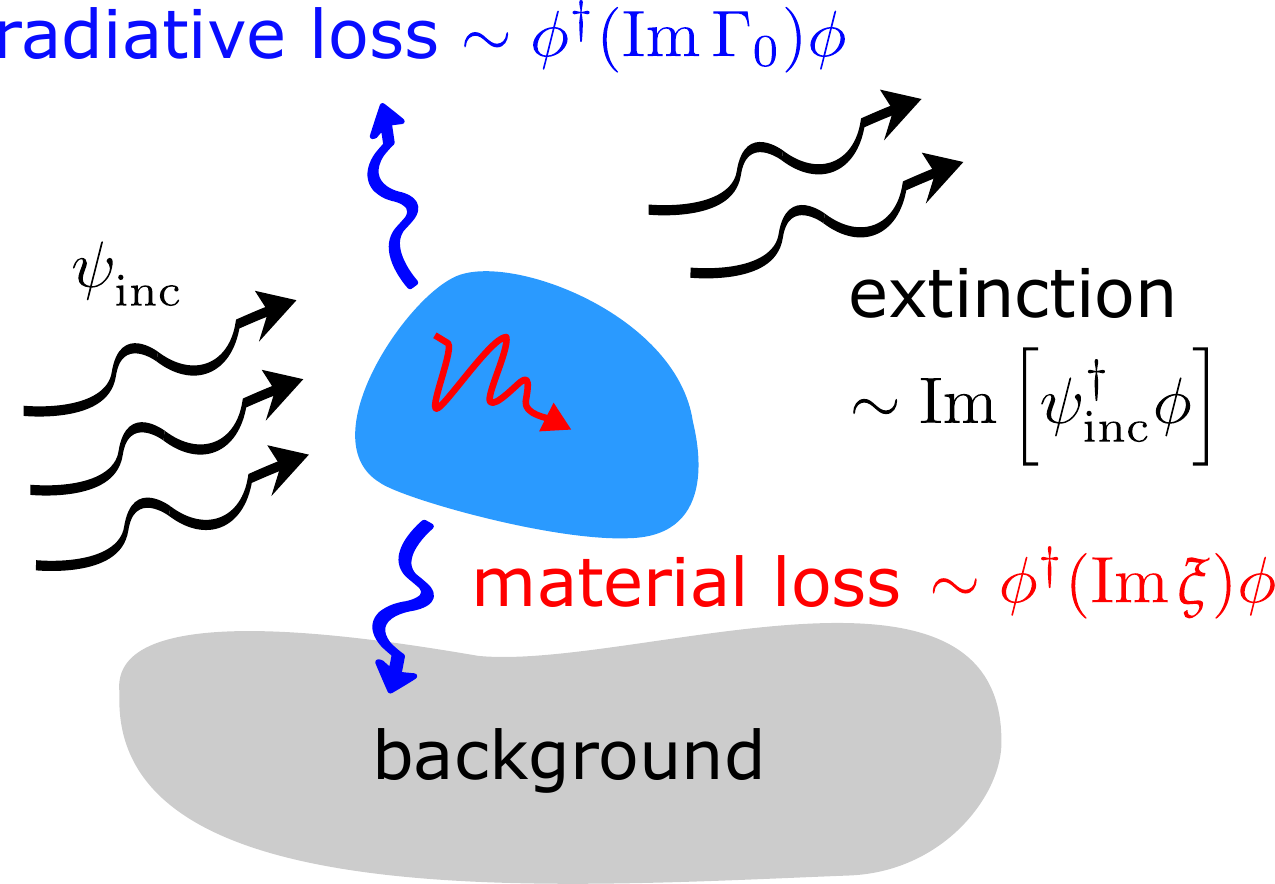}
        \caption{Illustration of the two loss mechanisms in electromagnetic scattering. An incident field $\psi_{\rm inc}$ induces polarization currents $\phi$ in the scatterer. Energy dissipated inside the material corresponds to material loss, determined by the operator $\Im{\xi}$, which equals $\Im \chi / |\chi|^2$ for a linear isotropic susceptibility $\chi$. Energy coupled to the background, into far-field or near-field power exchange, corresponds to radiative loss, determined by the operator $\Im{\Gamma_0}$, where $\Gamma_0$ represents the background (e.g. free-space) Green's function. Total extinction is the sum of the two and is linear in $\phi$, as dictated by the optical theorem.} 
	\label{fig:figure1}
\end{figure} 

Given the variety of bounds in Refs.~\cite{Yaghjian1996,hamam_coupled-mode_2007,kwon2009optimal, ruan2011design, liberal2014least, liberal2014upper,miller_fundamental_2016,miller2017limits,yang2017low,pendry1999radiative, thongrattanasiri_complete_2012,hugonin_fundamental_2015,miller2015shape, rahimzadegan2017fundamental,liu2018optimal,yang2018maximal,michon2019limits, nordebo2019optimal, shim2019fundamental, ivanenko2019optical,dias_fundamental_2019}, as well as those contained here, a natural question is whether the bounds we present here are the ``best possible'' bounds, or whether they will be ``superseded'' later. We argue that ultimately there will be no ``best'' single bound, but rather a general theory comprising different bounds at different levels of a priori information that is known about a given problem. Useful analogies can be made to information theory, where Shannon's bounds~\cite{Shannon1948,Shannon1949} were not a final conclusion but instead initiated an entire field of inquiry~\cite{Cover1999}, as well as the theory of composite materials, where early studies into properties of simple isotropic composites~\cite{Hashin1962} blossomed into a broad theoretical framework with bounds that vary with the amount of information known about the problem of interest~\cite{Milton2002,Bergman1980,Milton1980,Gibiansky1993,Cherkaeva1998,Kern2020}. In electromagnetism and optics, previous bounds~\cite{Yaghjian1996,hamam_coupled-mode_2007,kwon2009optimal, ruan2011design, liberal2014least, liberal2014upper,miller_fundamental_2016,miller2017limits,yang2017low,pendry1999radiative, thongrattanasiri_complete_2012,hugonin_fundamental_2015,miller2015shape, rahimzadegan2017fundamental,liu2018optimal,yang2018maximal,michon2019limits, nordebo2019optimal, shim2019fundamental, ivanenko2019optical,dias_fundamental_2019} utilized information about \emph{either} the number of available scattering channels \emph{or} the material loss rate; in this work, we present the first bounds that combine the two, unifying the previous disconnected threads. A useful indicator of whether future bounds, with possibly more known information, will significantly alter these results is to test whether physical designs can approach these bounds, as it can almost never be guaranteed (in any field) whether given bounds are precisely achievable by real physical implementations. As we show in \secref{plane-wave}, in the quest for ultrathin perfect absorbers, physical designs can approach the new bounds within a factor of two, suggesting minimal opportunity for later revision.

\section{General Formalism}
\label{sec:GF}
Our central finding is a set of upper bounds to maximal single-frequency response. The problem of interest is to optimize any electromagnetic response function $f$ subject only to Maxwell's equations, while allowing for arbitrary patterning within a prescribed region of space. However, Maxwell's equations represent a nonconvex and highly complex constraint for which global bounds are not known. Instead, we use the optical theorem, and in particular a volume-integral formulation of the optical theorem, as a simple quadratic constraint for which global bounds can be derived. We start with the volume-integral version of Maxwell's equations, which provide a simple and direct starting point to derive the optical theorem (Sec. IIA). The optical-theorem constraint is quadratic, and we discuss how many previous result can be derived from weaker forms of the constraint. Then in Sec. IIB we use the formalism of Lagrangian duality to derive a single general bound expression, \eqref{gen_bound}, from which many specialized results follow. In Sec. IIC we consider canonical electromagnetic response functions: absorption, scattering, extinction, and local density of states. Throughout, for compact general expressions we use six-vector notation with Greek letters denoting vectors and tensors: $\psi$ for fields, $\phi$ for polarization currents, $\chi$ for the susceptibility tensor (which in its most general form can be a nonlocal, inhomogeneous, bianisotropic, 6$\times$6 tensor operator~\cite{chew1995waves}), and we use dimensionless units for which the vacuum permittivity and permeability equal 1, $\varepsilon_0 = \mu_0 = 1$. The six-vector fields and polarization currents are given by
\begin{align}
\psi = \begin{pmatrix} \Ev \\ \Hv \end{pmatrix}, \quad
\phi = \begin{pmatrix} \Pv \\ \Mv \end{pmatrix}.
\end{align}

\subsection{Optical Theorem Constraint}
The optical theorem manifests energy conservation: the total power taken from an incident field must equal the sum of the powers absorbed and scattered. As discussed below, the key version of the optical theorem that enables a meaningful constraint is the version that arises from the volume equivalence principle. This principle enables the transformation of the differential Maxwell equations to a volume-integral form. It states that any scattering problem can be separated into a background material distribution (not necessarily homogeneous), and an additional distributed ``scatterer'' susceptibility. The total fields $\psi$ are given by the fields incident within the background, $\psi_{\rm inc}$, plus scattered fields $\Gamma_0 \phi$ that arise from polarization currents $\phi$ induced in the volume of the scatterer, where $\Gamma_0$ is the background-Green's-function convolution operator. For simplicity in the optical theorem below, we define a variable $\xi$ that is the negative inverse of the susceptibility operator, $\xi = -\chi^{-1}$. With this notation, the statement that the total field equals the sum of the incident and scattered fields can be written: $-\xi \phi = \psi_{\rm inc} + \Gamma_0 \phi$. Rearranging to have the unknown variables on the left-hand side and the known variables on the right-hand side yields the volume-integral equation (VIE),
\begin{align}
\left[ \Gamma_0 + \xi \right] \phi = -\psi_{\rm inc}.
\label{eq:vie}
\end{align}
We generally allow for $\chi$ to be nonlocal, as arises in the extreme near field~\cite{Yang2019} and in 2D materials~\cite{Fallahi2015}; when $\chi$ is local and can be written $\chi(\vect{x,x'}) = \chi(\vect{x})\delta(\vect{x}-\vect{x'})$, \eqref{vie} becomes a standard VIE~\cite{chew1995waves}: $\int_V\Gamma_0(\vect{x},\vect{x'})\phi(\vect{x'})\text{d}\vect{x'} -\chi^{-1}(\vect{x})\phi(\vect{x}) = - \psi_{\rm inc}(\vect{x})$, where $V$ is the volume of the scatterer.

The volume-integral-equation optical theorem can be derived from \eqref{vie} by taking the inner product of \eqref{vie} with $\phi$ (denoted $\phi^\dagger$), multiplying by $\omega/2$, and taking the imaginary part of both sides of the equation, yielding:
\begin{align}
\underbrace{\frac{\omega}{2} \phi^\dagger \left(\ImGO\right) \phi}_{\Pscat} + \underbrace{\frac{\omega}{2} \phi^\dagger \left(\Im \xi \right) \phi}_{\Pabs}  = \underbrace{\frac{\omega}{2} \Im \left(\psi_{\rm inc}^\dagger \phi\right)}_{\Pext},
\label{eq:optthm}
\end{align}
where the inner product is the integral over the volume of the scatterer. Within the optical theorem of \eqref{optthm}, we identify the three terms as scattered, absorbed, and extinguished power, respectively~\cite{polimeridis2014computation,kong1972theorems}, as depicted in \figref{figure1}. The operator $\ImGO$ represents power radiated into the background, into near-field or, more typically, far-field scattering channels. For any background materials, $\ImGO$ can be computed by standard volume-integral (or discrete-dipole-approximation) techniques~\cite{chew1995waves,Purcell1973}, and when the background is lossless over the scatterer domain it is nonsingular and simpler to compute~\cite{Reid2017subm}. In vacuum, the operator can be written analytically for high-symmetry domains. It is a positive semidefinite operator because the power radiated by any polarization currents must be nonnegative in a passive system. The second term with $\Im\xi$ represents absorbed power: work done by the polarization currents on the total fields. In terms of the susceptibility, $\Im \xi = \chi^{-1} \left(\Im \chi\right) \left(\chi^\dagger\right)^{-1}$; for scalar material permittivities, it simplifies to $\Im \chi/|\chi|^2$, which is the inverse of a material ``figure of merit'' that has appeared in many material-loss bounds~\cite{miller_fundamental_2016, shim2019fundamental, miller2017limits}. The operator $\Im \xi$ is positive definite for any material without gain~\cite{chew1995waves,welters2014speed}. Finally, the third term is the imaginary part of the overlap between the incident field and the induced currents, which corresponds to extinction (total power taken from the incident fields). 

While no simplification of Maxwell's equations will contain every possible constraint, the optical theorem of \eqref{optthm} has four key features: (1) it contains \emph{both} the powers radiated ($\Pscat$) and absorbed ($\Pabs$) by the polarization currents in a single expression, (2) it is a quadratic constraint that is known to have ``hidden'' convexity for any quadratic objective function~\cite{ben-tal_hidden_1996}, (3) it enforces power conservation in the scattering body, and (4) it incorporates information about the material composition of the scatterer, and possibly a bounding volume containing it, while independent of any other patterning details.

The optical-theorem constraint of \eqref{optthm} constrains the polarization-current vector $\phi$ to lie on the surface of a high-dimensional ellipsoid whose principal axes are the eigenvectors of $\ImGO+\Im\xi$ and whose radii are constrained by the norm of $\psiinc$. In the {\SM} we show that all previous channel or material-loss bounds discussed in the Introduction can be derived by applying weaker versions of \eqref{optthm}. Channel bounds can be derived by loosening \eqref{optthm} to the inequality $\Pscat \leq \Pext$, without the absorption term (but implicitly using the fact that absorbed power is nonnegative). Material-loss bounds can be derived by loosening \eqref{optthm} to the inequality $\Pabs \leq \Pext$, without the scattered-power term (but using the fact that scattered power is nonnegative). Of course, including both constraints simultaneously can only result in equal or tighter bounds.

\subsection{Optimization Formalism}
\label{sec:OF}
Any electromagnetic power-flow objective function $f$ is either linear or quadratic in the polarization currents $\phi$. Under a given basis, it can be generically written as $f(\phi) = \phi^\dagger \mathbb{A} \phi + \Im \left(\beta^\dagger \phi\right)$, where $\mathbb{A}$ is a Hermitian matrix and $\beta$ is any six-vector field on the scatterer domain. The same basis is used to discretize $\psi_{\rm inc}$, $\Im\xi$, and $\ImGO$, where the last two are now positive semi-definite matrices. Then the maximal $f$ that is possible for any scatterer is given by the optimization problem:                
\begin{equation}
\begin{aligned}
& \underset{\phi}{\text{maximize}}
& & f(\phi) = \phi^\dagger \mathbb{A} \phi +  \Im\left(\beta^\dagger \phi\right) \\
& \text{subject to}
& & \phi^\dagger \left\{\Im \xi + \ImGO \right\} \phi = \Im \left(\psi_{\rm inc}^\dagger \phi\right).
\end{aligned}
\label{eq:general-formalism}
\end{equation}
This is a quadratic objective with a single quadratic constraint, which is known to have strong duality~\cite{boyd2004convex}. If we follow standard convex-optimization conventions and consider as our ``primal'' problem that of \eqref{general-formalism}, but instead written as a minimization over the negative of $f(\phi)$, then strong duality implies that the maximum of the corresponding Lagrangian dual functions equals the minimum of the primal problem, and thus the maximum of \eqref{general-formalism}. By straightforward calculations, the dual function is
\begin{align}
g(\nu)=
\begin{cases}
-\frac{1}{4}(\beta + \nu \psiinc)^\dagger \BB^{-1}(\nu)(\beta + \nu \psiinc) & \nu > \nu_0 
\\
-\infty, & \nu< \nu_0
\end{cases}
\label{eq:gnu}
\end{align}
where $\nu$ is the dual variable, $\BB(\nu) = -\mathbb{A}+\nu(\Im\xi+\ImGO)$ and $\nu_0$ is the value of $\nu$ for which the minimum eigenvalue of $\BB(\nu_0)$ is zero. (The definiteness of $\ImGO$ and $\Im\xi$ ensure there is only one $\nu_0$, cf. {\SM}.) At $\nu=\nu_0$, some care is needed to evaluate $g(\nu_0)$ because the inverse of $\BB(\nu_0)$ does not exist (due to the 0 eigenvalue). If $\beta + \nu_0\psiinc$ is in the range of $\BB(\nu_0)$, then $g(\nu_0)$ takes the value of the first case in \eqref{gnu} with the inverse operator replaced by the pseudo-inverse; if not, then $g(\nu_0)\rightarrow -\infty$. (Each scenario arises in the examples below.) By the strong duality of \eqref{general-formalism}, the optimal value of the dual function, \eqref{gnu}, gives the optimal value of the ``primal" problem, \eqref{general-formalism} (accounting for the sign changes in converting the maximization to minimization). In the {\SM} we identify the only two possible optimal values of $\nu$: $\nu_0$, defined above, or $\nu_1$, which is the stationary point for $\nu>\nu_0$ at which the derivative of $g(\nu)$ equals zero. Denoting this optimal value $\nu^*$, we can write the maximal response as:
\begin{equation}
f_{\rm max} = \frac{1}{4}(\beta + \nu^* \psiinc)^\dagger  \left[-\mathbb{A}+\nu^*(\Im\xi+\ImGO)\right]^{-1}(\beta + \nu^* \psiinc).
\label{eq:gen_bound}
\end{equation}
Although \eqref{gen_bound} may appear abstract, it is a general bound that applies for any linear or quadratic electromagnetic response function, from which more domain-specific specialized results follow.

\subsection{Power quantities and LDOS}
If one wants to maximize one of the terms already present in the constraint, i.e. absorption, scattered power, or extinction, then the $\mathbb{A}$ and $\beta$ terms take particularly simple forms (cf. {\SM}), leading to the bounds:
\begin{align}
\Pext &\leq \frac{\omega}{2} \psiinc^\dagger \left(\Im\xi+\ImGO\right)^{-1} \psiinc \label{eq:ext-gen}\\
\Pabs &\leq \frac{\omega}{2}\frac{\nu^{*2}}{4} \psiinc^\dagger[(\nu^*-1)\Im\xi+\nu^*\ImGO]^{-1} \psiinc \label{eq:abs-gen} \\
\Pscat &\leq \frac{\omega}{2}\frac{\nu^{*2}}{4} \psiinc^\dagger[\nu^*\Im\xi + (\nu^* - 1)\ImGO]^{-1} \psiinc. \label{eq:sca-gen}
\end{align}
where $\nu^*$ is the dual-variable numerical constant (\SM).

Bounds on LDOS represent maximal spontaneous-emission enhancements~\cite{novotny2012principles, liang2013formulation, purcell1946resonance, taflove2013advances,xu2000quantum}. Total (electric) LDOS, $\rho_{\rm tot}$, is proportional to the averaged power emitted by three orthogonally polarized and uncorrelated unit electric dipoles~\cite{wijnands1997green, martin_electromagnetic_1998, d2004electromagnetic, joulain2003definition}. It can be separated into a radiative part, $\rho_{\rm rad}$, for far-field radiation, and a non-radiative part, $\rho_{\rm nr}$, that is absorbed by the scatterer~\cite{jackson1999classical}. Exact but somewhat cumbersome LDOS bounds for arbitrary materials are derived from \eqref{gen_bound} in the {\SM}; for nonmagnetic materials, the bounds simplify to expressions related to the maximum power quantities given in \eqrefrange{ext-gen}{sca-gen}: 
\begin{align}
\rho_{\rm tot} &\leq \frac{2}{\pi\omega^2}\sum_j P^{\rm max}_{\text{ext}, j} + \rho_0 \label{eq:rhotot_bnd} \\ 
\rho_{\rm nr} &\leq \frac{2}{\pi\omega^2}\sum_j P^{\rm max}_{\text{abs}, j} \\
\rho_{\rm rad} &\leq \frac{2}{\pi\omega^2}\sum_j P^{\rm max}_{\text{sca}, j} + \rho_0,
\label{eq:rhorad_bnd}
\end{align}
where $\rho_0$ is the electric LDOS of the background material, and takes the value of $\frac{\omega^2}{2\pi^2c^3}$ for a scatterer in vacuum~\cite{joulain2005surface}. The summation over $j={1,2,3}$ accounts for three orthogonally polarized unit dipoles.  As shown in the \SM, our bound is tighter than previous bounds on LDOS \cite{miller_fundamental_2016}. In the extreme near field, where material loss dominates, our bound agrees with the known material-loss bound~\cite{miller_fundamental_2016}.

The bounds of \eqrefrange{gen_bound}{rhorad_bnd} are sufficiently general to allow for arbitrary material composition (inhomogeneous, nonlocal, etc.), in which case the bounds require computations involving the $\ImGO$ and $\Im\xi$ matrices. In the {\SM}, we provide a sequence of simplifications, showing step-by-step the increasingly simplified bounds that arise under restrictions of the incident field, material, or bounding volumes involved. In the next section, we consider the important case in which a plane wave is incident upon an isotropic nonmagnetic medium.

\begin{figure*}[htb]
	\includegraphics[width=\textwidth]{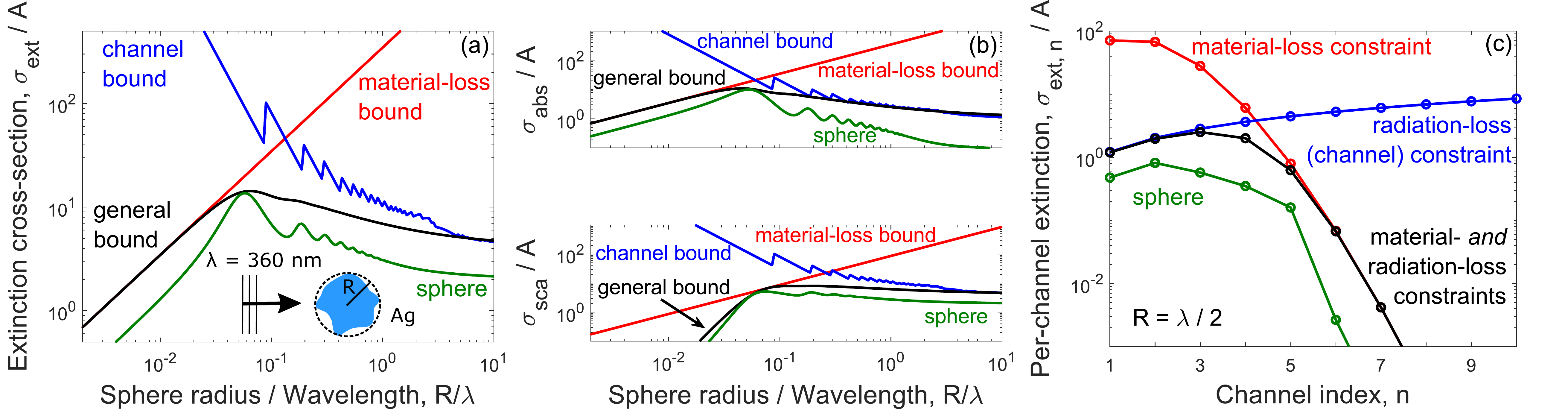}
	\centering
	\caption{Plane wave of wavelength $\lambda=360$ nm scattering from a finite Ag \cite{johnson1972optical} scatterer, enclosed by a spherical bounding volume with radius $R$. The channel bound is heuristically regularized by ignoring small-scattering high-order channels. All cross-sections are normalized by geometric cross-section $A$. (a). Bound of extinction cross-section for different $R$. The general bound regularizes divergence in previous bounds and are tighter for wavelength-scale sizes. (b) Similar behavior is observed in the bounds for scattering and absorption cross-sections. (c) Per-channel extinction cross section $\sigma_{\text{ext},n}$ (defined in SM) for $R=\lambda/2$. Low-order scattering channels are dominated by radiative loss, while high-order scattering channels are dominated by material loss.}
	\label{fig:figure2}
\end{figure*}

\section{Plane-wave scattering}
\label{sec:plane-wave}
A prototypical scattering problem is that of a plane wave in free space incident upon an isotropic (scalar susceptibility), nonmagnetic scatterer. The assumption of a scalar susceptibility introduces important simplifications into the bounds. The matrix $\Im \xi$ is then a scalar multiple of the identity matrix $\mathbb{I}$,
\begin{align}
\Im \xi = \frac{\Im \chi}{|\chi|^2} \mathbb{I},
\end{align}
and is therefore diagonal in any basis that diagonalizes $\ImGO$, simplifying the matrix-inverse expressions in the bounds of \eqrefrange{gen_bound}{rhorad_bnd}. For nonmagnetic materials, the polarization currents $\phi$ comprise nonzero electric polarization currents $\vect{P}$ only, such that the 6$\times$6 Green's tensor $\Gamma_0$ can replaced by its 3$\times$3 electric-field-from-electric-current sub-block $\Gbb^{\rm EE}_0$, and only the electric part $\vect{E}_{\rm inc}$ of the incident field $\psi_{\rm inc}$ enters the bounds of \eqrefrange{ext-gen}{sca-gen}. Because $\Im \Gbb^{\rm EE}_0$ is positive-definite, we can simplify its eigendecomposition to write $\Im\Gbb^{\rm EE}_0 = \Vbb\Vbb^\dagger$, where the columns of $\Vbb$, which we denote $\vv_i$, form an orthogonal basis of polarization currents. They are normalized such that the set of $\vv_i^\dagger \vv_i$ are the eigenvalues of $\Im \Gbb^{\rm EE}_0$ and represent the powers radiated by unit-normalization polarization currents. More simply, the $\vv_i$ span the space of scattering channels and the eigenvalues $\rho_i$ represent corresponding radiated powers.  

An incident propagating plane wave (or any wave incident from the far field, cf. {\SM}) can be decomposed in the basis $\Vbb$. We write the expansion as $\vect{E}_{\rm inc} = \frac{1}{k^{3/2}}\sum_i e_i \vv_i$, where the $e_i$ are the expansion coefficients, and we factor out the free-space wavenumber $k$ to simplify the expressions below. Inserting the eigendecomposition of $\Im \Gbb^{\rm EE}_0$ and the plane-wave expansion in this basis into \eqrefrange{ext-gen}{sca-gen} gives general power bounds for plane-wave scattering:
\begin{align}
P_{\rm ext} &\leq \frac{\lambda^2}{8\pi^2}\sum_i |e_i|^2\frac{\rho_i}{\Im{\xi} + \rho_i} \label{eq:ext-cha-bound} \\
P_{\rm abs} &\leq \frac{\lambda^2}{8\pi^2}\frac{\nu^{*2}}{4}\sum_i |e_i|^2\frac{\rho_i}{(\nu^*-1)\Im{\xi}+\nu^*\rho_i} \label{eq:abs-cha-bound} \\
P_{\rm sca} &\leq \frac{\lambda^2}{8\pi^2}\frac{\nu^{*2}}{4}\sum_i |e_i|^2\frac{\rho_i}{\nu^*\Im{\xi}+(\nu^*-1)\rho_i}. \label{eq:sca-cha-bound}
\end{align}
The variable $\nu^*$ is the optimal dual variable discussed above; its value can be found computationally via a transcendental equation given in the {\SM}. The bounds of \eqrefrange{ext-cha-bound}{sca-cha-bound} naturally generalize previous channel bounds ($\sim \lambda^2$) and material-absorption bounds ($\sim 1/\Im \xi = |\chi|^2 / \Im \chi$); in the {\SM}, we prove that removing either dissipation pathway results in the previous expressions. 

The bounds of \eqrefrange{ext-cha-bound}{sca-cha-bound} require knowledge of the eigenvalues of $\Im\Gbb^{\rm EE}_0$, and thus the exact shape of the scattering body, to compute the values of $\rho_i$. However, analytical expressions for $\rho_i$ are known for high-symmetry geometries, and a useful property of the optimization problem of \eqref{general-formalism} is that its value is bounded above by the same problem embedded in a larger bounding domain. (It is always possible for the currents in the ``excess'' region to be zero.) In the following two sub-sections we consider the two possible scenarios one can encounter: (a) scattering by finite-sized objects, which can be enclosed in spherical bounding surfaces, and (b) scattering by extended (e.g. periodic) objects, which can be enclosed in planar bounding surfaces.

\subsection{Finite-sized scatterers}
\label{sec:fss}
Finite-sized scatterers can be enclosed by a minimal bounding sphere with radius $R$, as in the inset of Fig. 2(a). The basis functions $\vv_i$ are vector spherical waves (VSWs), representing orthogonal scattering channels, with exact expressions given in the {\SM}. The state labels $i$ can be indexed by the triplet $i=\{n,m,j\}$ where $n=1,2,...$ is the total angular momentum, $m=-n,...,n$ is the $z$-directed angular momentum, and $j=1,2$ labels two polarizations. In this basis the expansion coefficients of a plane wave are given by $|e_i|^2=\pi(2n+1)\delta_{m,\pm 1}|E_0|^2$, where $E_0$ is the plane-wave amplitude. We show in the {\SM} that the values $\rho_i$ are given by integrals of spherical Bessel functions. With these expressions, bounds for extinction, scattering, and absorption cross-sections are easily determined from \eqrefrange{ext-cha-bound}{sca-cha-bound} after normalization by plane wave intensity $|E_0|^2/2$.

In \figref{figure2}, we compare cross-section bounds derived from \eqrefrange{ext-cha-bound}{sca-cha-bound} to the actual scattering properties of a silver sphere (permittivity data from \citeasnoun{johnson1972optical}) at wavelength $\lambda = \SI{360}{nm}$. We choose \SI{360}{nm} wavelength because it is close to the surface-plasmon resonance of a silver sphere, simplifying comparisons (instead of requiring inverse design for every data point). We also include the previously derived channel~\cite{ruan2011design} and material-absorption \cite{miller_fundamental_2016} bounds for comparison, and in each case one can see that our general bounds are significantly ``tighter'' (smaller) than the previous bounds, except in the expected small- and large-size asymptotic limits. At a particular radius, the scattering response even reaches the general bound. In \figref{figure2}(c), we fix the radius at a half-wavelength and depict the per-channel contributions to the extinction bounds in the  radiation-loss-only, material-loss-only, and tandem-loss constraint cases. Higher-order channels have increasingly smaller radiative losses (causing unphysical divergences discussed below), such that material loss is the dominant dissipation channel. Conversely, material-loss-only constraints are inefficient for lower-order channels where radiative losses dominate. Incorporating both loss mechanisms removes the unphysical divergence, accounts for radiative losses, and sets the tightest bound among the three across all channels. 

For structures smaller than roughly \SI{10}{nm}, instead of bulk permittivity data one must employ a nonlocal model of the permittivity~\cite{Yang2019}, which can still be subjected to bounds but requires modified techniques for modeling the polarization currents~\cite{miller2017limits}. We retain small ratios of size to wavelength throughout the paper, such as in \figref{figure2}, in order to observe the relevant scalings of the classical model, and because for mid-infrared plasmonic materials the lineshapes are quite similar while all sizes are scaled beyond \SI{10}{nm}.

Technically, the channel bound diverges for any finite-sized scatterer, and the blue solid line in \figref{figure2}(a) should be infinitely high. To obtain a reasonable finite value, we only incorporate channels for which the sphere scattering contributions are greater than 1\% of the maximal response. Yet requiring knowledge of the specific scattering structure to compute the upper limit highlights a key drawback of the channel bounds. This empirical threshold is responsible for two artifacts in the presented channel bounds. First, it results in a step-like behavior which is most prominent at small radii, where only a handful channels contribute. At each radius where a new channel is introduced for consideration (based on this threshold), there is an unphysical increase in the bound due to the larger power available for scattering, absorption, etc.. Such behavior is somewhat smoothed at large radii, where the contribution from each new channel is subsumed by the large number of existing channels. Second, as we show in the SM, there can potentially be large contributions from channels beyond this threshold. The arbitrary cut-off results in inaccurate and unphysical \emph{under}estimates of the cross-sections, which is mostly noticeable in the large size limit of Figs.~\ref{fig:figure2}(a,b), where the channel bound appears to be slightly smaller than the general bound. The only way to avoid such artifacts would be to include all channels, in which case the channel bounds trivialize to infinite value for any radius.

\begin{figure*}[htb]
	\includegraphics[width=1\textwidth]{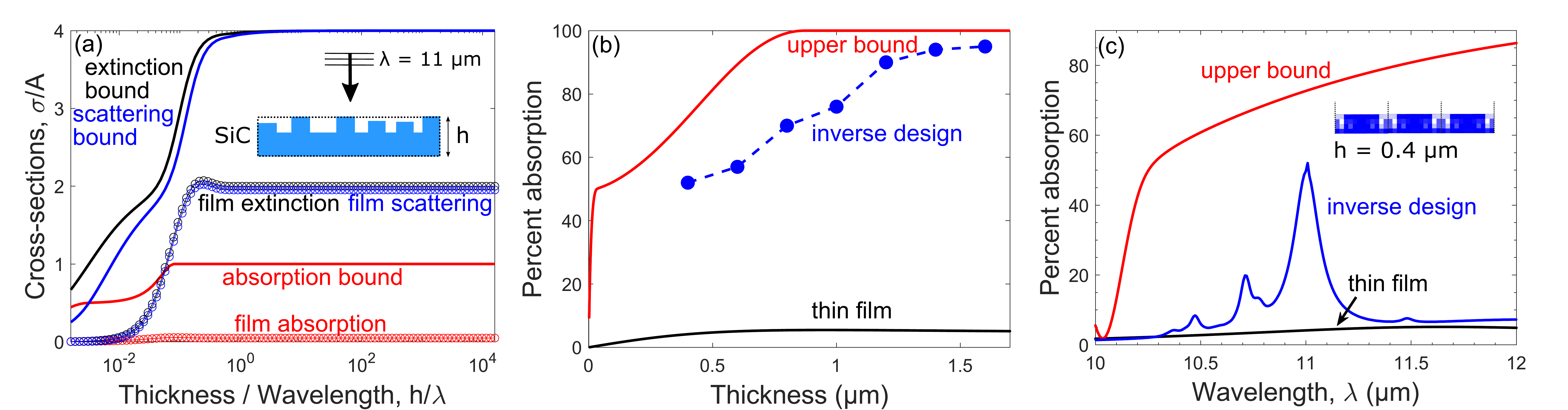}
	\centering
        \caption{Arbitrarily patterned SiC scatterer with maximum thickness $h$ excited by a plane wave at normal incidence and $\lambda=\SI{11}{\micro\meter}$ wavelength, where SiC is polaritonic. (a) Bounds for extinction, scattering, and absorption, compared to their values for a planar SiC \cite{francoeur2010spectral} film. (b) Inverse-designed SiC metasurfaces (blue markers), at varying thicknesses, achieve absorption levels at 64--95\% of the global bounds (red), suggesting the bounds are ``tight'' or nearly so. (c) Absorption spectrum of ultra-thin absorber from (b) with thickness $h=\SI{0.4}{\micro\meter}$. (Inset: inverse-design structure; blue represents SiC, white represents air.) At the target wavelength, the absorption of the inverse-designed structure is more than ten times that of the thin film, and reaches 72\% of the bound.}
	\label{fig:figure3}
\end{figure*}

\subsection{Extended scatterers}
The second possible scenario is scattering from an infinitely extended (e.g. periodic) scatterer. Such scatterers can always be enclosed by a minimal planar ``film'' bounding volume with thickness $h$, as in the inset of Fig. 3(a). Then the basis functions $\vv_i$ of $\Im\Gbb^{\rm EE}_0$ are known to be propagating plane waves with wave vector $\vect{k} = k_x\vect{\hat{x}} + k_y\vect{\hat{y}} + k_z\vect{\hat{z}}$. Now the index $i$ maps to the triplet $i=\{s,p,\vect{k}_\parallel\}$, where $s=\pm$ denotes even and odd modes, $p=M,N$ denotes TE and TM polarizations, and $\vect{k}_\parallel=k_x\vect{\hat{x}} + k_y\vect{\hat{y}}$ denotes the surface-parallel wave vector. In the {\SM} we provide the expressions for $\vv_i$, and show that the eigenvalues $\rho_i$ are given by
\begin{align}
\rho_{\pm, s}(\vect{k}_\parallel) = 
\begin{cases}
\frac{k^2h}{4k_z}(1\pm\frac{\sin(k_zh)}{k_zh}) \quad &s = \textrm{TE}\\
\frac{k^2h}{4k_z}(1\pm\frac{\sin(k_zh)}{k_zh})\mp\frac{\sin(k_zh)}{2}. \quad &s = \textrm{TM}
\end{cases}
\end{align}
The incident wave itself has nonzero expansion coefficients for basis functions with the same parallel wave vector, and is straightforward to expand: $|e_i|^2=2k_zk\delta_{p,p'}|E_0|^2$, where $p'$ is the incident polarization, $E_0$ is the plane wave amplitude, and $k=|\vect{k}|$. The optimal polarization currents only comprise waves with parallel wave vector identical to that of the incident wave, simplifying the final bounds. Normalizing the bounds of \eqrefrange{ext-cha-bound}{sca-cha-bound} by the $z$-directed plane-wave intensity, $|E_0|^2k_z/2k$, gives cross-sections bounds for extended structures:
\begin{align}
\sigma_{\rm ext} / A &\leq 2\sum_{s=\pm} \frac{\rho_{s, p'}}{\Im{\xi} + \rho_{s, p'}} \label{eq:ext-cha-extend} \\
\sigma_{\rm abs} / A &\leq \frac{\left(\nu^{*}\right)^2}{2}\sum_{s=\pm}\frac{\rho_{s, p'}}{(\nu^*-1)\Im{\xi}+\nu^*\rho_{s, p'}} \label{eq:abs-cha-extend} \\
\sigma_{\rm sca} / A &\leq \frac{\left(\nu^{*}\right)^2}{2}\sum_{s=\pm}\frac{\rho_{s, p'}}{\nu^*\Im{\xi}+(\nu^*-1)\rho_{s, p'}}, \label{eq:sca-cha-extend}
\end{align}
where $A$ is the total surface area and $\rho_{s,p'}$ denotes the radiation loss by a scattering channel with parity $s$, polarization $p'$, and parallel wave vector $\vect{k}_\parallel$. Again, the use of a high-symmetry bounding volume results in analytical expressions that are easy to compute.

Figure \ref{fig:figure3}(a) compares the upper bounds for the normalized cross-sections with the cross-sections of SiC thin films at normal incidence and wavelength $\lambda=\SI{11}{\micro\meter}$, where SiC supports phonon-polariton modes. One can see that the bounds indicate that scattering, absorption, and extinction must all be small at sufficiently small thicknesses, and crossover to near-maximal possible values at roughly one-tenth of the wavelength.

A key question for any bound is whether it is achievable with physical design. In order to test the feasibility of our bounds, we utilize inverse design~\cite{Jameson1998,Sigmund2003,Lu2011,Jensen2011,Miller2012a,Lalau-Keraly2013,Ganapati2014,aage2017giga}, a large-scale computational optimization technique for discovering optimal configurations of many design parameters, to design patterned SiC films that approach their bounds. We use a standard ``topology-optimization'' approach~\cite{Sigmund2003,Miller2012a} in which the material is represented by a grayscale density function ranging from 0 (air) to 1 (SiC) at every point, and derivatives of the objective function (absorption, in this case) are computed using adjoint sensitivities. We prioritize feasibility tests---are the bounds achievable, in theory?---over the design of easy-to-fabricate structures. To this end we utilize grayscale permittivity distributions, which in theory can be mimicked by highly subwavelength patterns of holes, but in practice would be difficult to fabricate. Recently developed techniques~\cite{Christiansen2019} are able to identify binary polaritonic structures that come quite close to their grayscale counterparts for many applications, and give confidence that binary structures with similar performance levels to those presented here can be discovered. We give algorithmic details for our inverse-design procedure in the {\SM}.

\begin{figure*}[htb]
	\includegraphics[width=\textwidth]{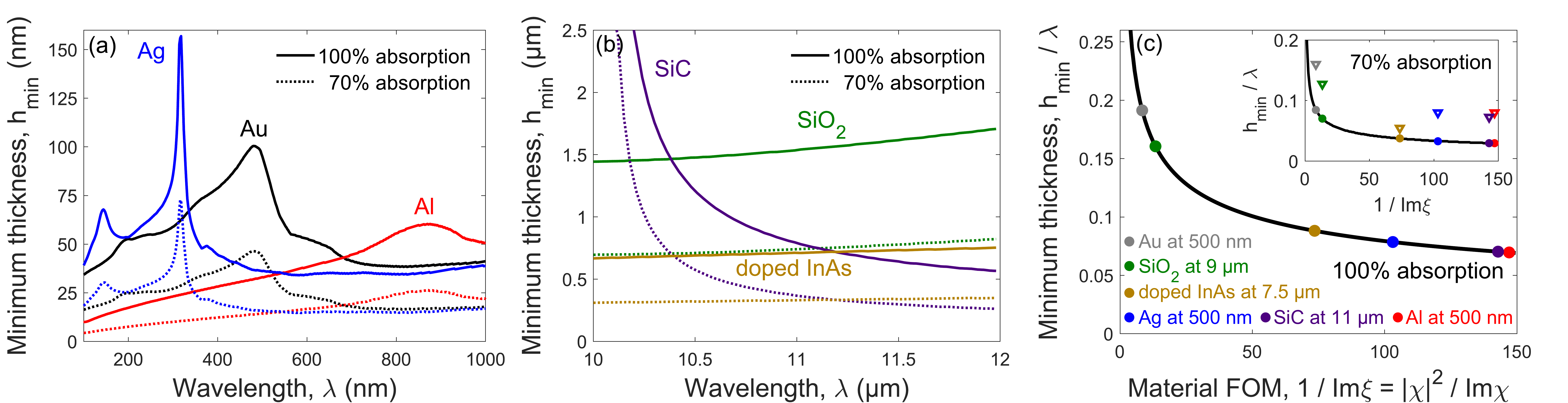}
	\centering
        \caption{Minimum thickness required for a perfect absorber to reach 70\% and 100\% absorption rate under normal incidence for typical materials that are polaritonic at (a) visible~\cite{palik1998handbook} and (b) infrared wavelengths~\cite{law2013all,francoeur2010spectral,popova1972optical}. (c) A universal curve showing minimum possible thicknesses for 100\% absorption as a function of perfect-absorber material figures of merit (FOM), given by $1/\Im\xi = |\chi|^2 / \Im \chi$. The same curve is shown in the inset for 70\% absorption where inverse-designed structures (triangular markers) demonstrate thicknesses within 1.5--2.7X of the bound.}
	\label{fig:figure4}
\end{figure*}    
    
    \Figref{figure3}(b) depicts the bounds (red solid line) and the performance of thin films (black solid line) as a function of thickness, as well as six different inverse-design structures that bridge most of the gap from the thin films to the bounds. The incident wavelength is $\SI{11}{\micro\meter}$ and the period is $\SI{1.1}{\micro\meter}$, with minimum feature size $\SI{0.1}{\micro\meter}$. For an ultrathin absorber with thickness $\SI{0.4}{\micro\meter}$, the inverse-designed metasurface can reach 72\% of the global bound. In \figref{figure3}(c) we isolate the design at this smallest thickness and show its spectral absorption percentage, as well as its geometrical design (inset). Detail of the inverse design are given in the \SM. Since the objective is to compare against the global, we do not impose binarization, lithographic, or other fabrication constraints. It is apparent that inverse design can come rather close to the bounds, suggesting they may be ``tight'' or nearly so.

An important ramification of the bounds of \eqrefrange{ext-cha-extend}{sca-cha-extend} is that they can be used to find the minimum thickness of any patterned ``perfect absorber''~\cite{Watts2012,lee2016metamaterials,landy_perfect_2008-1}, achieving 100\% absorption or close to it. Such absorbers are particularly useful for sensing applications~\cite{lee2016metamaterials,Liu2010} and \hl{the design of ultra-thin solar cells~\cite{Yu2010,cui_ultrabroadband_2012,massiot_nanopatterned_2012}} . Absorption cross-section per area, $\sigma_{\rm abs}/A$, is the percentage absorption, while the bound on the right-hand side of \eqref{abs-cha-extend} is a function only of the incident angle, the absorber thickness (defined as the thickness of its minimum bounding film), and its material susceptibility $\chi(\omega)$. For normally incident waves, we show in the {\SM} that the minimum thickness $h_{\rm min}$ to achieve 100\% absorption is given by the self-consistent equation
\begin{align}
h_{\rm min} = \left(\frac{2\lambda}{\pi}\right) \frac{\Im \xi}{1-\operatorname{sinc}^2(kh_{\rm min})}.
\label{eq:hmin}
\end{align}
\Figref{figure4}(a,b) shows the minimum thicknesses (solid lines) for 100\% absorption in common metallic and polar-dielectric materials. It is perhaps surprising how large the thicknesses are, averaging on the order of \SI{50}{nm} for metals~\cite{palik1998handbook} at visible wavelengths and 1 $\mu$m for polar dielectrics~\cite{law2013all,francoeur2010spectral,popova1972optical} at infrared wavelengths. The only previous bounds that could predict a minimal thickness for perfect absorption are the material-loss bounds~\cite{miller_fundamental_2016}, which predict minimal thicknesses on the order of \SI{0.5}{nm} and \SI{10}{nm} for the same materials and wavelengths, respectively. Also included in the figures are the minimal thicknesses for 70\% absorption, which are about a factor of two smaller than the 100\%-absorption curves. In the {\SM}, we present further analysis suggesting two points: first, that the minimum thickness is typically larger than the skin depth, and can be arbitrarily larger; second, that the nearly linear dependence of Aluminum's minimal thickness relative to wavelength indicates Drude-like permittivity, in contrast to highly non-Drude-like behavior for Ag and Au.  In \figref{figure4}(c) we present universal curves on which all perfect-absorber materials can be judged, showing the minimum thickness relative to the wavelength as a function of the inverse of material loss, $1/\Im\xi = |\chi|^2 / \Im \chi$, which is a material ``figure of merit'' (FOM) as discussed above~\cite{miller_fundamental_2016}. Using the same inverse-design techniques described above, we discovered ultra-thin absorbers with 70\% absorption rate using both the metals and polar dielectrics presented in \figref{figure4} (a,b). The grayscale design voxels are specified in the \SM. As shown in the inset, all of the materials achieve 70\% absorption at thicknesses within a factor of 1.5--2.7 of the bound. In the SM we show that in the highly subwavelength limit, the minimum thickness of a perfect-absorber scales with material FOM as $h_{\rm min} / \lambda  \sim (1 / \Im{\xi})^{-1/3}$. The inverse-cubic scaling means that there are diminishing returns to further reductions in loss, and explains the flattening of the curves on the right-hand side of \figref{figure4}(c).

\begin{figure*}[htb]
	\includegraphics[width=0.9\textwidth, trim={0 3.5cm 5cm 3.9cm},clip]{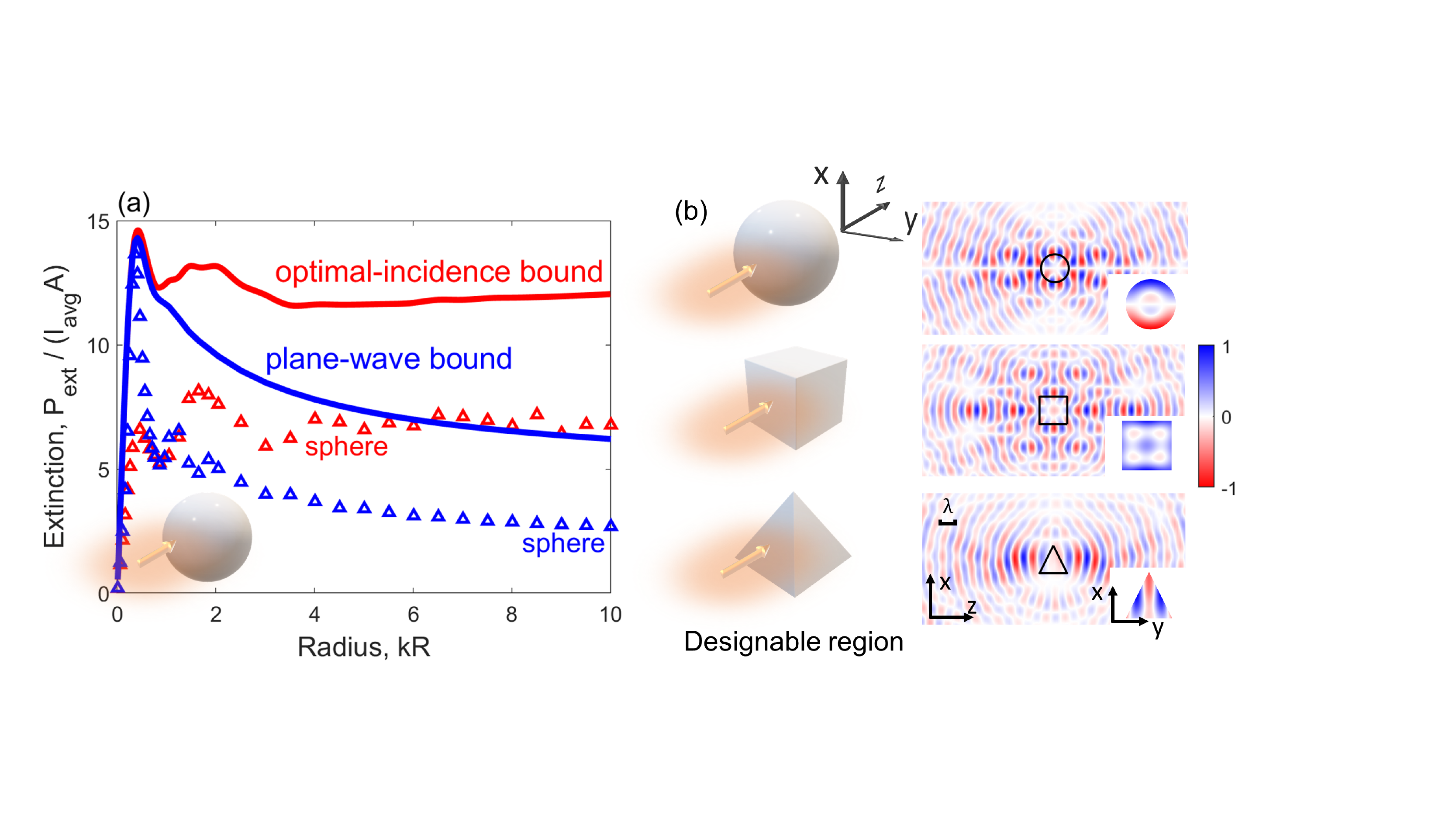}
	\centering
    \caption{Maximum extinction $P_{\rm ext}$ for arbitrary patterning and illumination, normalized by average field intensity $I_{\rm avg}$ and geometric cross section $A$ of the bounding sphere of radius $R$. The solid red line in (a) shows the maximal extinction that can possibly be obtained by the optimal incident field, as compared to the simple plane wave incidence shown by the solid blue line. The triangular markers gives the attained extinction from an unpatterned silver sphere of radius $R$ under either optimal incidence (red triangles) or plane wave illumination (blue triangles). (b) Three possible design regions (sphere, cube, and pyramid) and the corresponding optimal illumination fields ($\Im E_x$) in the x-z plane and x-y plane (inset).}
	\label{fig:figure5}
\end{figure*}

\section{Optimal Illumination Fields}
\label{sec:IF}

In this section, we identify the \emph{incident waves} that maximize the response bounds of \eqrefrange{ext-gen}{rhorad_bnd}. There is significant interest in such wavefront shaping~\cite{Popoff2014,Vellekoop2015,Horstmeyer2015,Jang2018}, in particular for the question of identifying optimal illumination fields~\cite{Polin2005,Mazilu2011,Taylor2015,Lee2017,Taylor2017,Fernandez-Corbaton2017,Liu2019}, and yet every current approach identifies optimal fields for a given scatterer. Using the framework developed above, we can instead only specify a designable region, and identify the optimal illumination field that maximizes the bound over all possible scatterers.

    To start, we assume that there is a basis $\Phi$ comprising accessible far-field illumination channels, such as plane waves, vector spherical waves, Bessel beams, excitations from a spatial light modulator, or any other basis~\cite{Levy2016}. Then the incident field can be written
\begin{align}
    \psiinc = \Phi \cinc,
\end{align}
where $\cinc$ is the vector of basis coefficients to be optimized. The objective is to maximize any of the response bounds, \eqrefrange{ext-gen}{rhorad_bnd}, subject to some constraint on the incoming wave. The absorption and scattering bounds, and their near-field counterparts, have a complex dependence on $\psi_{\rm inc}$ due to the presence of the dual variable $\nu^*$, which has a nonlinear dependence on $\psiinc$. Each of these quantities can be locally optimized using any gradient-based optimization method~\cite{Nocedal2006}. Extinction, as well as total near-field local density of states, have analytic forms that lead to simple formulations of \emph{global} bounds over all incident fields. Inserting the incident-wave basis into the extinction bound, \eqref{ext-gen}, one finds that the extinction bound can be written as
\begin{align}
    P_{\rm ext}^{\rm bound} = \frac{\omega}{2} \cinc^\dagger \Phi^\dagger \left(\Im\xi+\ImGO\right)^{-1} \Phi \cinc,
    \label{eq:Pextbound} 
\end{align}
which is a simple quadratic function of $\cinc$. This quantity should be maximized subject to an intensity or power constraint on the fields. Such a constraint would be of the form $\cinc^\dagger \mathbb{W} \cinc \leq 1$, where $\mathbb{W}$ is a positive-definite Hermitian matrix representing a power-flow measure of $\cinc$. Since the objective and constraint are both positive-definite quadratic forms, the optimal incident-wave coefficients are given by an extremal eigenvector~\cite{Trefethen1997}: the eigenvector(s) corresponding to the largest eigenvalue(s) $\lambda_{\rm max}$ of the generalized eigenproblem
\begin{align}
    \Phi^\dagger \left(\Im\xi+\ImGO\right)^{-1} \Phi \cinc = \lambda_{\rm max} \mathbb{W} \cinc.
    \label{eq:OptIF}
\end{align}
The solution of \eqref{OptIF} offers the largest upper bound of all possible incident fields.

\Figref{figure5}(a) demonstrates the utility of optimizing over incident fields. We consider incident fields impinging upon a finite silver scatterer within a bounding sphere of radius R at wavelength $\lambda = \SI{360}{nm}$ (as in \figref{figure2}, near the surface-plasmon resonance). We consider incident fields originating from one half-space, as might be typical in an experimental setup, and use as our basis 441 plane waves with wave vectors $\vect{k}$ whose evenly spaced transverse components range from $-0.8k$ to $0.8k$, where $k =2\pi/\lambda$ is the total wave number. The $0.8$ wave-vector cutoff corresponds to incident-field control over a solid angle of approximately \SI{2.5}{sr}, and can be matched to the specifics of any experimental setup. We impose the constraint that the average intensity over a region  that has twice the radius of the sphere must be equal to that of a unit-amplitude plane wave. \Figref{figure5}(a) shows the extinction bound evaluated for a plane wave (blue solid), as well as that for the optimal incident field (red solid). As the radius increases, incident-field shaping can have a substantial effect and yield bounds that are almost twice as large as those for plane waves (1.94$\times$ exactly). (Each quantity is normalized by average field intensity $I_{\rm avg}$ and the geometric cross-section $A = \pi R^2$, which is why the extinction bounds may decrease with increasing radius.) Intriguingly, we show that even an unpatterned sphere (red triangles) shows performance trending with that of the bound, and for the larger radii the unpatterned sphere under the optimal illumination field exhibits extinction values larger than the plane-wave bounds. This illustrates a key benefit of bounds: one can now conclude that an unpatterned sphere with optimal illumination fields can achieve extinction values that cannot possibly be achieved by \emph{any} structure under plane-wave illumination.

\Figref{figure5}(b) further extends the optimal-illumination results, considering three designable regions: a sphere, a cube, and a pyramid. The optimal illumination patterns are shown in two-dimensional cross-sections outside and within the designable regions. The sphere has a radius of one free-space wavelength, while the cube and pyramid have side lengths equal to twice the free-space wavelength. Within each domain the optimal illumination fields exhibit interesting patterns that seem to put field nodes (zeros) in the interior, with the largest field amplitudes around the walls of the domains. This can be explained physically: the optimal incident fields will be those that couple most strongly to the polarization currents that exhibit the smallest radiative losses. The polarization currents that have the smallest radiative losses will tend to have oscillations with far-field radiation patterns that cancel each other, as occurs for oscillating currents along structural boundaries, such as whispering-gallery modes~\cite{Vahala2003,He2013}. This procedure can be implemented for a beam generated by almost any means, e.g., and incident wave passing through a scatterer with a complex structural profile~\cite{Levy2007,Wei2013,KerenZur2016}, precisely controlled spatial light modulators~\cite{Weiner2000,Chattrapiban2003,Guo2007,Zhu2014}, or a light source with a complex spatial emission profile~\cite{Lodahl2004,Ringler2008,Bleuse2011}.

\section{Discussion and extensions}
\label{sec:Disc}
In this Article, we have shown that an energy-conservation law, arising as a generalized optical theorem, enables identification of maximal electromagnetic response at a single frequency. We considered: arbitrary linear and quadratic response functions, \eqref{gen_bound}, power-flow quantities such as absorption and scattering, \eqreftwo{abs-gen}{sca-gen}, and LDOS, \eqrefrange{rhotot_bnd}{rhorad_bnd}, more specific scenarios such as plane-wave scattering and perfect absorbers, \eqrefrange{ext-cha-bound}{hmin}, and optimal illumination fields, \eqref{OptIF}. In this section, we briefly touch on numerous other applications where this formalism can be seamlessly applied.

One important application is to understand the largest thermal absorption and emission of structured material. A direct consequence of the incoherent nature of the thermal source is that an upper bound to the average absorptivity/emissivity is given by
the average of the bounds for each independent incident field in an orthogonal basis, such as vector spherical waves for a finite scatterer. As detailed in the SM, a straightforward implementation of the our formalism leads to even a tighter bound than the recently published $\mathbb{T}$-operator bound of Ref. \cite{molesky2019fundamental}.

A natural extension of this work is to the emergent field of 2D materials~\cite{Novoselov2005,Geim2007,Koppens2011,Basov2016,Low2016}. From a theoretical perspective the only difference with a 2D material is that the induced polarization currents exist on a two-dimensional surfaces instead of within a three-dimensional volume, which would change the interpretation of $\phi$ in \eqref{general-formalism}, and would change the domain of the Green's function $\Gamma_0$, but otherwise the remainder of the derivation is identical. Instead of rederiving the bounds in a 2D domain, however, a simpler approach is to substitute the bulk susceptibility $\chi$ by the expression $\chi \rightarrow i\sigma_{\rm 2D}/\omega h$, where $\sigma_{\rm 2D}$ is the 2D-material conductivity and $h$ is an infinitesimal thickness going to zero. (The bounds do not diverge because the geoemtric or bounding volume is also proportional to $h$, canceling the $1/h$ divergence in the material parameter.) Then, all of the bounds derived herein apply to 2D materials as well.

Another important extension is to problems of field concentration \emph{away} from the scatterer itself. In surface-enhanced Raman scattering~\cite{Moskovits1985,Nie1997,Stiles2008}, for example, where recently material-loss bounds have been derived~\cite{michon2019limits}, it is important to maximize average field enhancement over a plane close to but not overlapping the scatterer itself. In this case the objective might be the integral of the scattered-field intensity over a plane $P$, i.e. $\int_P \psiscat^\dagger \psiscat$. The scattering field is the convolution of the background Green's function with the polarization fields $\phi$, such that this objective is a quadratic function of the polarization fields: $\phi^\dagger \left[\int_P \Gamma_0^\dagger \Gamma_0\right] \phi$, which is exactly of the form required by \eqref{general-formalism} and thus is bounded above by \eqref{gen_bound}. 

Similarly, cross density of states~\cite{caze_pierrat_carminati_2013} measures the coupling strength between dipoles at two spatial locations, typically coupled via near-field interactions, for applications including F{\"o}rster energy transfer~\cite{gonzaga-galeana_zurita-sanchez_2013} and quantum entanglement~\cite{gonzalez-tudela_martin-cano_2011,lassalle_long-lifetime_2020}. Such coupling effectively reduces to optimizing the field strength at one location from a point source at another location, mapping identically to the field concentration problem.

Maximizing optical forces and torques has been a topic of substantial interest~\cite{Mazilu2011,rahimzadegan2017fundamental,Lee2017,Liu2019,Horodynski2019}, and is one that our framework applies to very naturally. One can compute force and torque via surface integrals of quantities related to the Maxwell stress tensor, which is a quadratic function of the electric and magnetic fields. By the same connection of the scattered fields to the induced polarization fields, it is possible to write any force/torque optimization function as a sum of quadratic- and linear-in-polarization terms, thereby equivalent to \eqref{general-formalism} and subject to the bounds of \eqref{gen_bound}.

During the preparation of this manuscript, \hl{two preprints appeared~\cite{gustafsson2019upper,Molesky2020a}} that contain similar ideas to those contained here. It is recognized in Refs.~\cite{gustafsson2019upper,Molesky2020a} that one can utilize the equality of absorption plus scattering and extinction, i.e. \eqref{optthm}, as a quadratic electromagnetic constraint. They further show that an additional constraint can be identified; essentially, the real-part analog of \eqref{optthm}. In this context they provide bounds very similar to ours for power-flow quantities and local density of states, \citeasnoun{gustafsson2019upper} considers the problem of directional scattering, and they \hl{both} show a two-parameter dual formulation for incorporating the second constraint. Conversely, they do not have bounds for arbitrary linear and quadratic functions, i.e. our \eqref{gen_bound}, or for non-scalar or nonlocal susceptibility operators, nor do they consider \hl{the possibility of bounds over all incoming wavefronts.} And they do not identify the optimal value of the dual variable $\nu^*$, which is important, for example, in determining the analytical bound of \eqref{hmin}. Without an analytical value for $\nu^*$, it is not possible to identify the minimum thickness of a perfect absorber.

\hl{More recent preprints have shown that one can generate an infinite set of (mostly nonconvex) constraints from spatially localized versions of the optical theorem~\cite{Kuang2020,Molesky2020c}. There are advantages and drawbacks to such an approach relative to the one we presented here. With more constraints, one can potentially identify tighter bounds. But since most of the constraints are nonconvex, global optima are only identifiable through convex relaxations~\cite{Luo2010}, which introduce two disadvantages to the computational approach. First, the bounds are numerical in nature and do not offer the intuition of semi-analytical bounds (as presented here). Second, they are computationally expensive and thus currently limited to wavelength-scale device sizes. Moreover, the non-analytical nature of the bounds precludes explicit identification of the dependence of the bounds on the incident fields, which enabled the wavefront-shaping results of \secref{IF}, and which appears to not be possible in the approaches of Refs.~\cite{Kuang2020,Molesky2020c}. Thus, the framework in this manuscript is complementary to that of Refs.~\cite{Kuang2020,Molesky2020c}, with each offering unique comparative advantages.}

Looking forward, the energy-conservation approach developed here provides a framework for further generalizations and unifications. The incorporation of multiple constraints naturally leads to connections to the optimization field of semidefinite programming~\cite{Luo2010}, as utilized in \citeasnoun{Shim2019subm}, where rapid global-optimization computational techniques are well-established~\cite{Nocedal2006}. Away from single-frequency problems, the question of how to incorporate nonzero bandwidth in a bound framework would have important ramifications. As shown in \citeasnoun{shim2019fundamental}, it may be possible to do so through generalized quadratic constraints based on causality. Finally, a key variable missing from \hl{semi-analytical,} conservation-law-based bounds is the refractive index of a transparent medium, which \hl{does} appear in bounds pertaining to the broadband absorption of sunlight~\cite{Yablonovitch1982,Yu2010,buddhiraju_theory_2017,Benzaouia2019}. Accounting for refractive index may require a unification of conservation-law approaches with, perhaps, those based on Lagrangian duality~\cite{Angeris2019}, \hl{or on sophisticated approaches developed in the theory of composite materials~\cite{Bergman1981,Milton1981,Milton2002,Kern2020}.} With such generalizations and unifications, it may be possible to understand the extreme limits of electromagnetic response in any scenario.

\section{Acknowledgments}
Z.K. and L.Z. were supported by the Army Research Office under grant number W911NF-19-1-0279. O.D.M. was partially supported by the Army Research Office under grant number W911NF-19-1-0279 and partially supported by the Air Force Office of Scientific Research under award number FA9550-17-1-0093. 

\bibliography{final_bib}

\providecommand{\noopsort}[1]{}\providecommand{\singleletter}[1]{#1}%
\begin{thebibliography}{143}%
\makeatletter
\providecommand \@ifxundefined [1]{%
 \@ifx{#1\undefined}
}%
\providecommand \@ifnum [1]{%
 \ifnum #1\expandafter \@firstoftwo
 \else \expandafter \@secondoftwo
 \fi
}%
\providecommand \@ifx [1]{%
 \ifx #1\expandafter \@firstoftwo
 \else \expandafter \@secondoftwo
 \fi
}%
\providecommand \natexlab [1]{#1}%
\providecommand \enquote  [1]{``#1''}%
\providecommand \bibnamefont  [1]{#1}%
\providecommand \bibfnamefont [1]{#1}%
\providecommand \citenamefont [1]{#1}%
\providecommand \href@noop [0]{\@secondoftwo}%
\providecommand \href [0]{\begingroup \@sanitize@url \@href}%
\providecommand \@href[1]{\@@startlink{#1}\@@href}%
\providecommand \@@href[1]{\endgroup#1\@@endlink}%
\providecommand \@sanitize@url [0]{\catcode `\\12\catcode `\$12\catcode
  `\&12\catcode `\#12\catcode `\^12\catcode `\_12\catcode `\%12\relax}%
\providecommand \@@startlink[1]{}%
\providecommand \@@endlink[0]{}%
\providecommand \url  [0]{\begingroup\@sanitize@url \@url }%
\providecommand \@url [1]{\endgroup\@href {#1}{\urlprefix }}%
\providecommand \urlprefix  [0]{URL }%
\providecommand \Eprint [0]{\href }%
\providecommand \doibase [0]{https://doi.org/}%
\providecommand \selectlanguage [0]{\@gobble}%
\providecommand \bibinfo  [0]{\@secondoftwo}%
\providecommand \bibfield  [0]{\@secondoftwo}%
\providecommand \translation [1]{[#1]}%
\providecommand \BibitemOpen [0]{}%
\providecommand \bibitemStop [0]{}%
\providecommand \bibitemNoStop [0]{.\EOS\space}%
\providecommand \EOS [0]{\spacefactor3000\relax}%
\providecommand \BibitemShut  [1]{\csname bibitem#1\endcsname}%
\let\auto@bib@innerbib\@empty
\bibitem [{\citenamefont {Yaghjian}(1996)}]{Yaghjian1996}%
  \BibitemOpen
  \bibfield  {author} {\bibinfo {author} {\bibfnamefont {A.~D.}\ \bibnamefont
  {Yaghjian}},\ }\bibfield  {title} {\bibinfo {title} {{Sampling criteria for
  resonant antennas and scatterers}},\ }\href
  {https://doi.org/10.1063/1.362683} {\bibfield  {journal} {\bibinfo  {journal}
  {J. Appl. Phys.}\ }\textbf {\bibinfo {volume} {79}},\ \bibinfo {pages} {7474}
  (\bibinfo {year} {1996})}\BibitemShut {NoStop}%
\bibitem [{\citenamefont {Hamam}\ \emph {et~al.}(2007)\citenamefont {Hamam},
  \citenamefont {Karalis}, \citenamefont {Joannopoulos},\ and\ \citenamefont
  {Soljačić}}]{hamam_coupled-mode_2007}%
  \BibitemOpen
  \bibfield  {author} {\bibinfo {author} {\bibfnamefont {R.~E.}\ \bibnamefont
  {Hamam}}, \bibinfo {author} {\bibfnamefont {A.}~\bibnamefont {Karalis}},
  \bibinfo {author} {\bibfnamefont {J.~D.}\ \bibnamefont {Joannopoulos}},\ and\
  \bibinfo {author} {\bibfnamefont {M.}~\bibnamefont {Soljačić}},\ }\bibfield
   {title} {\bibinfo {title} {Coupled-mode theory for general free-space
  resonant scattering of waves},\ }\href
  {https://doi.org/10.1103/PhysRevA.75.053801} {\bibfield  {journal} {\bibinfo
  {journal} {Physical Review A}\ }\textbf {\bibinfo {volume} {75}},\ \bibinfo
  {pages} {053801} (\bibinfo {year} {2007})}\BibitemShut {NoStop}%
\bibitem [{\citenamefont {Kwon}\ and\ \citenamefont
  {Pozar}(2009)}]{kwon2009optimal}%
  \BibitemOpen
  \bibfield  {author} {\bibinfo {author} {\bibfnamefont {D.-H.}\ \bibnamefont
  {Kwon}}\ and\ \bibinfo {author} {\bibfnamefont {D.~M.}\ \bibnamefont
  {Pozar}},\ }\bibfield  {title} {\bibinfo {title} {Optimal characteristics of
  an arbitrary receive antenna},\ }\href
  {https://doi.org/10.1109/TAP.2009.2025975} {\bibfield  {journal} {\bibinfo
  {journal} {IEEE Transactions on Antennas and Propagation}\ }\textbf {\bibinfo
  {volume} {57}},\ \bibinfo {pages} {3720} (\bibinfo {year}
  {2009})}\BibitemShut {NoStop}%
\bibitem [{\citenamefont {Ruan}\ and\ \citenamefont
  {Fan}(2011)}]{ruan2011design}%
  \BibitemOpen
  \bibfield  {author} {\bibinfo {author} {\bibfnamefont {Z.}~\bibnamefont
  {Ruan}}\ and\ \bibinfo {author} {\bibfnamefont {S.}~\bibnamefont {Fan}},\
  }\bibfield  {title} {\bibinfo {title} {Design of subwavelength
  superscattering nanospheres},\ }\href {https://doi.org/10.1063/1.3536475}
  {\bibfield  {journal} {\bibinfo  {journal} {Applied Physics Letters}\
  }\textbf {\bibinfo {volume} {98}},\ \bibinfo {pages} {043101} (\bibinfo
  {year} {2011})}\BibitemShut {NoStop}%
\bibitem [{\citenamefont {Liberal}\ \emph
  {et~al.}(2014{\natexlab{a}})\citenamefont {Liberal}, \citenamefont {Ra'di},
  \citenamefont {Gonzalo}, \citenamefont {Ederra}, \citenamefont {Tretyakov},\
  and\ \citenamefont {Ziolkowski}}]{liberal2014least}%
  \BibitemOpen
  \bibfield  {author} {\bibinfo {author} {\bibfnamefont {I.}~\bibnamefont
  {Liberal}}, \bibinfo {author} {\bibfnamefont {Y.}~\bibnamefont {Ra'di}},
  \bibinfo {author} {\bibfnamefont {R.}~\bibnamefont {Gonzalo}}, \bibinfo
  {author} {\bibfnamefont {I.}~\bibnamefont {Ederra}}, \bibinfo {author}
  {\bibfnamefont {S.~A.}\ \bibnamefont {Tretyakov}},\ and\ \bibinfo {author}
  {\bibfnamefont {R.~W.}\ \bibnamefont {Ziolkowski}},\ }\bibfield  {title}
  {\bibinfo {title} {Least upper bounds of the powers extracted and scattered
  by bi-anisotropic particles},\ }\href
  {https://doi.org/10.1109/TAP.2014.2330620} {\bibfield  {journal} {\bibinfo
  {journal} {IEEE Transactions on Antennas and Propagation}\ }\textbf {\bibinfo
  {volume} {62}},\ \bibinfo {pages} {4726} (\bibinfo {year}
  {2014}{\natexlab{a}})}\BibitemShut {NoStop}%
\bibitem [{\citenamefont {Liberal}\ \emph
  {et~al.}(2014{\natexlab{b}})\citenamefont {Liberal}, \citenamefont {Ederra},
  \citenamefont {Gonzalo},\ and\ \citenamefont
  {Ziolkowski}}]{liberal2014upper}%
  \BibitemOpen
  \bibfield  {author} {\bibinfo {author} {\bibfnamefont {I.}~\bibnamefont
  {Liberal}}, \bibinfo {author} {\bibfnamefont {I.}~\bibnamefont {Ederra}},
  \bibinfo {author} {\bibfnamefont {R.}~\bibnamefont {Gonzalo}},\ and\ \bibinfo
  {author} {\bibfnamefont {R.~W.}\ \bibnamefont {Ziolkowski}},\ }\bibfield
  {title} {\bibinfo {title} {Upper bounds on scattering processes and
  metamaterial-inspired structures that reach them},\ }\href
  {https://doi.org/10.1109/TAP.2014.2359206} {\bibfield  {journal} {\bibinfo
  {journal} {IEEE Transactions on Antennas and Propagation}\ }\textbf {\bibinfo
  {volume} {62}},\ \bibinfo {pages} {6344} (\bibinfo {year}
  {2014}{\natexlab{b}})}\BibitemShut {NoStop}%
\bibitem [{\citenamefont {Miller}\ \emph {et~al.}(2016)\citenamefont {Miller},
  \citenamefont {Polimeridis}, \citenamefont {Reid}, \citenamefont {Hsu},
  \citenamefont {DeLacy}, \citenamefont {Joannopoulos}, \citenamefont
  {Solja{\v{c}}i{\'c}},\ and\ \citenamefont
  {Johnson}}]{miller_fundamental_2016}%
  \BibitemOpen
  \bibfield  {author} {\bibinfo {author} {\bibfnamefont {O.~D.}\ \bibnamefont
  {Miller}}, \bibinfo {author} {\bibfnamefont {A.~G.}\ \bibnamefont
  {Polimeridis}}, \bibinfo {author} {\bibfnamefont {M.~H.}\ \bibnamefont
  {Reid}}, \bibinfo {author} {\bibfnamefont {C.~W.}\ \bibnamefont {Hsu}},
  \bibinfo {author} {\bibfnamefont {B.~G.}\ \bibnamefont {DeLacy}}, \bibinfo
  {author} {\bibfnamefont {J.~D.}\ \bibnamefont {Joannopoulos}}, \bibinfo
  {author} {\bibfnamefont {M.}~\bibnamefont {Solja{\v{c}}i{\'c}}},\ and\
  \bibinfo {author} {\bibfnamefont {S.~G.}\ \bibnamefont {Johnson}},\
  }\bibfield  {title} {\bibinfo {title} {Fundamental limits to optical response
  in absorptive systems},\ }\href {https://doi.org/10.1364/OE.24.003329}
  {\bibfield  {journal} {\bibinfo  {journal} {Optics express}\ }\textbf
  {\bibinfo {volume} {24}},\ \bibinfo {pages} {3329} (\bibinfo {year}
  {2016})}\BibitemShut {NoStop}%
\bibitem [{\citenamefont {Miller}\ \emph {et~al.}(2017)\citenamefont {Miller},
  \citenamefont {Ilic}, \citenamefont {Christensen}, \citenamefont {Reid},
  \citenamefont {Atwater}, \citenamefont {Joannopoulos}, \citenamefont
  {Solja{\v c}i{\' c}},\ and\ \citenamefont {Johnson}}]{miller2017limits}%
  \BibitemOpen
  \bibfield  {author} {\bibinfo {author} {\bibfnamefont {O.~D.}\ \bibnamefont
  {Miller}}, \bibinfo {author} {\bibfnamefont {O.}~\bibnamefont {Ilic}},
  \bibinfo {author} {\bibfnamefont {T.}~\bibnamefont {Christensen}}, \bibinfo
  {author} {\bibfnamefont {M.~H.}\ \bibnamefont {Reid}}, \bibinfo {author}
  {\bibfnamefont {H.~A.}\ \bibnamefont {Atwater}}, \bibinfo {author}
  {\bibfnamefont {J.~D.}\ \bibnamefont {Joannopoulos}}, \bibinfo {author}
  {\bibfnamefont {M.}~\bibnamefont {Solja{\v c}i{\' c}}},\ and\ \bibinfo
  {author} {\bibfnamefont {S.~G.}\ \bibnamefont {Johnson}},\ }\bibfield
  {title} {\bibinfo {title} {Limits to the optical response of graphene and
  two-dimensional materials},\ }\href
  {https://doi.org/10.1021/acs.nanolett.7b02007} {\bibfield  {journal}
  {\bibinfo  {journal} {Nano letters}\ }\textbf {\bibinfo {volume} {17}},\
  \bibinfo {pages} {5408} (\bibinfo {year} {2017})}\BibitemShut {NoStop}%
\bibitem [{\citenamefont {Yang}\ \emph {et~al.}(2017)\citenamefont {Yang},
  \citenamefont {Miller}, \citenamefont {Christensen}, \citenamefont
  {Joannopoulos},\ and\ \citenamefont {Soljacic}}]{yang2017low}%
  \BibitemOpen
  \bibfield  {author} {\bibinfo {author} {\bibfnamefont {Y.}~\bibnamefont
  {Yang}}, \bibinfo {author} {\bibfnamefont {O.~D.}\ \bibnamefont {Miller}},
  \bibinfo {author} {\bibfnamefont {T.}~\bibnamefont {Christensen}}, \bibinfo
  {author} {\bibfnamefont {J.~D.}\ \bibnamefont {Joannopoulos}},\ and\ \bibinfo
  {author} {\bibfnamefont {M.}~\bibnamefont {Soljacic}},\ }\bibfield  {title}
  {\bibinfo {title} {Low-loss plasmonic dielectric nanoresonators},\ }\href
  {https://pubs.acs.org/doi/abs/10.1021/acs.nanolett.7b00852} {\bibfield
  {journal} {\bibinfo  {journal} {Nano letters}\ }\textbf {\bibinfo {volume}
  {17}},\ \bibinfo {pages} {3238} (\bibinfo {year} {2017})}\BibitemShut
  {NoStop}%
\bibitem [{\citenamefont {Pendry}(1999)}]{pendry1999radiative}%
  \BibitemOpen
  \bibfield  {author} {\bibinfo {author} {\bibfnamefont {J.}~\bibnamefont
  {Pendry}},\ }\bibfield  {title} {\bibinfo {title} {Radiative exchange of heat
  between nanostructures},\ }\href
  {https://doi.org/10.1088/0953-8984/11/35/301} {\bibfield  {journal} {\bibinfo
   {journal} {Journal of Physics: Condensed Matter}\ }\textbf {\bibinfo
  {volume} {11}},\ \bibinfo {pages} {6621} (\bibinfo {year}
  {1999})}\BibitemShut {NoStop}%
\bibitem [{\citenamefont {Thongrattanasiri}\ \emph {et~al.}(2012)\citenamefont
  {Thongrattanasiri}, \citenamefont {Koppens},\ and\ \citenamefont
  {De~Abajo}}]{thongrattanasiri_complete_2012}%
  \BibitemOpen
  \bibfield  {author} {\bibinfo {author} {\bibfnamefont {S.}~\bibnamefont
  {Thongrattanasiri}}, \bibinfo {author} {\bibfnamefont {F.~H.}\ \bibnamefont
  {Koppens}},\ and\ \bibinfo {author} {\bibfnamefont {F.~J.~G.}\ \bibnamefont
  {De~Abajo}},\ }\bibfield  {title} {\bibinfo {title} {Complete optical
  absorption in periodically patterned graphene},\ }\href@noop {} {\bibfield
  {journal} {\bibinfo  {journal} {Physical review letters}\ }\textbf {\bibinfo
  {volume} {108}},\ \bibinfo {pages} {047401} (\bibinfo {year}
  {2012})}\BibitemShut {NoStop}%
\bibitem [{\citenamefont {Hugonin}\ \emph {et~al.}(2015)\citenamefont
  {Hugonin}, \citenamefont {Besbes},\ and\ \citenamefont
  {Ben-Abdallah}}]{hugonin_fundamental_2015}%
  \BibitemOpen
  \bibfield  {author} {\bibinfo {author} {\bibfnamefont {J.-P.}\ \bibnamefont
  {Hugonin}}, \bibinfo {author} {\bibfnamefont {M.}~\bibnamefont {Besbes}},\
  and\ \bibinfo {author} {\bibfnamefont {P.}~\bibnamefont {Ben-Abdallah}},\
  }\bibfield  {title} {{\selectlanguage {en}\bibinfo {title} {Fundamental
  limits for light absorption and scattering induced by cooperative
  electromagnetic interactions}},\ }\href
  {https://doi.org/10.1103/PhysRevB.91.180202} {\bibfield  {journal} {\bibinfo
  {journal} {Physical Review B}\ }\textbf {\bibinfo {volume} {91}},\ \bibinfo
  {pages} {180202} (\bibinfo {year} {2015})}\BibitemShut {NoStop}%
\bibitem [{\citenamefont {Miller}\ \emph {et~al.}(2015)\citenamefont {Miller},
  \citenamefont {Johnson},\ and\ \citenamefont {Rodriguez}}]{miller2015shape}%
  \BibitemOpen
  \bibfield  {author} {\bibinfo {author} {\bibfnamefont {O.~D.}\ \bibnamefont
  {Miller}}, \bibinfo {author} {\bibfnamefont {S.~G.}\ \bibnamefont
  {Johnson}},\ and\ \bibinfo {author} {\bibfnamefont {A.~W.}\ \bibnamefont
  {Rodriguez}},\ }\bibfield  {title} {\bibinfo {title} {Shape-independent
  limits to near-field radiative heat transfer},\ }\href
  {https://doi.org/10.1103/PhysRevLett.115.204302} {\bibfield  {journal}
  {\bibinfo  {journal} {Physical review letters}\ }\textbf {\bibinfo {volume}
  {115}},\ \bibinfo {pages} {204302} (\bibinfo {year} {2015})}\BibitemShut
  {NoStop}%
\bibitem [{\citenamefont {Rahimzadegan}\ \emph {et~al.}(2017)\citenamefont
  {Rahimzadegan}, \citenamefont {Alaee}, \citenamefont {Fernandez-Corbaton},\
  and\ \citenamefont {Rockstuhl}}]{rahimzadegan2017fundamental}%
  \BibitemOpen
  \bibfield  {author} {\bibinfo {author} {\bibfnamefont {A.}~\bibnamefont
  {Rahimzadegan}}, \bibinfo {author} {\bibfnamefont {R.}~\bibnamefont {Alaee}},
  \bibinfo {author} {\bibfnamefont {I.}~\bibnamefont {Fernandez-Corbaton}},\
  and\ \bibinfo {author} {\bibfnamefont {C.}~\bibnamefont {Rockstuhl}},\
  }\bibfield  {title} {\bibinfo {title} {Fundamental limits of optical force
  and torque},\ }\href {https://doi.org/10.1103/PhysRevB.95.035106} {\bibfield
  {journal} {\bibinfo  {journal} {Physical Review B}\ }\textbf {\bibinfo
  {volume} {95}},\ \bibinfo {pages} {035106} (\bibinfo {year}
  {2017})}\BibitemShut {NoStop}%
\bibitem [{\citenamefont {Liu}\ \emph {et~al.}(2018)\citenamefont {Liu},
  \citenamefont {Fan}, \citenamefont {Lee}, \citenamefont {Fang}, \citenamefont
  {Johnson},\ and\ \citenamefont {Miller}}]{liu2018optimal}%
  \BibitemOpen
  \bibfield  {author} {\bibinfo {author} {\bibfnamefont {Y.}~\bibnamefont
  {Liu}}, \bibinfo {author} {\bibfnamefont {L.}~\bibnamefont {Fan}}, \bibinfo
  {author} {\bibfnamefont {Y.~E.}\ \bibnamefont {Lee}}, \bibinfo {author}
  {\bibfnamefont {N.~X.}\ \bibnamefont {Fang}}, \bibinfo {author}
  {\bibfnamefont {S.~G.}\ \bibnamefont {Johnson}},\ and\ \bibinfo {author}
  {\bibfnamefont {O.~D.}\ \bibnamefont {Miller}},\ }\bibfield  {title}
  {\bibinfo {title} {Optimal nanoparticle forces, torques, and illumination
  fields},\ }\href {https://doi.org/10.1021/acsphotonics.8b01263} {\bibfield
  {journal} {\bibinfo  {journal} {ACS Photonics}\ }\textbf {\bibinfo {volume}
  {6}},\ \bibinfo {pages} {395} (\bibinfo {year} {2018})}\BibitemShut {NoStop}%
\bibitem [{\citenamefont {Yang}\ \emph {et~al.}(2018)\citenamefont {Yang},
  \citenamefont {Massuda}, \citenamefont {Roques-Carmes}, \citenamefont {Kooi},
  \citenamefont {Christensen}, \citenamefont {Johnson}, \citenamefont
  {Joannopoulos}, \citenamefont {Miller}, \citenamefont {Kaminer},\ and\
  \citenamefont {Solja{\v{c}}i{\'c}}}]{yang2018maximal}%
  \BibitemOpen
  \bibfield  {author} {\bibinfo {author} {\bibfnamefont {Y.}~\bibnamefont
  {Yang}}, \bibinfo {author} {\bibfnamefont {A.}~\bibnamefont {Massuda}},
  \bibinfo {author} {\bibfnamefont {C.}~\bibnamefont {Roques-Carmes}}, \bibinfo
  {author} {\bibfnamefont {S.~E.}\ \bibnamefont {Kooi}}, \bibinfo {author}
  {\bibfnamefont {T.}~\bibnamefont {Christensen}}, \bibinfo {author}
  {\bibfnamefont {S.~G.}\ \bibnamefont {Johnson}}, \bibinfo {author}
  {\bibfnamefont {J.~D.}\ \bibnamefont {Joannopoulos}}, \bibinfo {author}
  {\bibfnamefont {O.~D.}\ \bibnamefont {Miller}}, \bibinfo {author}
  {\bibfnamefont {I.}~\bibnamefont {Kaminer}},\ and\ \bibinfo {author}
  {\bibfnamefont {M.}~\bibnamefont {Solja{\v{c}}i{\'c}}},\ }\bibfield  {title}
  {\bibinfo {title} {Maximal spontaneous photon emission and energy loss from
  free electrons},\ }\href {https://doi.org/10.1038/s41567-018-0180-2}
  {\bibfield  {journal} {\bibinfo  {journal} {Nature Physics}\ }\textbf
  {\bibinfo {volume} {14}},\ \bibinfo {pages} {894} (\bibinfo {year}
  {2018})}\BibitemShut {NoStop}%
\bibitem [{\citenamefont {Michon}\ \emph {et~al.}(2019)\citenamefont {Michon},
  \citenamefont {Benzaouia}, \citenamefont {Yao}, \citenamefont {Miller},\ and\
  \citenamefont {Johnson}}]{michon2019limits}%
  \BibitemOpen
  \bibfield  {author} {\bibinfo {author} {\bibfnamefont {J.}~\bibnamefont
  {Michon}}, \bibinfo {author} {\bibfnamefont {M.}~\bibnamefont {Benzaouia}},
  \bibinfo {author} {\bibfnamefont {W.}~\bibnamefont {Yao}}, \bibinfo {author}
  {\bibfnamefont {O.~D.}\ \bibnamefont {Miller}},\ and\ \bibinfo {author}
  {\bibfnamefont {S.~G.}\ \bibnamefont {Johnson}},\ }\bibfield  {title}
  {\bibinfo {title} {Limits to surface-enhanced raman scattering near
  arbitrary-shape scatterers},\ }\href {https://doi.org/10.1364/OE.27.035189}
  {\bibfield  {journal} {\bibinfo  {journal} {Optics Express}\ }\textbf
  {\bibinfo {volume} {27}},\ \bibinfo {pages} {35189} (\bibinfo {year}
  {2019})}\BibitemShut {NoStop}%
\bibitem [{\citenamefont {Nordebo}\ \emph {et~al.}(2019)\citenamefont
  {Nordebo}, \citenamefont {Kristensson}, \citenamefont {Mirmoosa},\ and\
  \citenamefont {Tretyakov}}]{nordebo2019optimal}%
  \BibitemOpen
  \bibfield  {author} {\bibinfo {author} {\bibfnamefont {S.}~\bibnamefont
  {Nordebo}}, \bibinfo {author} {\bibfnamefont {G.}~\bibnamefont
  {Kristensson}}, \bibinfo {author} {\bibfnamefont {M.}~\bibnamefont
  {Mirmoosa}},\ and\ \bibinfo {author} {\bibfnamefont {S.}~\bibnamefont
  {Tretyakov}},\ }\bibfield  {title} {\bibinfo {title} {Optimal plasmonic
  multipole resonances of a sphere in lossy media},\ }\href
  {https://doi.org/10.1103/PhysRevB.99.054301} {\bibfield  {journal} {\bibinfo
  {journal} {Physical Review B}\ }\textbf {\bibinfo {volume} {99}},\ \bibinfo
  {pages} {054301} (\bibinfo {year} {2019})}\BibitemShut {NoStop}%
\bibitem [{\citenamefont {Shim}\ \emph
  {et~al.}(2019{\natexlab{a}})\citenamefont {Shim}, \citenamefont {Fan},
  \citenamefont {Johnson},\ and\ \citenamefont {Miller}}]{shim2019fundamental}%
  \BibitemOpen
  \bibfield  {author} {\bibinfo {author} {\bibfnamefont {H.}~\bibnamefont
  {Shim}}, \bibinfo {author} {\bibfnamefont {L.}~\bibnamefont {Fan}}, \bibinfo
  {author} {\bibfnamefont {S.~G.}\ \bibnamefont {Johnson}},\ and\ \bibinfo
  {author} {\bibfnamefont {O.~D.}\ \bibnamefont {Miller}},\ }\bibfield  {title}
  {\bibinfo {title} {Fundamental limits to near-field optical response over any
  bandwidth},\ }\href {https://doi.org/10.1103/PhysRevX.9.011043} {\bibfield
  {journal} {\bibinfo  {journal} {Physical Review X}\ }\textbf {\bibinfo
  {volume} {9}},\ \bibinfo {pages} {011043} (\bibinfo {year}
  {2019}{\natexlab{a}})}\BibitemShut {NoStop}%
\bibitem [{\citenamefont {Ivanenko}\ \emph {et~al.}(2019)\citenamefont
  {Ivanenko}, \citenamefont {Gustafsson},\ and\ \citenamefont
  {Nordebo}}]{ivanenko2019optical}%
  \BibitemOpen
  \bibfield  {author} {\bibinfo {author} {\bibfnamefont {Y.}~\bibnamefont
  {Ivanenko}}, \bibinfo {author} {\bibfnamefont {M.}~\bibnamefont
  {Gustafsson}},\ and\ \bibinfo {author} {\bibfnamefont {S.}~\bibnamefont
  {Nordebo}},\ }\bibfield  {title} {\bibinfo {title} {Optical theorems and
  physical bounds on absorption in lossy media},\ }\href
  {https://doi.org/10.1364/OE.27.034323} {\bibfield  {journal} {\bibinfo
  {journal} {Optics Express}\ }\textbf {\bibinfo {volume} {27}},\ \bibinfo
  {pages} {34323} (\bibinfo {year} {2019})}\BibitemShut {NoStop}%
\bibitem [{\citenamefont {Dias}\ and\ \citenamefont {García~de
  Abajo}(2019)}]{dias_fundamental_2019}%
  \BibitemOpen
  \bibfield  {author} {\bibinfo {author} {\bibfnamefont {E.~J.~C.}\
  \bibnamefont {Dias}}\ and\ \bibinfo {author} {\bibfnamefont {F.~J.}\
  \bibnamefont {García~de Abajo}},\ }\bibfield  {title} {\bibinfo {title}
  {Fundamental limits to the coupling between light and 2d polaritons by small
  scatterers},\ }\href {https://doi.org/10.1021/acsnano.8b09283} {\bibfield
  {journal} {\bibinfo  {journal} {{ACS} Nano}\ }\textbf {\bibinfo {volume}
  {13}},\ \bibinfo {pages} {5184} (\bibinfo {year} {2019})}\BibitemShut
  {NoStop}%
\bibitem [{\citenamefont {Jackson}(1999)}]{jackson1999classical}%
  \BibitemOpen
  \bibfield  {author} {\bibinfo {author} {\bibfnamefont {J.~D.}\ \bibnamefont
  {Jackson}},\ }\href@noop {} {\bibinfo {title} {Classical electrodynamics}}
  (\bibinfo {year} {1999})\BibitemShut {NoStop}%
\bibitem [{\citenamefont {Newton}(1976)}]{newton1976optical}%
  \BibitemOpen
  \bibfield  {author} {\bibinfo {author} {\bibfnamefont {R.~G.}\ \bibnamefont
  {Newton}},\ }\bibfield  {title} {\bibinfo {title} {Optical theorem and
  beyond},\ }\href {https://doi.org/10.1119/1.10324} {\bibfield  {journal}
  {\bibinfo  {journal} {American Journal of Physics}\ }\textbf {\bibinfo
  {volume} {44}},\ \bibinfo {pages} {639} (\bibinfo {year} {1976})}\BibitemShut
  {NoStop}%
\bibitem [{\citenamefont {Bohren}\ and\ \citenamefont
  {Huffman}(2008)}]{bohren2008absorption}%
  \BibitemOpen
  \bibfield  {author} {\bibinfo {author} {\bibfnamefont {C.~F.}\ \bibnamefont
  {Bohren}}\ and\ \bibinfo {author} {\bibfnamefont {D.~R.}\ \bibnamefont
  {Huffman}},\ }\href@noop {} {\emph {\bibinfo {title} {Absorption and
  scattering of light by small particles}}}\ (\bibinfo  {publisher} {John Wiley
  \& Sons},\ \bibinfo {year} {2008})\BibitemShut {NoStop}%
\bibitem [{\citenamefont {Carney}\ \emph {et~al.}(2004)\citenamefont {Carney},
  \citenamefont {Schotland},\ and\ \citenamefont
  {Wolf}}]{carney2004generalized}%
  \BibitemOpen
  \bibfield  {author} {\bibinfo {author} {\bibfnamefont {P.~S.}\ \bibnamefont
  {Carney}}, \bibinfo {author} {\bibfnamefont {J.~C.}\ \bibnamefont
  {Schotland}},\ and\ \bibinfo {author} {\bibfnamefont {E.}~\bibnamefont
  {Wolf}},\ }\bibfield  {title} {\bibinfo {title} {Generalized optical theorem
  for reflection, transmission, and extinction of power for scalar fields},\
  }\href {https://doi.org/10.1103/PhysRevE.70.036611} {\bibfield  {journal}
  {\bibinfo  {journal} {Physical Review E}\ }\textbf {\bibinfo {volume} {70}},\
  \bibinfo {pages} {036611} (\bibinfo {year} {2004})}\BibitemShut {NoStop}%
\bibitem [{\citenamefont {Popoff}\ \emph {et~al.}(2014)\citenamefont {Popoff},
  \citenamefont {Goetschy}, \citenamefont {Liew}, \citenamefont {Stone},\ and\
  \citenamefont {Cao}}]{Popoff2014}%
  \BibitemOpen
  \bibfield  {author} {\bibinfo {author} {\bibfnamefont {S.~M.}\ \bibnamefont
  {Popoff}}, \bibinfo {author} {\bibfnamefont {A.}~\bibnamefont {Goetschy}},
  \bibinfo {author} {\bibfnamefont {S.~F.}\ \bibnamefont {Liew}}, \bibinfo
  {author} {\bibfnamefont {A.~D.}\ \bibnamefont {Stone}},\ and\ \bibinfo
  {author} {\bibfnamefont {H.}~\bibnamefont {Cao}},\ }\bibfield  {title}
  {\bibinfo {title} {{Coherent control of total transmission of light through
  disordered media}},\ }\href {https://doi.org/10.1103/PhysRevLett.112.133903}
  {\bibfield  {journal} {\bibinfo  {journal} {Phys. Rev. Lett.}\ }\textbf
  {\bibinfo {volume} {112}},\ \bibinfo {pages} {1} (\bibinfo {year} {2014})},\
  \Eprint {https://arxiv.org/abs/1308.0781} {1308.0781} \BibitemShut {NoStop}%
\bibitem [{\citenamefont {Vellekoop}(2015)}]{Vellekoop2015}%
  \BibitemOpen
  \bibfield  {author} {\bibinfo {author} {\bibfnamefont {I.~M.}\ \bibnamefont
  {Vellekoop}},\ }\bibfield  {title} {\bibinfo {title} {{Feedback-based
  wavefront shaping}},\ }\href {https://doi.org/10.1364/oe.23.012189}
  {\bibfield  {journal} {\bibinfo  {journal} {Opt. Express}\ }\textbf {\bibinfo
  {volume} {23}},\ \bibinfo {pages} {12189} (\bibinfo {year} {2015})},\ \Eprint
  {https://arxiv.org/abs/arXiv:1310.5736} {arXiv:1310.5736} \BibitemShut
  {NoStop}%
\bibitem [{\citenamefont {Horstmeyer}\ \emph {et~al.}(2015)\citenamefont
  {Horstmeyer}, \citenamefont {Ruan},\ and\ \citenamefont
  {Yang}}]{Horstmeyer2015}%
  \BibitemOpen
  \bibfield  {author} {\bibinfo {author} {\bibfnamefont {R.}~\bibnamefont
  {Horstmeyer}}, \bibinfo {author} {\bibfnamefont {H.}~\bibnamefont {Ruan}},\
  and\ \bibinfo {author} {\bibfnamefont {C.}~\bibnamefont {Yang}},\ }\bibfield
  {title} {\bibinfo {title} {{Guidestar-assisted wavefront-shaping methods for
  focusing light into biological tissue}},\ }\href
  {https://doi.org/10.1038/nphoton.2015.140} {\bibfield  {journal} {\bibinfo
  {journal} {Nat. Photonics}\ }\textbf {\bibinfo {volume} {9}},\ \bibinfo
  {pages} {563} (\bibinfo {year} {2015})}\BibitemShut {NoStop}%
\bibitem [{\citenamefont {Jang}\ \emph {et~al.}(2018)\citenamefont {Jang},
  \citenamefont {Horie}, \citenamefont {Shibukawa}, \citenamefont {Brake},
  \citenamefont {Liu}, \citenamefont {Kamali}, \citenamefont {Arbabi},
  \citenamefont {Ruan}, \citenamefont {Faraon},\ and\ \citenamefont
  {Yang}}]{Jang2018}%
  \BibitemOpen
  \bibfield  {author} {\bibinfo {author} {\bibfnamefont {M.}~\bibnamefont
  {Jang}}, \bibinfo {author} {\bibfnamefont {Y.}~\bibnamefont {Horie}},
  \bibinfo {author} {\bibfnamefont {A.}~\bibnamefont {Shibukawa}}, \bibinfo
  {author} {\bibfnamefont {J.}~\bibnamefont {Brake}}, \bibinfo {author}
  {\bibfnamefont {Y.}~\bibnamefont {Liu}}, \bibinfo {author} {\bibfnamefont
  {S.~M.}\ \bibnamefont {Kamali}}, \bibinfo {author} {\bibfnamefont
  {A.}~\bibnamefont {Arbabi}}, \bibinfo {author} {\bibfnamefont
  {H.}~\bibnamefont {Ruan}}, \bibinfo {author} {\bibfnamefont {A.}~\bibnamefont
  {Faraon}},\ and\ \bibinfo {author} {\bibfnamefont {C.}~\bibnamefont {Yang}},\
  }\bibfield  {title} {\bibinfo {title} {{Wavefront shaping with
  disorder-engineered metasurfaces}},\ }\href
  {https://doi.org/10.1038/s41566-017-0078-z} {\bibfield  {journal} {\bibinfo
  {journal} {Nat. Photonics}\ }\textbf {\bibinfo {volume} {12}},\ \bibinfo
  {pages} {84} (\bibinfo {year} {2018})}\BibitemShut {NoStop}%
\bibitem [{\citenamefont {Polder}\ and\ \citenamefont {{Van
  Hove}}(1971)}]{Polder1971}%
  \BibitemOpen
  \bibfield  {author} {\bibinfo {author} {\bibfnamefont {D.}~\bibnamefont
  {Polder}}\ and\ \bibinfo {author} {\bibfnamefont {M.}~\bibnamefont {{Van
  Hove}}},\ }\bibfield  {title} {\bibinfo {title} {{Theory of Radiative Heat
  Transfer between Closely Spaced Bodies}},\ }\href
  {https://doi.org/10.1103/PhysRevB.4.3303} {\bibfield  {journal} {\bibinfo
  {journal} {Phys. Rev. B}\ }\textbf {\bibinfo {volume} {4}},\ \bibinfo {pages}
  {3303} (\bibinfo {year} {1971})}\BibitemShut {NoStop}%
\bibitem [{\citenamefont {Otey}\ \emph {et~al.}(2014)\citenamefont {Otey},
  \citenamefont {Zhu}, \citenamefont {Sandhu},\ and\ \citenamefont
  {Fan}}]{Otey2014}%
  \BibitemOpen
  \bibfield  {author} {\bibinfo {author} {\bibfnamefont {C.~R.}\ \bibnamefont
  {Otey}}, \bibinfo {author} {\bibfnamefont {L.}~\bibnamefont {Zhu}}, \bibinfo
  {author} {\bibfnamefont {S.}~\bibnamefont {Sandhu}},\ and\ \bibinfo {author}
  {\bibfnamefont {S.}~\bibnamefont {Fan}},\ }\bibfield  {title} {\bibinfo
  {title} {{Fluctuational electrodynamics calculations of near-field heat
  transfer in non-planar geometries: A brief overview}},\ }\href
  {https://doi.org/10.1016/j.jqsrt.2013.04.017} {\bibfield  {journal} {\bibinfo
   {journal} {J. Quant. Spectrosc. Radiat. Transf.}\ }\textbf {\bibinfo
  {volume} {132}},\ \bibinfo {pages} {3} (\bibinfo {year} {2014})}\BibitemShut
  {NoStop}%
\bibitem [{\citenamefont {Rodriguez}\ \emph {et~al.}(2012)\citenamefont
  {Rodriguez}, \citenamefont {Reid},\ and\ \citenamefont
  {Johnson}}]{rodriguez_fluctuating-surface-current_2012}%
  \BibitemOpen
  \bibfield  {author} {\bibinfo {author} {\bibfnamefont {A.~W.}\ \bibnamefont
  {Rodriguez}}, \bibinfo {author} {\bibfnamefont {M.~T.~H.}\ \bibnamefont
  {Reid}},\ and\ \bibinfo {author} {\bibfnamefont {S.~G.}\ \bibnamefont
  {Johnson}},\ }\bibfield  {title} {{\selectlanguage {en}\bibinfo {title}
  {Fluctuating-surface-current formulation of radiative heat transfer for
  arbitrary geometries}},\ }\href {https://doi.org/10.1103/PhysRevB.86.220302}
  {\bibfield  {journal} {\bibinfo  {journal} {Physical Review B}\ }\textbf
  {\bibinfo {volume} {86}},\ \bibinfo {pages} {220302} (\bibinfo {year}
  {2012})}\BibitemShut {NoStop}%
\bibitem [{\citenamefont {Mazilu}\ \emph {et~al.}(2011)\citenamefont {Mazilu},
  \citenamefont {Baumgartl}, \citenamefont {Kosmeier},\ and\ \citenamefont
  {Dholakia}}]{Mazilu2011}%
  \BibitemOpen
  \bibfield  {author} {\bibinfo {author} {\bibfnamefont {M.}~\bibnamefont
  {Mazilu}}, \bibinfo {author} {\bibfnamefont {J.}~\bibnamefont {Baumgartl}},
  \bibinfo {author} {\bibfnamefont {S.}~\bibnamefont {Kosmeier}},\ and\
  \bibinfo {author} {\bibfnamefont {K.}~\bibnamefont {Dholakia}},\ }\bibfield
  {title} {\bibinfo {title} {{Optical Eigenmodes; exploiting the quadratic
  nature of the light-matter interaction}},\ }\href
  {https://doi.org/10.1364/OE.19.000933} {\bibfield  {journal} {\bibinfo
  {journal} {Opt. Express}\ }\textbf {\bibinfo {volume} {19}},\ \bibinfo
  {pages} {933} (\bibinfo {year} {2011})}\BibitemShut {NoStop}%
\bibitem [{\citenamefont {Lee}\ \emph {et~al.}(2017)\citenamefont {Lee},
  \citenamefont {Miller}, \citenamefont {Reid}, \citenamefont {Johnson},\ and\
  \citenamefont {Fang}}]{Lee2017}%
  \BibitemOpen
  \bibfield  {author} {\bibinfo {author} {\bibfnamefont {Y.~E.}\ \bibnamefont
  {Lee}}, \bibinfo {author} {\bibfnamefont {O.~D.}\ \bibnamefont {Miller}},
  \bibinfo {author} {\bibfnamefont {M.~T.~H.}\ \bibnamefont {Reid}}, \bibinfo
  {author} {\bibfnamefont {S.~G.}\ \bibnamefont {Johnson}},\ and\ \bibinfo
  {author} {\bibfnamefont {N.~X.}\ \bibnamefont {Fang}},\ }\bibfield  {title}
  {\bibinfo {title} {{Computational inverse design of non-intuitive
  illumination patterns to maximize optical force or torque}},\ }\href
  {https://doi.org/10.1364/OE.25.006757} {\bibfield  {journal} {\bibinfo
  {journal} {Opt. Express}\ }\textbf {\bibinfo {volume} {25}},\ \bibinfo
  {pages} {6757} (\bibinfo {year} {2017})}\BibitemShut {NoStop}%
\bibitem [{\citenamefont {Liu}\ \emph {et~al.}(2019)\citenamefont {Liu},
  \citenamefont {Fan}, \citenamefont {Lee}, \citenamefont {Fang}, \citenamefont
  {Johnson},\ and\ \citenamefont {Miller}}]{Liu2019}%
  \BibitemOpen
  \bibfield  {author} {\bibinfo {author} {\bibfnamefont {Y.}~\bibnamefont
  {Liu}}, \bibinfo {author} {\bibfnamefont {L.}~\bibnamefont {Fan}}, \bibinfo
  {author} {\bibfnamefont {Y.~E.}\ \bibnamefont {Lee}}, \bibinfo {author}
  {\bibfnamefont {N.~X.}\ \bibnamefont {Fang}}, \bibinfo {author}
  {\bibfnamefont {S.~G.}\ \bibnamefont {Johnson}},\ and\ \bibinfo {author}
  {\bibfnamefont {O.~D.}\ \bibnamefont {Miller}},\ }\bibfield  {title}
  {\bibinfo {title} {Optimal nanoparticle forces, torques, and illumination
  fields},\ }\href {https://doi.org/10.1021/acsphotonics.8b01263} {\bibfield
  {journal} {\bibinfo  {journal} {ACS Photonics}\ }\textbf {\bibinfo {volume}
  {6}},\ \bibinfo {pages} {395} (\bibinfo {year} {2019})},\ \Eprint
  {https://arxiv.org/abs/1805.11471} {1805.11471} \BibitemShut {NoStop}%
\bibitem [{\citenamefont {Horodynski}\ \emph {et~al.}(2019)\citenamefont
  {Horodynski}, \citenamefont {K{\"{u}}hmayer}, \citenamefont
  {Brandst{\"{o}}tter}, \citenamefont {Pichler}, \citenamefont {Fyodorov},
  \citenamefont {Kuhl},\ and\ \citenamefont {Rotter}}]{Horodynski2019}%
  \BibitemOpen
  \bibfield  {author} {\bibinfo {author} {\bibfnamefont {M.}~\bibnamefont
  {Horodynski}}, \bibinfo {author} {\bibfnamefont {M.}~\bibnamefont
  {K{\"{u}}hmayer}}, \bibinfo {author} {\bibfnamefont {A.}~\bibnamefont
  {Brandst{\"{o}}tter}}, \bibinfo {author} {\bibfnamefont {K.}~\bibnamefont
  {Pichler}}, \bibinfo {author} {\bibfnamefont {Y.~V.}\ \bibnamefont
  {Fyodorov}}, \bibinfo {author} {\bibfnamefont {U.}~\bibnamefont {Kuhl}},\
  and\ \bibinfo {author} {\bibfnamefont {S.}~\bibnamefont {Rotter}},\
  }\bibfield  {title} {\bibinfo {title} {{Optimal wave fields for
  micromanipulation in complex scattering environments}},\ }\bibfield
  {journal} {\bibinfo  {journal} {Nat. Photonics}\ }\href
  {https://doi.org/10.1038/s41566-019-0550-z} {10.1038/s41566-019-0550-z}
  (\bibinfo {year} {2019})\BibitemShut {NoStop}%
\bibitem [{\citenamefont {Lalanne}\ \emph {et~al.}(1999)\citenamefont
  {Lalanne}, \citenamefont {Astilean}, \citenamefont {Chavel}, \citenamefont
  {Cambril},\ and\ \citenamefont {Launois}}]{lalanne_design_1999}%
  \BibitemOpen
  \bibfield  {author} {\bibinfo {author} {\bibfnamefont {P.}~\bibnamefont
  {Lalanne}}, \bibinfo {author} {\bibfnamefont {S.}~\bibnamefont {Astilean}},
  \bibinfo {author} {\bibfnamefont {P.}~\bibnamefont {Chavel}}, \bibinfo
  {author} {\bibfnamefont {E.}~\bibnamefont {Cambril}},\ and\ \bibinfo {author}
  {\bibfnamefont {H.}~\bibnamefont {Launois}},\ }\bibfield  {title}
  {{\selectlanguage {EN}\bibinfo {title} {Design and fabrication of blazed
  binary diffractive elements with sampling periods smaller than the structural
  cutoff}},\ }\href {https://doi.org/10.1364/JOSAA.16.001143} {\bibfield
  {journal} {\bibinfo  {journal} {JOSA A}\ }\textbf {\bibinfo {volume} {16}},\
  \bibinfo {pages} {1143} (\bibinfo {year} {1999})},\ \bibinfo {note}
  {publisher: Optical Society of America}\BibitemShut {NoStop}%
\bibitem [{\citenamefont {Lalanne}\ and\ \citenamefont
  {Chavel}(2017)}]{lalanne_metalenses_2017-1}%
  \BibitemOpen
  \bibfield  {author} {\bibinfo {author} {\bibfnamefont {P.}~\bibnamefont
  {Lalanne}}\ and\ \bibinfo {author} {\bibfnamefont {P.}~\bibnamefont
  {Chavel}},\ }\bibfield  {title} {\bibinfo {title} {Metalenses at visible
  wavelengths: past, present, perspectives},\ }\href
  {https://doi.org/https://doi.org/10.1002/lpor.201600295} {\bibfield
  {journal} {\bibinfo  {journal} {Laser \& Photonics Reviews}\ }\textbf
  {\bibinfo {volume} {11}},\ \bibinfo {pages} {1600295} (\bibinfo {year}
  {2017})},\ \bibinfo {note} {\_eprint:
  https://onlinelibrary.wiley.com/doi/pdf/10.1002/lpor.201600295}\BibitemShut
  {NoStop}%
\bibitem [{\citenamefont {Chung}\ \emph {et~al.}(2020)\citenamefont {Chung},
  \citenamefont {Chung}, \citenamefont {Miller},\ and\ \citenamefont
  {Miller}}]{chung_high-na_2020}%
  \BibitemOpen
  \bibfield  {author} {\bibinfo {author} {\bibfnamefont {H.}~\bibnamefont
  {Chung}}, \bibinfo {author} {\bibfnamefont {H.}~\bibnamefont {Chung}},
  \bibinfo {author} {\bibfnamefont {O.~D.}\ \bibnamefont {Miller}},\ and\
  \bibinfo {author} {\bibfnamefont {O.~D.}\ \bibnamefont {Miller}},\ }\bibfield
   {title} {{\selectlanguage {EN}\bibinfo {title} {High-{NA} achromatic
  metalenses by inverse design}},\ }\href {https://doi.org/10.1364/OE.385440}
  {\bibfield  {journal} {\bibinfo  {journal} {Optics Express}\ }\textbf
  {\bibinfo {volume} {28}},\ \bibinfo {pages} {6945} (\bibinfo {year}
  {2020})},\ \bibinfo {note} {publisher: Optical Society of
  America}\BibitemShut {NoStop}%
\bibitem [{\citenamefont {Miller}(2019)}]{Miller2019}%
  \BibitemOpen
  \bibfield  {author} {\bibinfo {author} {\bibfnamefont {D.~A.~B.}\
  \bibnamefont {Miller}},\ }\bibfield  {title} {\bibinfo {title} {{Waves,
  modes, communications, and optics: a tutorial}},\ }\href
  {https://doi.org/10.1364/aop.11.000679} {\bibfield  {journal} {\bibinfo
  {journal} {Adv. Opt. Photonics}\ }\textbf {\bibinfo {volume} {11}},\ \bibinfo
  {pages} {679} (\bibinfo {year} {2019})}\BibitemShut {NoStop}%
\bibitem [{\citenamefont {Newton}(2013)}]{newton2013scattering}%
  \BibitemOpen
  \bibfield  {author} {\bibinfo {author} {\bibfnamefont {R.~G.}\ \bibnamefont
  {Newton}},\ }\href@noop {} {\emph {\bibinfo {title} {Scattering theory of
  waves and particles}}}\ (\bibinfo  {publisher} {Springer Science \& Business
  Media},\ \bibinfo {year} {2013})\BibitemShut {NoStop}%
\bibitem [{\citenamefont {Mahaux}\ and\ \citenamefont
  {Weidenm{\"u}ller}(1969)}]{mahaux1969shell}%
  \BibitemOpen
  \bibfield  {author} {\bibinfo {author} {\bibfnamefont {C.}~\bibnamefont
  {Mahaux}}\ and\ \bibinfo {author} {\bibfnamefont {H.~A.}\ \bibnamefont
  {Weidenm{\"u}ller}},\ }\bibfield  {title} {\bibinfo {title} {Shell-model
  approach to nuclear reactions.},\ }\href@noop {} {\bibfield  {journal}
  {\bibinfo  {journal} {Soft Matter}\ } (\bibinfo {year} {1969})}\BibitemShut
  {NoStop}%
\bibitem [{\citenamefont {Jalas}\ \emph {et~al.}(2013)\citenamefont {Jalas},
  \citenamefont {Petrov}, \citenamefont {Eich}, \citenamefont {Freude},
  \citenamefont {Fan}, \citenamefont {Yu}, \citenamefont {Baets}, \citenamefont
  {Popovi{\'c}}, \citenamefont {Melloni}, \citenamefont {Joannopoulos} \emph
  {et~al.}}]{jalas2013and}%
  \BibitemOpen
  \bibfield  {author} {\bibinfo {author} {\bibfnamefont {D.}~\bibnamefont
  {Jalas}}, \bibinfo {author} {\bibfnamefont {A.}~\bibnamefont {Petrov}},
  \bibinfo {author} {\bibfnamefont {M.}~\bibnamefont {Eich}}, \bibinfo {author}
  {\bibfnamefont {W.}~\bibnamefont {Freude}}, \bibinfo {author} {\bibfnamefont
  {S.}~\bibnamefont {Fan}}, \bibinfo {author} {\bibfnamefont {Z.}~\bibnamefont
  {Yu}}, \bibinfo {author} {\bibfnamefont {R.}~\bibnamefont {Baets}}, \bibinfo
  {author} {\bibfnamefont {M.}~\bibnamefont {Popovi{\'c}}}, \bibinfo {author}
  {\bibfnamefont {A.}~\bibnamefont {Melloni}}, \bibinfo {author} {\bibfnamefont
  {J.~D.}\ \bibnamefont {Joannopoulos}}, \emph {et~al.},\ }\bibfield  {title}
  {\bibinfo {title} {What is---and what is not---an optical isolator},\ }\href
  {https://doi.org/10.1038/nphoton.2013.185} {\bibfield  {journal} {\bibinfo
  {journal} {Nature Photonics}\ }\textbf {\bibinfo {volume} {7}},\ \bibinfo
  {pages} {579} (\bibinfo {year} {2013})}\BibitemShut {NoStop}%
\bibitem [{\citenamefont {Rotter}\ and\ \citenamefont
  {Gigan}(2017)}]{rotter2017light}%
  \BibitemOpen
  \bibfield  {author} {\bibinfo {author} {\bibfnamefont {S.}~\bibnamefont
  {Rotter}}\ and\ \bibinfo {author} {\bibfnamefont {S.}~\bibnamefont {Gigan}},\
  }\bibfield  {title} {\bibinfo {title} {Light fields in complex media:
  Mesoscopic scattering meets wave control},\ }\href
  {https://doi.org/10.1103/RevModPhys.89.015005} {\bibfield  {journal}
  {\bibinfo  {journal} {Reviews of Modern Physics}\ }\textbf {\bibinfo {volume}
  {89}},\ \bibinfo {pages} {015005} (\bibinfo {year} {2017})}\BibitemShut
  {NoStop}%
\bibitem [{\citenamefont {Stutzman}\ and\ \citenamefont
  {Thiele}(2012)}]{Stutzman2012}%
  \BibitemOpen
  \bibfield  {author} {\bibinfo {author} {\bibfnamefont {W.~L.}\ \bibnamefont
  {Stutzman}}\ and\ \bibinfo {author} {\bibfnamefont {G.~A.}\ \bibnamefont
  {Thiele}},\ }\href@noop {} {\emph {\bibinfo {title} {{Antenna theory and
  design}}}},\ \bibinfo {edition} {3rd}\ ed.\ (\bibinfo  {publisher} {John
  Wiley {\&} Sons},\ \bibinfo {year} {2012})\BibitemShut {NoStop}%
\bibitem [{\citenamefont {Loudon}(2000)}]{Loudon2000}%
  \BibitemOpen
  \bibfield  {author} {\bibinfo {author} {\bibfnamefont {R.}~\bibnamefont
  {Loudon}},\ }\href@noop {} {\emph {\bibinfo {title} {{The Quantum Theory of
  Light}}}},\ \bibinfo {edition} {3rd}\ ed.\ (\bibinfo  {publisher} {Oxford
  University Press},\ \bibinfo {address} {New York},\ \bibinfo {year}
  {2000})\BibitemShut {NoStop}%
\bibitem [{\citenamefont {Molesky}\ \emph
  {et~al.}(2019{\natexlab{a}})\citenamefont {Molesky}, \citenamefont {Jin},
  \citenamefont {Venkataram},\ and\ \citenamefont {Rodriguez}}]{molesky2019t}%
  \BibitemOpen
  \bibfield  {author} {\bibinfo {author} {\bibfnamefont {S.}~\bibnamefont
  {Molesky}}, \bibinfo {author} {\bibfnamefont {W.}~\bibnamefont {Jin}},
  \bibinfo {author} {\bibfnamefont {P.~S.}\ \bibnamefont {Venkataram}},\ and\
  \bibinfo {author} {\bibfnamefont {A.~W.}\ \bibnamefont {Rodriguez}},\
  }\bibfield  {title} {\bibinfo {title} {T operator bounds on angle-integrated
  absorption and thermal radiation for arbitrary objects},\ }\href
  {https://doi.org/10.1103/PhysRevLett.123.257401} {\bibfield  {journal}
  {\bibinfo  {journal} {Physical Review Letters}\ }\textbf {\bibinfo {volume}
  {123}},\ \bibinfo {pages} {257401} (\bibinfo {year}
  {2019}{\natexlab{a}})}\BibitemShut {NoStop}%
\bibitem [{\citenamefont {Molesky}\ \emph
  {et~al.}(2019{\natexlab{b}})\citenamefont {Molesky}, \citenamefont
  {Venkataram}, \citenamefont {Jin},\ and\ \citenamefont
  {Rodriguez}}]{molesky2019fundamental}%
  \BibitemOpen
  \bibfield  {author} {\bibinfo {author} {\bibfnamefont {S.}~\bibnamefont
  {Molesky}}, \bibinfo {author} {\bibfnamefont {P.~S.}\ \bibnamefont
  {Venkataram}}, \bibinfo {author} {\bibfnamefont {W.}~\bibnamefont {Jin}},\
  and\ \bibinfo {author} {\bibfnamefont {A.~W.}\ \bibnamefont {Rodriguez}},\
  }\bibfield  {title} {\bibinfo {title} {Fundamental limits to radiative heat
  transfer: theory},\ }\href {http://arxiv.org/abs/1907.03000} {\bibfield
  {journal} {\bibinfo  {journal} {arXiv preprint arXiv:1907.03000}\ } (\bibinfo
  {year} {2019}{\natexlab{b}})}\BibitemShut {NoStop}%
\bibitem [{\citenamefont {Venkataram}\ \emph {et~al.}(2019)\citenamefont
  {Venkataram}, \citenamefont {Molesky}, \citenamefont {Chao},\ and\
  \citenamefont {Rodriguez}}]{venkataram2019fundamental}%
  \BibitemOpen
  \bibfield  {author} {\bibinfo {author} {\bibfnamefont {P.~S.}\ \bibnamefont
  {Venkataram}}, \bibinfo {author} {\bibfnamefont {S.}~\bibnamefont {Molesky}},
  \bibinfo {author} {\bibfnamefont {P.}~\bibnamefont {Chao}},\ and\ \bibinfo
  {author} {\bibfnamefont {A.~W.}\ \bibnamefont {Rodriguez}},\ }\bibfield
  {title} {\bibinfo {title} {Fundamental limits to attractive and repulsive
  casimir--polder forces},\ }\href {https://arxiv.org/abs/1911.10295}
  {\bibfield  {journal} {\bibinfo  {journal} {arXiv preprint arXiv:1911.10295}\
  } (\bibinfo {year} {2019})}\BibitemShut {NoStop}%
\bibitem [{\citenamefont {Shannon}(1948)}]{Shannon1948}%
  \BibitemOpen
  \bibfield  {author} {\bibinfo {author} {\bibfnamefont {C.~E.}\ \bibnamefont
  {Shannon}},\ }\bibfield  {title} {\bibinfo {title} {{A Mathematical Theory of
  Communication}},\ }\href@noop {} {\bibfield  {journal} {\bibinfo  {journal}
  {Bell Syst. Tech. J.}\ }\textbf {\bibinfo {volume} {XXVII}},\ \bibinfo
  {pages} {379} (\bibinfo {year} {1948})}\BibitemShut {NoStop}%
\bibitem [{\citenamefont {Shannon}\ and\ \citenamefont
  {Weaver}(1949)}]{Shannon1949}%
  \BibitemOpen
  \bibfield  {author} {\bibinfo {author} {\bibfnamefont {C.~E.}\ \bibnamefont
  {Shannon}}\ and\ \bibinfo {author} {\bibfnamefont {W.}~\bibnamefont
  {Weaver}},\ }\href@noop {} {\emph {\bibinfo {title} {{The Mathematical Theory
  of Communication}}}}\ (\bibinfo  {publisher} {Univ. of Illinois Press},\
  \bibinfo {address} {Urbana, IL},\ \bibinfo {year} {1949})\BibitemShut
  {NoStop}%
\bibitem [{\citenamefont {Cover}(1999)}]{Cover1999}%
  \BibitemOpen
  \bibfield  {author} {\bibinfo {author} {\bibfnamefont {T.~M.}\ \bibnamefont
  {Cover}},\ }\href@noop {} {\emph {\bibinfo {title} {{Elements of information
  theory}}}}\ (\bibinfo  {publisher} {John Wiley {\&} Sons},\ \bibinfo {year}
  {1999})\BibitemShut {NoStop}%
\bibitem [{\citenamefont {Hashin}\ and\ \citenamefont
  {Shtrikman}(1962)}]{Hashin1962}%
  \BibitemOpen
  \bibfield  {author} {\bibinfo {author} {\bibfnamefont {Z.}~\bibnamefont
  {Hashin}}\ and\ \bibinfo {author} {\bibfnamefont {S.}~\bibnamefont
  {Shtrikman}},\ }\bibfield  {title} {\bibinfo {title} {{A Variational approach
  to the theory of the effective magnetic permeability of multiphase
  materials}},\ }\href {https://doi.org/10.1063/1.1728579} {\bibfield
  {journal} {\bibinfo  {journal} {J. Appl. Phys.}\ }\textbf {\bibinfo {volume}
  {33}},\ \bibinfo {pages} {3125} (\bibinfo {year} {1962})}\BibitemShut
  {NoStop}%
\bibitem [{\citenamefont {Milton}(2002)}]{Milton2002}%
  \BibitemOpen
  \bibfield  {author} {\bibinfo {author} {\bibfnamefont {G.~W.}\ \bibnamefont
  {Milton}},\ }\href@noop {} {\emph {\bibinfo {title} {{The Theory of
  Composites}}}}\ (\bibinfo  {publisher} {Cambridge University Press},\
  \bibinfo {year} {2002})\BibitemShut {NoStop}%
\bibitem [{\citenamefont {Bergman}(1980)}]{Bergman1980}%
  \BibitemOpen
  \bibfield  {author} {\bibinfo {author} {\bibfnamefont {D.~J.}\ \bibnamefont
  {Bergman}},\ }\bibfield  {title} {\bibinfo {title} {{Exactly Solvable
  Microscopic Geometries and Rigorous Bounds for the Complex Dielectric
  Constant of a Two-Component Composite Material}},\ }\href@noop {} {\bibfield
  {journal} {\bibinfo  {journal} {Phys. Rev. Lett.}\ }\textbf {\bibinfo
  {volume} {44}},\ \bibinfo {pages} {1285} (\bibinfo {year}
  {1980})}\BibitemShut {NoStop}%
\bibitem [{\citenamefont {Milton}(1980)}]{Milton1980}%
  \BibitemOpen
  \bibfield  {author} {\bibinfo {author} {\bibfnamefont {G.~W.}\ \bibnamefont
  {Milton}},\ }\bibfield  {title} {\bibinfo {title} {{Bounds on the complex
  dielectric constant of a composite material}},\ }\href
  {https://doi.org/10.1063/1.91895} {\bibfield  {journal} {\bibinfo  {journal}
  {Appl. Phys. Lett.}\ }\textbf {\bibinfo {volume} {37}},\ \bibinfo {pages}
  {300} (\bibinfo {year} {1980})}\BibitemShut {NoStop}%
\bibitem [{\citenamefont {Gibiansky}\ and\ \citenamefont
  {Milton}(1993)}]{Gibiansky1993}%
  \BibitemOpen
  \bibfield  {author} {\bibinfo {author} {\bibfnamefont {L.~V.}\ \bibnamefont
  {Gibiansky}}\ and\ \bibinfo {author} {\bibfnamefont {G.~W.}\ \bibnamefont
  {Milton}},\ }\bibfield  {title} {\bibinfo {title} {{On the effective
  viscoelastic moduli of two-phase media. I. Rigorous bounds on the Complex
  Bulk Modulus}},\ }\href {https://doi.org/10.1098/rspa.1999.0395} {\bibfield
  {journal} {\bibinfo  {journal} {Proc. R. Soc. A Math. Phys. Eng. Sci.}\
  }\textbf {\bibinfo {volume} {440}},\ \bibinfo {pages} {163} (\bibinfo {year}
  {1993})}\BibitemShut {NoStop}%
\bibitem [{\citenamefont {Cherkaeva}\ and\ \citenamefont
  {Golden}(1998)}]{Cherkaeva1998}%
  \BibitemOpen
  \bibfield  {author} {\bibinfo {author} {\bibfnamefont {E.}~\bibnamefont
  {Cherkaeva}}\ and\ \bibinfo {author} {\bibfnamefont {K.~M.}\ \bibnamefont
  {Golden}},\ }\bibfield  {title} {\bibinfo {title} {{Inverse bounds for
  microstructural parameters of composite media derived from complex
  permittivity measurements}},\ }\href
  {https://doi.org/10.1088/0959-7174/8/4/004} {\bibfield  {journal} {\bibinfo
  {journal} {Waves in Random Media}\ }\textbf {\bibinfo {volume} {8}},\
  \bibinfo {pages} {437} (\bibinfo {year} {1998})}\BibitemShut {NoStop}%
\bibitem [{\citenamefont {Kern}\ \emph {et~al.}(2020)\citenamefont {Kern},
  \citenamefont {Miller},\ and\ \citenamefont {Milton}}]{Kern2020}%
  \BibitemOpen
  \bibfield  {author} {\bibinfo {author} {\bibfnamefont {C.}~\bibnamefont
  {Kern}}, \bibinfo {author} {\bibfnamefont {O.~D.}\ \bibnamefont {Miller}},\
  and\ \bibinfo {author} {\bibfnamefont {G.~W.}\ \bibnamefont {Milton}},\
  }\bibfield  {title} {\bibinfo {title} {{On the Range of Effective Complex
  Electrical Permittivities of Isotropic Composite Materials}},\ }\href@noop {}
  {\bibfield  {journal} {\bibinfo  {journal} {arXiv:2006.03830}\ } (\bibinfo
  {year} {2020})},\ \Eprint {https://arxiv.org/abs/arXiv:2006.03830}
  {arXiv:arXiv:2006.03830} \BibitemShut {NoStop}%
\bibitem [{\citenamefont {Chew}(1995)}]{chew1995waves}%
  \BibitemOpen
  \bibfield  {author} {\bibinfo {author} {\bibfnamefont {W.~C.}\ \bibnamefont
  {Chew}},\ }\href@noop {} {\emph {\bibinfo {title} {Waves and fields in
  inhomogeneous media}}}\ (\bibinfo  {publisher} {IEEE press},\ \bibinfo {year}
  {1995})\BibitemShut {NoStop}%
\bibitem [{\citenamefont {Yang}\ \emph {et~al.}(2019)\citenamefont {Yang},
  \citenamefont {Zhu}, \citenamefont {Yan}, \citenamefont {Agarwal},
  \citenamefont {Zheng}, \citenamefont {Joannopoulos}, \citenamefont {Lalanne},
  \citenamefont {Christensen}, \citenamefont {Berggren},\ and\ \citenamefont
  {Solja{\v{c}}i{\'{c}}}}]{Yang2019}%
  \BibitemOpen
  \bibfield  {author} {\bibinfo {author} {\bibfnamefont {Y.}~\bibnamefont
  {Yang}}, \bibinfo {author} {\bibfnamefont {D.}~\bibnamefont {Zhu}}, \bibinfo
  {author} {\bibfnamefont {W.}~\bibnamefont {Yan}}, \bibinfo {author}
  {\bibfnamefont {A.}~\bibnamefont {Agarwal}}, \bibinfo {author} {\bibfnamefont
  {M.}~\bibnamefont {Zheng}}, \bibinfo {author} {\bibfnamefont {J.~D.}\
  \bibnamefont {Joannopoulos}}, \bibinfo {author} {\bibfnamefont
  {P.}~\bibnamefont {Lalanne}}, \bibinfo {author} {\bibfnamefont
  {T.}~\bibnamefont {Christensen}}, \bibinfo {author} {\bibfnamefont {K.~K.}\
  \bibnamefont {Berggren}},\ and\ \bibinfo {author} {\bibfnamefont
  {M.}~\bibnamefont {Solja{\v{c}}i{\'{c}}}},\ }\bibfield  {title} {\bibinfo
  {title} {{A general theoretical and experimental framework for nanoscale
  electromagnetism}},\ }\href {https://doi.org/10.1038/s41586-019-1803-1}
  {\bibfield  {journal} {\bibinfo  {journal} {Nature}\ }\textbf {\bibinfo
  {volume} {576}},\ \bibinfo {pages} {248} (\bibinfo {year}
  {2019})}\BibitemShut {NoStop}%
\bibitem [{\citenamefont {Fallahi}\ \emph {et~al.}(2015)\citenamefont
  {Fallahi}, \citenamefont {Low}, \citenamefont {Tamagnone},\ and\
  \citenamefont {Perruisseau-Carrier}}]{Fallahi2015}%
  \BibitemOpen
  \bibfield  {author} {\bibinfo {author} {\bibfnamefont {A.}~\bibnamefont
  {Fallahi}}, \bibinfo {author} {\bibfnamefont {T.}~\bibnamefont {Low}},
  \bibinfo {author} {\bibfnamefont {M.}~\bibnamefont {Tamagnone}},\ and\
  \bibinfo {author} {\bibfnamefont {J.}~\bibnamefont {Perruisseau-Carrier}},\
  }\bibfield  {title} {\bibinfo {title} {{Nonlocal electromagnetic response of
  graphene nanostructures}},\ }\href
  {https://doi.org/10.1103/PhysRevB.91.121405} {\bibfield  {journal} {\bibinfo
  {journal} {Phys. Rev. B}\ }\textbf {\bibinfo {volume} {91}},\ \bibinfo
  {pages} {121405(R)} (\bibinfo {year} {2015})}\BibitemShut {NoStop}%
\bibitem [{\citenamefont {Polimeridis}\ \emph {et~al.}(2014)\citenamefont
  {Polimeridis}, \citenamefont {Reid}, \citenamefont {Johnson}, \citenamefont
  {White},\ and\ \citenamefont {Rodriguez}}]{polimeridis2014computation}%
  \BibitemOpen
  \bibfield  {author} {\bibinfo {author} {\bibfnamefont {A.~G.}\ \bibnamefont
  {Polimeridis}}, \bibinfo {author} {\bibfnamefont {M.~H.}\ \bibnamefont
  {Reid}}, \bibinfo {author} {\bibfnamefont {S.~G.}\ \bibnamefont {Johnson}},
  \bibinfo {author} {\bibfnamefont {J.~K.}\ \bibnamefont {White}},\ and\
  \bibinfo {author} {\bibfnamefont {A.~W.}\ \bibnamefont {Rodriguez}},\
  }\bibfield  {title} {\bibinfo {title} {On the computation of power in volume
  integral equation formulations},\ }\href
  {https://doi.org/10.1109/TAP.2014.2382636} {\bibfield  {journal} {\bibinfo
  {journal} {IEEE Transactions on Antennas and Propagation}\ }\textbf {\bibinfo
  {volume} {63}},\ \bibinfo {pages} {611} (\bibinfo {year} {2014})}\BibitemShut
  {NoStop}%
\bibitem [{\citenamefont {Kong}(1972)}]{kong1972theorems}%
  \BibitemOpen
  \bibfield  {author} {\bibinfo {author} {\bibfnamefont {J.~A.}\ \bibnamefont
  {Kong}},\ }\bibfield  {title} {\bibinfo {title} {Theorems of bianisotropic
  media},\ }\href {https://doi.org/10.1109/PROC.1972.8851} {\bibfield
  {journal} {\bibinfo  {journal} {Proceedings of the IEEE}\ }\textbf {\bibinfo
  {volume} {60}},\ \bibinfo {pages} {1036} (\bibinfo {year}
  {1972})}\BibitemShut {NoStop}%
\bibitem [{\citenamefont {Purcell}\ and\ \citenamefont
  {Pennypacker}(1973)}]{Purcell1973}%
  \BibitemOpen
  \bibfield  {author} {\bibinfo {author} {\bibfnamefont {E.~M.}\ \bibnamefont
  {Purcell}}\ and\ \bibinfo {author} {\bibfnamefont {C.~R.}\ \bibnamefont
  {Pennypacker}},\ }\bibfield  {title} {\bibinfo {title} {{Scattering and
  Absorption of Light by Nonspherical Dielectric Grains}},\ }\href
  {https://doi.org/10.1086/152538} {\bibfield  {journal} {\bibinfo  {journal}
  {Astrophys. J.}\ }\textbf {\bibinfo {volume} {186}},\ \bibinfo {pages} {705}
  (\bibinfo {year} {1973})}\BibitemShut {NoStop}%
\bibitem [{\citenamefont {Reid}\ \emph {et~al.}(2017)\citenamefont {Reid},
  \citenamefont {Miller}, \citenamefont {Polimeridis}, \citenamefont
  {Rodriguez}, \citenamefont {Tomlinson},\ and\ \citenamefont
  {Johnson}}]{Reid2017subm}%
  \BibitemOpen
  \bibfield  {author} {\bibinfo {author} {\bibfnamefont {M.~T.~H.}\
  \bibnamefont {Reid}}, \bibinfo {author} {\bibfnamefont {O.~D.}\ \bibnamefont
  {Miller}}, \bibinfo {author} {\bibfnamefont {A.~G.}\ \bibnamefont
  {Polimeridis}}, \bibinfo {author} {\bibfnamefont {A.~W.}\ \bibnamefont
  {Rodriguez}}, \bibinfo {author} {\bibfnamefont {E.~M.}\ \bibnamefont
  {Tomlinson}},\ and\ \bibinfo {author} {\bibfnamefont {S.~G.}\ \bibnamefont
  {Johnson}},\ }\bibfield  {title} {\bibinfo {title} {Photon torpedoes and
  rytov pinwheels: Integral-equation modeling of non-equilibrium
  fluctuation-induced forces and torques on nanoparticles},\ }\href@noop {}
  {\bibfield  {journal} {\bibinfo  {journal} {arXiv:1708.01985}\ } (\bibinfo
  {year} {2017})}\BibitemShut {NoStop}%
\bibitem [{\citenamefont {Welters}\ \emph {et~al.}(2014)\citenamefont
  {Welters}, \citenamefont {Avniel},\ and\ \citenamefont
  {Johnson}}]{welters2014speed}%
  \BibitemOpen
  \bibfield  {author} {\bibinfo {author} {\bibfnamefont {A.}~\bibnamefont
  {Welters}}, \bibinfo {author} {\bibfnamefont {Y.}~\bibnamefont {Avniel}},\
  and\ \bibinfo {author} {\bibfnamefont {S.~G.}\ \bibnamefont {Johnson}},\
  }\bibfield  {title} {\bibinfo {title} {Speed-of-light limitations in passive
  linear media},\ }\href {https://doi.org/10.1103/PhysRevA.90.023847}
  {\bibfield  {journal} {\bibinfo  {journal} {Physical Review A}\ }\textbf
  {\bibinfo {volume} {90}},\ \bibinfo {pages} {023847} (\bibinfo {year}
  {2014})}\BibitemShut {NoStop}%
\bibitem [{\citenamefont {Ben-Tal}\ and\ \citenamefont
  {Teboulle}(1996)}]{ben-tal_hidden_1996}%
  \BibitemOpen
  \bibfield  {author} {\bibinfo {author} {\bibfnamefont {A.}~\bibnamefont
  {Ben-Tal}}\ and\ \bibinfo {author} {\bibfnamefont {M.}~\bibnamefont
  {Teboulle}},\ }\bibfield  {title} {\bibinfo {title} {Hidden convexity in some
  nonconvex quadratically constrained quadratic programming},\ }\href
  {https://doi.org/10.1007/BF02592331} {\bibfield  {journal} {\bibinfo
  {journal} {Mathematical Programming}\ }\textbf {\bibinfo {volume} {72}},\
  \bibinfo {pages} {51} (\bibinfo {year} {1996})}\BibitemShut {NoStop}%
\bibitem [{\citenamefont {Boyd}\ and\ \citenamefont
  {Vandenberghe}(2004)}]{boyd2004convex}%
  \BibitemOpen
  \bibfield  {author} {\bibinfo {author} {\bibfnamefont {S.}~\bibnamefont
  {Boyd}}\ and\ \bibinfo {author} {\bibfnamefont {L.}~\bibnamefont
  {Vandenberghe}},\ }\href@noop {} {\emph {\bibinfo {title} {Convex
  optimization}}}\ (\bibinfo  {publisher} {Cambridge university press},\
  \bibinfo {year} {2004})\BibitemShut {NoStop}%
\bibitem [{\citenamefont {Johnson}\ and\ \citenamefont
  {Christy}(1972)}]{johnson1972optical}%
  \BibitemOpen
  \bibfield  {author} {\bibinfo {author} {\bibfnamefont {P.~B.}\ \bibnamefont
  {Johnson}}\ and\ \bibinfo {author} {\bibfnamefont {R.-W.}\ \bibnamefont
  {Christy}},\ }\bibfield  {title} {\bibinfo {title} {Optical constants of the
  noble metals},\ }\href {https://doi.org/10.1103/PhysRevB.6.4370} {\bibfield
  {journal} {\bibinfo  {journal} {Physical review B}\ }\textbf {\bibinfo
  {volume} {6}},\ \bibinfo {pages} {4370} (\bibinfo {year} {1972})}\BibitemShut
  {NoStop}%
\bibitem [{\citenamefont {Novotny}\ and\ \citenamefont
  {Hecht}(2012)}]{novotny2012principles}%
  \BibitemOpen
  \bibfield  {author} {\bibinfo {author} {\bibfnamefont {L.}~\bibnamefont
  {Novotny}}\ and\ \bibinfo {author} {\bibfnamefont {B.}~\bibnamefont
  {Hecht}},\ }\href@noop {} {\emph {\bibinfo {title} {Principles of
  nano-optics}}}\ (\bibinfo  {publisher} {Cambridge university press},\
  \bibinfo {year} {2012})\BibitemShut {NoStop}%
\bibitem [{\citenamefont {Liang}\ and\ \citenamefont
  {Johnson}(2013)}]{liang2013formulation}%
  \BibitemOpen
  \bibfield  {author} {\bibinfo {author} {\bibfnamefont {X.}~\bibnamefont
  {Liang}}\ and\ \bibinfo {author} {\bibfnamefont {S.~G.}\ \bibnamefont
  {Johnson}},\ }\bibfield  {title} {\bibinfo {title} {Formulation for scalable
  optimization of microcavities via the frequency-averaged local density of
  states},\ }\href {https://doi.org/10.1364/OE.21.030812} {\bibfield  {journal}
  {\bibinfo  {journal} {Optics express}\ }\textbf {\bibinfo {volume} {21}},\
  \bibinfo {pages} {30812} (\bibinfo {year} {2013})}\BibitemShut {NoStop}%
\bibitem [{\citenamefont {Purcell}\ \emph {et~al.}(1946)\citenamefont
  {Purcell}, \citenamefont {Torrey},\ and\ \citenamefont
  {Pound}}]{purcell1946resonance}%
  \BibitemOpen
  \bibfield  {author} {\bibinfo {author} {\bibfnamefont {E.~M.}\ \bibnamefont
  {Purcell}}, \bibinfo {author} {\bibfnamefont {H.~C.}\ \bibnamefont
  {Torrey}},\ and\ \bibinfo {author} {\bibfnamefont {R.~V.}\ \bibnamefont
  {Pound}},\ }\bibfield  {title} {\bibinfo {title} {Resonance absorption by
  nuclear magnetic moments in a solid},\ }\href
  {https://doi.org/10.1103/PhysRev.69.37} {\bibfield  {journal} {\bibinfo
  {journal} {Physical review}\ }\textbf {\bibinfo {volume} {69}},\ \bibinfo
  {pages} {37} (\bibinfo {year} {1946})}\BibitemShut {NoStop}%
\bibitem [{\citenamefont {Taflove}\ \emph {et~al.}(2013)\citenamefont
  {Taflove}, \citenamefont {Oskooi},\ and\ \citenamefont
  {Johnson}}]{taflove2013advances}%
  \BibitemOpen
  \bibfield  {author} {\bibinfo {author} {\bibfnamefont {A.}~\bibnamefont
  {Taflove}}, \bibinfo {author} {\bibfnamefont {A.}~\bibnamefont {Oskooi}},\
  and\ \bibinfo {author} {\bibfnamefont {S.~G.}\ \bibnamefont {Johnson}},\
  }\href@noop {} {\emph {\bibinfo {title} {Advances in FDTD computational
  electrodynamics: photonics and nanotechnology}}}\ (\bibinfo  {publisher}
  {Artech house},\ \bibinfo {year} {2013})\ Chap.~\bibinfo {chapter} {4}, pp.\
  \bibinfo {pages} {65--100}\BibitemShut {NoStop}%
\bibitem [{\citenamefont {Xu}\ \emph {et~al.}(2000)\citenamefont {Xu},
  \citenamefont {Lee},\ and\ \citenamefont {Yariv}}]{xu2000quantum}%
  \BibitemOpen
  \bibfield  {author} {\bibinfo {author} {\bibfnamefont {Y.}~\bibnamefont
  {Xu}}, \bibinfo {author} {\bibfnamefont {R.~K.}\ \bibnamefont {Lee}},\ and\
  \bibinfo {author} {\bibfnamefont {A.}~\bibnamefont {Yariv}},\ }\bibfield
  {title} {\bibinfo {title} {Quantum analysis and the classical analysis of
  spontaneous emission in a microcavity},\ }\href
  {https://doi.org/10.1103/PhysRevA.61.033807} {\bibfield  {journal} {\bibinfo
  {journal} {Physical Review A}\ }\textbf {\bibinfo {volume} {61}},\ \bibinfo
  {pages} {033807} (\bibinfo {year} {2000})}\BibitemShut {NoStop}%
\bibitem [{\citenamefont {Wijnands}\ \emph {et~al.}(1997)\citenamefont
  {Wijnands}, \citenamefont {Pendry}, \citenamefont {Garcia-Vidal},
  \citenamefont {Bell}, \citenamefont {Roberts}, \citenamefont {Marti} \emph
  {et~al.}}]{wijnands1997green}%
  \BibitemOpen
  \bibfield  {author} {\bibinfo {author} {\bibfnamefont {F.}~\bibnamefont
  {Wijnands}}, \bibinfo {author} {\bibfnamefont {J.}~\bibnamefont {Pendry}},
  \bibinfo {author} {\bibfnamefont {F.}~\bibnamefont {Garcia-Vidal}}, \bibinfo
  {author} {\bibfnamefont {P.}~\bibnamefont {Bell}}, \bibinfo {author}
  {\bibfnamefont {P.}~\bibnamefont {Roberts}}, \bibinfo {author} {\bibfnamefont
  {L.}~\bibnamefont {Marti}}, \emph {et~al.},\ }\bibfield  {title} {\bibinfo
  {title} {Green's functions for maxwell's equations: application to
  spontaneous emission},\ }\href {https://doi.org/10.1023/A:1018506222632}
  {\bibfield  {journal} {\bibinfo  {journal} {Optical and Quantum Electronics}\
  }\textbf {\bibinfo {volume} {29}},\ \bibinfo {pages} {199} (\bibinfo {year}
  {1997})}\BibitemShut {NoStop}%
\bibitem [{\citenamefont {Martin}\ and\ \citenamefont
  {Piller}(1998)}]{martin_electromagnetic_1998}%
  \BibitemOpen
  \bibfield  {author} {\bibinfo {author} {\bibfnamefont {O.~J.~F.}\
  \bibnamefont {Martin}}\ and\ \bibinfo {author} {\bibfnamefont {N.~B.}\
  \bibnamefont {Piller}},\ }\bibfield  {title} {\bibinfo {title}
  {Electromagnetic scattering in polarizable backgrounds},\ }\href
  {https://doi.org/10.1103/PhysRevE.58.3909} {\bibfield  {journal} {\bibinfo
  {journal} {Physical Review E}\ }\textbf {\bibinfo {volume} {58}},\ \bibinfo
  {pages} {3909} (\bibinfo {year} {1998})}\BibitemShut {NoStop}%
\bibitem [{\citenamefont {D’Aguanno}\ \emph {et~al.}(2004)\citenamefont
  {D’Aguanno}, \citenamefont {Mattiucci}, \citenamefont {Centini},
  \citenamefont {Scalora},\ and\ \citenamefont
  {Bloemer}}]{d2004electromagnetic}%
  \BibitemOpen
  \bibfield  {author} {\bibinfo {author} {\bibfnamefont {G.}~\bibnamefont
  {D’Aguanno}}, \bibinfo {author} {\bibfnamefont {N.}~\bibnamefont
  {Mattiucci}}, \bibinfo {author} {\bibfnamefont {M.}~\bibnamefont {Centini}},
  \bibinfo {author} {\bibfnamefont {M.}~\bibnamefont {Scalora}},\ and\ \bibinfo
  {author} {\bibfnamefont {M.~J.}\ \bibnamefont {Bloemer}},\ }\bibfield
  {title} {\bibinfo {title} {Electromagnetic density of modes for a finite-size
  three-dimensional structure},\ }\href
  {https://doi.org/10.1103/PhysRevE.69.057601} {\bibfield  {journal} {\bibinfo
  {journal} {Physical Review E}\ }\textbf {\bibinfo {volume} {69}},\ \bibinfo
  {pages} {057601} (\bibinfo {year} {2004})}\BibitemShut {NoStop}%
\bibitem [{\citenamefont {Joulain}\ \emph {et~al.}(2003)\citenamefont
  {Joulain}, \citenamefont {Carminati}, \citenamefont {Mulet},\ and\
  \citenamefont {Greffet}}]{joulain2003definition}%
  \BibitemOpen
  \bibfield  {author} {\bibinfo {author} {\bibfnamefont {K.}~\bibnamefont
  {Joulain}}, \bibinfo {author} {\bibfnamefont {R.}~\bibnamefont {Carminati}},
  \bibinfo {author} {\bibfnamefont {J.-P.}\ \bibnamefont {Mulet}},\ and\
  \bibinfo {author} {\bibfnamefont {J.-J.}\ \bibnamefont {Greffet}},\
  }\bibfield  {title} {\bibinfo {title} {Definition and measurement of the
  local density of electromagnetic states close to an interface},\ }\href
  {https://doi.org/10.1103/PhysRevB.68.245405} {\bibfield  {journal} {\bibinfo
  {journal} {Physical Review B}\ }\textbf {\bibinfo {volume} {68}},\ \bibinfo
  {pages} {245405} (\bibinfo {year} {2003})}\BibitemShut {NoStop}%
\bibitem [{\citenamefont {Joulain}\ \emph {et~al.}(2005)\citenamefont
  {Joulain}, \citenamefont {Mulet}, \citenamefont {Marquier}, \citenamefont
  {Carminati},\ and\ \citenamefont {Greffet}}]{joulain2005surface}%
  \BibitemOpen
  \bibfield  {author} {\bibinfo {author} {\bibfnamefont {K.}~\bibnamefont
  {Joulain}}, \bibinfo {author} {\bibfnamefont {J.-P.}\ \bibnamefont {Mulet}},
  \bibinfo {author} {\bibfnamefont {F.}~\bibnamefont {Marquier}}, \bibinfo
  {author} {\bibfnamefont {R.}~\bibnamefont {Carminati}},\ and\ \bibinfo
  {author} {\bibfnamefont {J.-J.}\ \bibnamefont {Greffet}},\ }\bibfield
  {title} {\bibinfo {title} {Surface electromagnetic waves thermally excited:
  Radiative heat transfer, coherence properties and casimir forces revisited in
  the near field},\ }\href {https://doi.org/10.1016/j.surfrep.2004.12.002}
  {\bibfield  {journal} {\bibinfo  {journal} {Surface Science Reports}\
  }\textbf {\bibinfo {volume} {57}},\ \bibinfo {pages} {59} (\bibinfo {year}
  {2005})}\BibitemShut {NoStop}%
\bibitem [{\citenamefont {Francoeur}\ \emph {et~al.}(2010)\citenamefont
  {Francoeur}, \citenamefont {Meng{\"u}{\c{c}}},\ and\ \citenamefont
  {Vaillon}}]{francoeur2010spectral}%
  \BibitemOpen
  \bibfield  {author} {\bibinfo {author} {\bibfnamefont {M.}~\bibnamefont
  {Francoeur}}, \bibinfo {author} {\bibfnamefont {M.~P.}\ \bibnamefont
  {Meng{\"u}{\c{c}}}},\ and\ \bibinfo {author} {\bibfnamefont {R.}~\bibnamefont
  {Vaillon}},\ }\bibfield  {title} {\bibinfo {title} {Spectral tuning of
  near-field radiative heat flux between two thin silicon carbide films},\
  }\href {https://doi.org/10.1088/0022-3727/43/7/075501} {\bibfield  {journal}
  {\bibinfo  {journal} {Journal of Physics D: Applied Physics}\ }\textbf
  {\bibinfo {volume} {43}},\ \bibinfo {pages} {075501} (\bibinfo {year}
  {2010})}\BibitemShut {NoStop}%
\bibitem [{\citenamefont {Palik}(1998)}]{palik1998handbook}%
  \BibitemOpen
  \bibfield  {author} {\bibinfo {author} {\bibfnamefont {E.~D.}\ \bibnamefont
  {Palik}},\ }\href@noop {} {\emph {\bibinfo {title} {Handbook of optical
  constants of solids}}},\ Vol.~\bibinfo {volume} {3}\ (\bibinfo  {publisher}
  {Academic press},\ \bibinfo {year} {1998})\BibitemShut {NoStop}%
\bibitem [{\citenamefont {Law}\ \emph {et~al.}(2013)\citenamefont {Law},
  \citenamefont {Yu}, \citenamefont {Rosenberg},\ and\ \citenamefont
  {Wasserman}}]{law2013all}%
  \BibitemOpen
  \bibfield  {author} {\bibinfo {author} {\bibfnamefont {S.}~\bibnamefont
  {Law}}, \bibinfo {author} {\bibfnamefont {L.}~\bibnamefont {Yu}}, \bibinfo
  {author} {\bibfnamefont {A.}~\bibnamefont {Rosenberg}},\ and\ \bibinfo
  {author} {\bibfnamefont {D.}~\bibnamefont {Wasserman}},\ }\bibfield  {title}
  {\bibinfo {title} {All-semiconductor plasmonic nanoantennas for infrared
  sensing},\ }\href {https://doi.org/10.1021/nl402766t} {\bibfield  {journal}
  {\bibinfo  {journal} {Nano letters}\ }\textbf {\bibinfo {volume} {13}},\
  \bibinfo {pages} {4569} (\bibinfo {year} {2013})}\BibitemShut {NoStop}%
\bibitem [{\citenamefont {Popova}\ \emph {et~al.}(1972)\citenamefont {Popova},
  \citenamefont {Tolstykh},\ and\ \citenamefont {Vorobev}}]{popova1972optical}%
  \BibitemOpen
  \bibfield  {author} {\bibinfo {author} {\bibfnamefont {S.}~\bibnamefont
  {Popova}}, \bibinfo {author} {\bibfnamefont {T.}~\bibnamefont {Tolstykh}},\
  and\ \bibinfo {author} {\bibfnamefont {V.}~\bibnamefont {Vorobev}},\
  }\bibfield  {title} {\bibinfo {title} {Optical characteristics of amorphous
  quartz in the 1400-200 cm- 1 region},\ }\href@noop {} {\bibfield  {journal}
  {\bibinfo  {journal} {Opt. Spectrosc}\ }\textbf {\bibinfo {volume} {33}},\
  \bibinfo {pages} {444} (\bibinfo {year} {1972})}\BibitemShut {NoStop}%
\bibitem [{\citenamefont {Jameson}\ \emph {et~al.}(1998)\citenamefont
  {Jameson}, \citenamefont {Martinelli},\ and\ \citenamefont
  {Pierce}}]{Jameson1998}%
  \BibitemOpen
  \bibfield  {author} {\bibinfo {author} {\bibfnamefont {A.}~\bibnamefont
  {Jameson}}, \bibinfo {author} {\bibfnamefont {L.}~\bibnamefont
  {Martinelli}},\ and\ \bibinfo {author} {\bibfnamefont {N.~A.}\ \bibnamefont
  {Pierce}},\ }\bibfield  {title} {\bibinfo {title} {{Optimum Aerodynamic
  Design Using the Navier-Stokes Equations}},\ }\href
  {https://doi.org/10.1007/s001620050060} {\bibfield  {journal} {\bibinfo
  {journal} {Theor. Comput. Fluid Dyn.}\ }\textbf {\bibinfo {volume} {10}},\
  \bibinfo {pages} {213} (\bibinfo {year} {1998})}\BibitemShut {NoStop}%
\bibitem [{\citenamefont {Sigmund}\ and\ \citenamefont {{S{\o}ndergaard
  Jensen}}(2003)}]{Sigmund2003}%
  \BibitemOpen
  \bibfield  {author} {\bibinfo {author} {\bibfnamefont {O.}~\bibnamefont
  {Sigmund}}\ and\ \bibinfo {author} {\bibfnamefont {J.}~\bibnamefont
  {{S{\o}ndergaard Jensen}}},\ }\bibfield  {title} {\bibinfo {title}
  {{Systematic design of phononic band--gap materials and structures by
  topology optimization}},\ }\href
  {http://rsta.royalsocietypublishing.org/content/361/1806/1001.short}
  {\bibfield  {journal} {\bibinfo  {journal} {Philos. Trans. R. Soc. London.
  Ser. A Math. Phys. Eng. Sci.}\ }\textbf {\bibinfo {volume} {361}},\ \bibinfo
  {pages} {1001} (\bibinfo {year} {2003})}\BibitemShut {NoStop}%
\bibitem [{\citenamefont {Lu}\ \emph {et~al.}(2011)\citenamefont {Lu},
  \citenamefont {Boyd},\ and\ \citenamefont {Vuckovi{\'{c}}}}]{Lu2011}%
  \BibitemOpen
  \bibfield  {author} {\bibinfo {author} {\bibfnamefont {J.}~\bibnamefont
  {Lu}}, \bibinfo {author} {\bibfnamefont {S.}~\bibnamefont {Boyd}},\ and\
  \bibinfo {author} {\bibfnamefont {J.}~\bibnamefont {Vuckovi{\'{c}}}},\
  }\bibfield  {title} {\bibinfo {title} {{Inverse design of a three-dimensional
  nanophotonic resonator}},\ }\href
  {http://www.opticsinfobase.org/abstract.cfm?URI=oe-19-11-10563} {\bibfield
  {journal} {\bibinfo  {journal} {Opt. Express}\ }\textbf {\bibinfo {volume}
  {19}},\ \bibinfo {pages} {10563} (\bibinfo {year} {2011})}\BibitemShut
  {NoStop}%
\bibitem [{\citenamefont {Jensen}\ and\ \citenamefont
  {Sigmund}(2011)}]{Jensen2011}%
  \BibitemOpen
  \bibfield  {author} {\bibinfo {author} {\bibfnamefont {J.~S.}\ \bibnamefont
  {Jensen}}\ and\ \bibinfo {author} {\bibfnamefont {O.}~\bibnamefont
  {Sigmund}},\ }\bibfield  {title} {\bibinfo {title} {{Topology optimization
  for nano-photonics}},\ }\href {https://doi.org/10.1002/lpor.201000014}
  {\bibfield  {journal} {\bibinfo  {journal} {Laser {\&} Photonics Rev.}\
  }\textbf {\bibinfo {volume} {5}},\ \bibinfo {pages} {308} (\bibinfo {year}
  {2011})}\BibitemShut {NoStop}%
\bibitem [{\citenamefont {Miller}(2012)}]{Miller2012a}%
  \BibitemOpen
  \bibfield  {author} {\bibinfo {author} {\bibfnamefont {O.~D.}\ \bibnamefont
  {Miller}},\ }\emph {\bibinfo {title} {{Photonic Design: From Fundamental
  Solar Cell Physics to Computational Inverse Design}}},\ \href
  {http://arxiv.org/abs/1308.0212} {Ph.D. thesis},\ \bibinfo  {school}
  {University of California, Berkeley} (\bibinfo {year} {2012})\BibitemShut
  {NoStop}%
\bibitem [{\citenamefont {Lalau-Keraly}\ \emph {et~al.}(2013)\citenamefont
  {Lalau-Keraly}, \citenamefont {Bhargava}, \citenamefont {Miller},\ and\
  \citenamefont {Yablonovitch}}]{Lalau-Keraly2013}%
  \BibitemOpen
  \bibfield  {author} {\bibinfo {author} {\bibfnamefont {C.~M.}\ \bibnamefont
  {Lalau-Keraly}}, \bibinfo {author} {\bibfnamefont {S.}~\bibnamefont
  {Bhargava}}, \bibinfo {author} {\bibfnamefont {O.~D.}\ \bibnamefont
  {Miller}},\ and\ \bibinfo {author} {\bibfnamefont {E.}~\bibnamefont
  {Yablonovitch}},\ }\bibfield  {title} {\bibinfo {title} {Adjoint shape
  optimization applied to electromagnetic design},\ }\href
  {https://doi.org/10.1364/OE.21.021693} {\bibfield  {journal} {\bibinfo
  {journal} {Optics Express}\ }\textbf {\bibinfo {volume} {21}},\ \bibinfo
  {pages} {21693} (\bibinfo {year} {2013})}\BibitemShut {NoStop}%
\bibitem [{\citenamefont {Ganapati}\ \emph {et~al.}(2014)\citenamefont
  {Ganapati}, \citenamefont {Miller},\ and\ \citenamefont
  {Yablonovitch}}]{Ganapati2014}%
  \BibitemOpen
  \bibfield  {author} {\bibinfo {author} {\bibfnamefont {V.}~\bibnamefont
  {Ganapati}}, \bibinfo {author} {\bibfnamefont {O.~D.}\ \bibnamefont
  {Miller}},\ and\ \bibinfo {author} {\bibfnamefont {E.}~\bibnamefont
  {Yablonovitch}},\ }\bibfield  {title} {\bibinfo {title} {Light trapping
  textures designed by electromagnetic optimization for subwavelength thick
  solar cells},\ }\href {https://doi.org/10.1109/JPHOTOV.2013.2280340}
  {\bibfield  {journal} {\bibinfo  {journal} {IEEE Journal of Photovoltaics}\
  }\textbf {\bibinfo {volume} {4}},\ \bibinfo {pages} {175} (\bibinfo {year}
  {2014})}\BibitemShut {NoStop}%
\bibitem [{\citenamefont {Aage}\ \emph {et~al.}(2017)\citenamefont {Aage},
  \citenamefont {Andreassen}, \citenamefont {Lazarov},\ and\ \citenamefont
  {Sigmund}}]{aage2017giga}%
  \BibitemOpen
  \bibfield  {author} {\bibinfo {author} {\bibfnamefont {N.}~\bibnamefont
  {Aage}}, \bibinfo {author} {\bibfnamefont {E.}~\bibnamefont {Andreassen}},
  \bibinfo {author} {\bibfnamefont {B.~S.}\ \bibnamefont {Lazarov}},\ and\
  \bibinfo {author} {\bibfnamefont {O.}~\bibnamefont {Sigmund}},\ }\bibfield
  {title} {\bibinfo {title} {Giga-voxel computational morphogenesis for
  structural design},\ }\href {https://doi.org/10.1038/nature23911} {\bibfield
  {journal} {\bibinfo  {journal} {Nature}\ }\textbf {\bibinfo {volume} {550}},\
  \bibinfo {pages} {84} (\bibinfo {year} {2017})}\BibitemShut {NoStop}%
\bibitem [{\citenamefont {Christiansen}\ \emph {et~al.}(2019)\citenamefont
  {Christiansen}, \citenamefont {Vester-Petersen}, \citenamefont {Madsen},\
  and\ \citenamefont {Sigmund}}]{Christiansen2019}%
  \BibitemOpen
  \bibfield  {author} {\bibinfo {author} {\bibfnamefont {R.~E.}\ \bibnamefont
  {Christiansen}}, \bibinfo {author} {\bibfnamefont {J.}~\bibnamefont
  {Vester-Petersen}}, \bibinfo {author} {\bibfnamefont {S.~P.}\ \bibnamefont
  {Madsen}},\ and\ \bibinfo {author} {\bibfnamefont {O.}~\bibnamefont
  {Sigmund}},\ }\bibfield  {title} {\bibinfo {title} {{A non-linear material
  interpolation for design of metallic nano-particles using topology
  optimization}},\ }\href {https://doi.org/10.1016/j.cma.2018.08.034}
  {\bibfield  {journal} {\bibinfo  {journal} {Comput. Methods Appl. Mech.
  Eng.}\ }\textbf {\bibinfo {volume} {343}},\ \bibinfo {pages} {23} (\bibinfo
  {year} {2019})}\BibitemShut {NoStop}%
\bibitem [{\citenamefont {Watts}\ \emph {et~al.}(2012)\citenamefont {Watts},
  \citenamefont {Liu},\ and\ \citenamefont {Padilla}}]{Watts2012}%
  \BibitemOpen
  \bibfield  {author} {\bibinfo {author} {\bibfnamefont {C.~M.}\ \bibnamefont
  {Watts}}, \bibinfo {author} {\bibfnamefont {X.}~\bibnamefont {Liu}},\ and\
  \bibinfo {author} {\bibfnamefont {W.~J.}\ \bibnamefont {Padilla}},\
  }\bibfield  {title} {\bibinfo {title} {{Metamaterial electromagnetic wave
  absorbers}},\ }\bibfield  {journal} {\bibinfo  {journal} {Adv. Mater.}\
  }\textbf {\bibinfo {volume} {24}},\ \href
  {https://doi.org/10.1002/adma.201200674} {10.1002/adma.201200674} (\bibinfo
  {year} {2012})\BibitemShut {NoStop}%
\bibitem [{\citenamefont {Lee}\ \emph {et~al.}(2016)\citenamefont {Lee},
  \citenamefont {Rhee}, \citenamefont {Yoo},\ and\ \citenamefont
  {Kim}}]{lee2016metamaterials}%
  \BibitemOpen
  \bibfield  {author} {\bibinfo {author} {\bibfnamefont {Y.~P.}\ \bibnamefont
  {Lee}}, \bibinfo {author} {\bibfnamefont {J.~Y.}\ \bibnamefont {Rhee}},
  \bibinfo {author} {\bibfnamefont {Y.~J.}\ \bibnamefont {Yoo}},\ and\ \bibinfo
  {author} {\bibfnamefont {K.~W.}\ \bibnamefont {Kim}},\ }\href@noop {} {\emph
  {\bibinfo {title} {{Metamaterials for perfect absorption}}}},\ Vol.\ \bibinfo
  {volume} {236}\ (\bibinfo  {publisher} {Springer},\ \bibinfo {year}
  {2016})\BibitemShut {NoStop}%
\bibitem [{\citenamefont {Landy}\ \emph {et~al.}(2008)\citenamefont {Landy},
  \citenamefont {Sajuyigbe}, \citenamefont {Mock}, \citenamefont {Smith},\ and\
  \citenamefont {Padilla}}]{landy_perfect_2008-1}%
  \BibitemOpen
  \bibfield  {author} {\bibinfo {author} {\bibfnamefont {N.~I.}\ \bibnamefont
  {Landy}}, \bibinfo {author} {\bibfnamefont {S.}~\bibnamefont {Sajuyigbe}},
  \bibinfo {author} {\bibfnamefont {J.~J.}\ \bibnamefont {Mock}}, \bibinfo
  {author} {\bibfnamefont {D.~R.}\ \bibnamefont {Smith}},\ and\ \bibinfo
  {author} {\bibfnamefont {W.~J.}\ \bibnamefont {Padilla}},\ }\bibfield
  {title} {{\selectlanguage {en}\bibinfo {title} {Perfect {Metamaterial}
  {Absorber}}},\ }\bibfield  {journal} {\bibinfo  {journal} {Physical Review
  Letters}\ }\textbf {\bibinfo {volume} {100}},\ \href
  {https://doi.org/10.1103/PhysRevLett.100.207402}
  {10.1103/PhysRevLett.100.207402} (\bibinfo {year} {2008})\BibitemShut
  {NoStop}%
\bibitem [{\citenamefont {Liu}\ \emph {et~al.}(2010)\citenamefont {Liu},
  \citenamefont {Mesch}, \citenamefont {Weiss}, \citenamefont {Hentschel},\
  and\ \citenamefont {Giessen}}]{Liu2010}%
  \BibitemOpen
  \bibfield  {author} {\bibinfo {author} {\bibfnamefont {N.}~\bibnamefont
  {Liu}}, \bibinfo {author} {\bibfnamefont {M.}~\bibnamefont {Mesch}}, \bibinfo
  {author} {\bibfnamefont {T.}~\bibnamefont {Weiss}}, \bibinfo {author}
  {\bibfnamefont {M.}~\bibnamefont {Hentschel}},\ and\ \bibinfo {author}
  {\bibfnamefont {H.}~\bibnamefont {Giessen}},\ }\bibfield  {title} {\bibinfo
  {title} {{Infrared perfect absorber and its application as plasmonic
  sensor}},\ }\href {https://doi.org/10.1021/nl9041033} {\bibfield  {journal}
  {\bibinfo  {journal} {Nano Lett.}\ }\textbf {\bibinfo {volume} {10}},\
  \bibinfo {pages} {2342} (\bibinfo {year} {2010})}\BibitemShut {NoStop}%
\bibitem [{\citenamefont {Yu}\ \emph {et~al.}(2010)\citenamefont {Yu},
  \citenamefont {Raman},\ and\ \citenamefont {Fan}}]{Yu2010}%
  \BibitemOpen
  \bibfield  {author} {\bibinfo {author} {\bibfnamefont {Z.}~\bibnamefont
  {Yu}}, \bibinfo {author} {\bibfnamefont {A.}~\bibnamefont {Raman}},\ and\
  \bibinfo {author} {\bibfnamefont {S.}~\bibnamefont {Fan}},\ }\bibfield
  {title} {\bibinfo {title} {{Fundamental limit of nanophotonic light trapping
  in solar cells}},\ }\href {https://doi.org/10.1073/pnas.1008296107}
  {\bibfield  {journal} {\bibinfo  {journal} {Proc. Natl. Acad. Sci. U. S. A.}\
  }\textbf {\bibinfo {volume} {107}},\ \bibinfo {pages} {17491} (\bibinfo
  {year} {2010})}\BibitemShut {NoStop}%
\bibitem [{\citenamefont {Cui}\ \emph {et~al.}(2012)\citenamefont {Cui},
  \citenamefont {Fung}, \citenamefont {Xu}, \citenamefont {Ma}, \citenamefont
  {Jin}, \citenamefont {He},\ and\ \citenamefont
  {Fang}}]{cui_ultrabroadband_2012}%
  \BibitemOpen
  \bibfield  {author} {\bibinfo {author} {\bibfnamefont {Y.}~\bibnamefont
  {Cui}}, \bibinfo {author} {\bibfnamefont {K.~H.}\ \bibnamefont {Fung}},
  \bibinfo {author} {\bibfnamefont {J.}~\bibnamefont {Xu}}, \bibinfo {author}
  {\bibfnamefont {H.}~\bibnamefont {Ma}}, \bibinfo {author} {\bibfnamefont
  {Y.}~\bibnamefont {Jin}}, \bibinfo {author} {\bibfnamefont {S.}~\bibnamefont
  {He}},\ and\ \bibinfo {author} {\bibfnamefont {N.~X.}\ \bibnamefont {Fang}},\
  }\bibfield  {title} {\bibinfo {title} {Ultrabroadband {Light} {Absorption} by
  a {Sawtooth} {Anisotropic} {Metamaterial} {Slab}},\ }\href
  {https://doi.org/10.1021/nl204118h} {\bibfield  {journal} {\bibinfo
  {journal} {Nano Letters}\ }\textbf {\bibinfo {volume} {12}},\ \bibinfo
  {pages} {1443} (\bibinfo {year} {2012})},\ \bibinfo {note} {publisher:
  American Chemical Society}\BibitemShut {NoStop}%
\bibitem [{\citenamefont {Massiot}\ \emph {et~al.}(2012)\citenamefont
  {Massiot}, \citenamefont {Colin}, \citenamefont {Péré-Laperne},
  \citenamefont {Roca~i Cabarrocas}, \citenamefont {Sauvan}, \citenamefont
  {Lalanne}, \citenamefont {Pelouard},\ and\ \citenamefont
  {Collin}}]{massiot_nanopatterned_2012}%
  \BibitemOpen
  \bibfield  {author} {\bibinfo {author} {\bibfnamefont {I.}~\bibnamefont
  {Massiot}}, \bibinfo {author} {\bibfnamefont {C.}~\bibnamefont {Colin}},
  \bibinfo {author} {\bibfnamefont {N.}~\bibnamefont {Péré-Laperne}},
  \bibinfo {author} {\bibfnamefont {P.}~\bibnamefont {Roca~i Cabarrocas}},
  \bibinfo {author} {\bibfnamefont {C.}~\bibnamefont {Sauvan}}, \bibinfo
  {author} {\bibfnamefont {P.}~\bibnamefont {Lalanne}}, \bibinfo {author}
  {\bibfnamefont {J.-L.}\ \bibnamefont {Pelouard}},\ and\ \bibinfo {author}
  {\bibfnamefont {S.}~\bibnamefont {Collin}},\ }\bibfield  {title}
  {{\selectlanguage {en}\bibinfo {title} {Nanopatterned front contact for
  broadband absorption in ultra-thin amorphous silicon solar cells}},\ }\href
  {https://doi.org/10.1063/1.4758468} {\bibfield  {journal} {\bibinfo
  {journal} {Applied Physics Letters}\ }\textbf {\bibinfo {volume} {101}},\
  \bibinfo {pages} {163901} (\bibinfo {year} {2012})}\BibitemShut {NoStop}%
\bibitem [{\citenamefont {Polin}\ \emph {et~al.}(2005)\citenamefont {Polin},
  \citenamefont {Ladavac}, \citenamefont {Lee}, \citenamefont {Roichman},\ and\
  \citenamefont {Grier}}]{Polin2005}%
  \BibitemOpen
  \bibfield  {author} {\bibinfo {author} {\bibfnamefont {M.}~\bibnamefont
  {Polin}}, \bibinfo {author} {\bibfnamefont {K.}~\bibnamefont {Ladavac}},
  \bibinfo {author} {\bibfnamefont {S.-H.}\ \bibnamefont {Lee}}, \bibinfo
  {author} {\bibfnamefont {Y.}~\bibnamefont {Roichman}},\ and\ \bibinfo
  {author} {\bibfnamefont {D.~G.}\ \bibnamefont {Grier}},\ }\bibfield  {title}
  {\bibinfo {title} {{Optimized holographic optical traps}},\ }\href
  {https://doi.org/10.1364/OPEX.13.005831} {\bibfield  {journal} {\bibinfo
  {journal} {Opt. Express}\ }\textbf {\bibinfo {volume} {13}},\ \bibinfo
  {pages} {5831} (\bibinfo {year} {2005})}\BibitemShut {NoStop}%
\bibitem [{\citenamefont {Taylor}\ \emph {et~al.}(2015)\citenamefont {Taylor},
  \citenamefont {Waleed}, \citenamefont {Stilgoe}, \citenamefont
  {Rubinsztein-Dunlop},\ and\ \citenamefont {Bowen}}]{Taylor2015}%
  \BibitemOpen
  \bibfield  {author} {\bibinfo {author} {\bibfnamefont {M.~A.}\ \bibnamefont
  {Taylor}}, \bibinfo {author} {\bibfnamefont {M.}~\bibnamefont {Waleed}},
  \bibinfo {author} {\bibfnamefont {A.~B.}\ \bibnamefont {Stilgoe}}, \bibinfo
  {author} {\bibfnamefont {H.}~\bibnamefont {Rubinsztein-Dunlop}},\ and\
  \bibinfo {author} {\bibfnamefont {W.~P.}\ \bibnamefont {Bowen}},\ }\bibfield
  {title} {\bibinfo {title} {{Enhanced optical trapping via structured
  scattering}},\ }\href {https://doi.org/10.1038/nphoton.2015.160} {\bibfield
  {journal} {\bibinfo  {journal} {Nat. Photonics}\ }\textbf {\bibinfo {volume}
  {9}},\ \bibinfo {pages} {669} (\bibinfo {year} {2015})}\BibitemShut {NoStop}%
\bibitem [{\citenamefont {Taylor}(2017)}]{Taylor2017}%
  \BibitemOpen
  \bibfield  {author} {\bibinfo {author} {\bibfnamefont {M.~A.}\ \bibnamefont
  {Taylor}},\ }\bibfield  {title} {\bibinfo {title} {{Optimizing phase to
  enhance optical trap stiffness}},\ }\href
  {https://doi.org/10.1038/s41598-017-00762-z} {\bibfield  {journal} {\bibinfo
  {journal} {Sci. Rep.}\ }\textbf {\bibinfo {volume} {7}},\ \bibinfo {pages}
  {555} (\bibinfo {year} {2017})}\BibitemShut {NoStop}%
\bibitem [{\citenamefont {Fernandez-Corbaton}\ and\ \citenamefont
  {Rockstuhl}(2017)}]{Fernandez-Corbaton2017}%
  \BibitemOpen
  \bibfield  {author} {\bibinfo {author} {\bibfnamefont {I.}~\bibnamefont
  {Fernandez-Corbaton}}\ and\ \bibinfo {author} {\bibfnamefont
  {C.}~\bibnamefont {Rockstuhl}},\ }\bibfield  {title} {\bibinfo {title}
  {{Unified theory to describe and engineer conservation laws in light-matter
  interactions}},\ }\href {https://doi.org/10.1103/PhysRevA.95.053829}
  {\bibfield  {journal} {\bibinfo  {journal} {Phys. Rev. A}\ }\textbf {\bibinfo
  {volume} {95}},\ \bibinfo {pages} {1} (\bibinfo {year} {2017})}\BibitemShut
  {NoStop}%
\bibitem [{\citenamefont {Levy}\ \emph {et~al.}(2016)\citenamefont {Levy},
  \citenamefont {Derevyanko},\ and\ \citenamefont {Silberberg}}]{Levy2016}%
  \BibitemOpen
  \bibfield  {author} {\bibinfo {author} {\bibfnamefont {U.}~\bibnamefont
  {Levy}}, \bibinfo {author} {\bibfnamefont {S.}~\bibnamefont {Derevyanko}},\
  and\ \bibinfo {author} {\bibfnamefont {Y.}~\bibnamefont {Silberberg}},\
  }\bibfield  {title} {\bibinfo {title} {{Light Modes of Free Space}},\ }\href
  {https://doi.org/10.1016/bs.po.2015.10.001} {\bibfield  {journal} {\bibinfo
  {journal} {Progress in Optics}\ }\textbf {\bibinfo {volume} {61}},\ \bibinfo
  {pages} {237} (\bibinfo {year} {2016})}\BibitemShut {NoStop}%
\bibitem [{\citenamefont {Nocedal}\ and\ \citenamefont
  {Wright}(2006)}]{Nocedal2006}%
  \BibitemOpen
  \bibfield  {author} {\bibinfo {author} {\bibfnamefont {J.}~\bibnamefont
  {Nocedal}}\ and\ \bibinfo {author} {\bibfnamefont {S.~J.}\ \bibnamefont
  {Wright}},\ }\href@noop {} {\emph {\bibinfo {title} {{Numerical
  Optimization}}}},\ \bibinfo {edition} {2nd}\ ed.\ (\bibinfo  {publisher}
  {Springer},\ \bibinfo {address} {New York, NY},\ \bibinfo {year}
  {2006})\BibitemShut {NoStop}%
\bibitem [{\citenamefont {Trefethen}\ and\ \citenamefont
  {Bau}(1997)}]{Trefethen1997}%
  \BibitemOpen
  \bibfield  {author} {\bibinfo {author} {\bibfnamefont {L.~N.}\ \bibnamefont
  {Trefethen}}\ and\ \bibinfo {author} {\bibfnamefont {D.}~\bibnamefont
  {Bau}},\ }\href@noop {} {\emph {\bibinfo {title} {{Numerical Linear
  Algebra}}}}\ (\bibinfo  {publisher} {Society for Industrial and Applied
  Mathematics},\ \bibinfo {address} {Philadelphia, PA},\ \bibinfo {year}
  {1997})\BibitemShut {NoStop}%
\bibitem [{\citenamefont {Vahala}(2003)}]{Vahala2003}%
  \BibitemOpen
  \bibfield  {author} {\bibinfo {author} {\bibfnamefont {K.~J.}\ \bibnamefont
  {Vahala}},\ }\bibfield  {title} {\bibinfo {title} {{Optical microcavities}},\
  }\href@noop {} {\bibfield  {journal} {\bibinfo  {journal} {Nature}\ }\textbf
  {\bibinfo {volume} {424}},\ \bibinfo {pages} {839} (\bibinfo {year}
  {2003})}\BibitemShut {NoStop}%
\bibitem [{\citenamefont {He}\ \emph {et~al.}(2013)\citenamefont {He},
  \citenamefont {Ozdemir},\ and\ \citenamefont {Yang}}]{He2013}%
  \BibitemOpen
  \bibfield  {author} {\bibinfo {author} {\bibfnamefont {L.}~\bibnamefont
  {He}}, \bibinfo {author} {\bibfnamefont {S.~K.}\ \bibnamefont {Ozdemir}},\
  and\ \bibinfo {author} {\bibfnamefont {L.}~\bibnamefont {Yang}},\ }\bibfield
  {title} {\bibinfo {title} {{Whispering gallery microcavity lasers}},\ }\href
  {https://doi.org/10.1002/lpor.201100032} {\bibfield  {journal} {\bibinfo
  {journal} {Laser Photonics Rev.}\ }\textbf {\bibinfo {volume} {7}},\ \bibinfo
  {pages} {60} (\bibinfo {year} {2013})}\BibitemShut {NoStop}%
\bibitem [{\citenamefont {Levy}\ \emph {et~al.}(2007)\citenamefont {Levy},
  \citenamefont {Abashin}, \citenamefont {Ikeda}, \citenamefont
  {Krishnamoorthy}, \citenamefont {Cunningham},\ and\ \citenamefont
  {Fainman}}]{Levy2007}%
  \BibitemOpen
  \bibfield  {author} {\bibinfo {author} {\bibfnamefont {U.}~\bibnamefont
  {Levy}}, \bibinfo {author} {\bibfnamefont {M.}~\bibnamefont {Abashin}},
  \bibinfo {author} {\bibfnamefont {K.}~\bibnamefont {Ikeda}}, \bibinfo
  {author} {\bibfnamefont {A.}~\bibnamefont {Krishnamoorthy}}, \bibinfo
  {author} {\bibfnamefont {J.}~\bibnamefont {Cunningham}},\ and\ \bibinfo
  {author} {\bibfnamefont {Y.}~\bibnamefont {Fainman}},\ }\bibfield  {title}
  {\bibinfo {title} {{Inhomogenous dielectric metamaterials with space-variant
  polarizability}},\ }\href {https://doi.org/10.1103/PhysRevLett.98.243901}
  {\bibfield  {journal} {\bibinfo  {journal} {Physical Review Letters}\
  }\textbf {\bibinfo {volume} {98}},\ \bibinfo {pages} {243901} (\bibinfo
  {year} {2007})}\BibitemShut {NoStop}%
\bibitem [{\citenamefont {Wei}\ \emph {et~al.}(2013)\citenamefont {Wei},
  \citenamefont {Long}, \citenamefont {Gong}, \citenamefont {Li}, \citenamefont
  {Su},\ and\ \citenamefont {Cao}}]{Wei2013}%
  \BibitemOpen
  \bibfield  {author} {\bibinfo {author} {\bibfnamefont {Z.}~\bibnamefont
  {Wei}}, \bibinfo {author} {\bibfnamefont {Y.}~\bibnamefont {Long}}, \bibinfo
  {author} {\bibfnamefont {Z.}~\bibnamefont {Gong}}, \bibinfo {author}
  {\bibfnamefont {H.}~\bibnamefont {Li}}, \bibinfo {author} {\bibfnamefont
  {X.}~\bibnamefont {Su}},\ and\ \bibinfo {author} {\bibfnamefont
  {Y.}~\bibnamefont {Cao}},\ }\bibfield  {title} {\bibinfo {title} {{Highly
  efficient beam steering with a transparent metasurface}},\ }\href
  {https://doi.org/10.1364/oe.21.010739} {\bibfield  {journal} {\bibinfo
  {journal} {Optics Express}\ }\textbf {\bibinfo {volume} {21}},\ \bibinfo
  {pages} {10739} (\bibinfo {year} {2013})}\BibitemShut {NoStop}%
\bibitem [{\citenamefont {Keren-Zur}\ \emph {et~al.}(2016)\citenamefont
  {Keren-Zur}, \citenamefont {Avayu}, \citenamefont {Michaeli},\ and\
  \citenamefont {Ellenbogen}}]{KerenZur2016}%
  \BibitemOpen
  \bibfield  {author} {\bibinfo {author} {\bibfnamefont {S.}~\bibnamefont
  {Keren-Zur}}, \bibinfo {author} {\bibfnamefont {O.}~\bibnamefont {Avayu}},
  \bibinfo {author} {\bibfnamefont {L.}~\bibnamefont {Michaeli}},\ and\
  \bibinfo {author} {\bibfnamefont {T.}~\bibnamefont {Ellenbogen}},\ }\bibfield
   {title} {\bibinfo {title} {{Nonlinear Beam Shaping with Plasmonic
  Metasurfaces}},\ }\href {https://doi.org/10.1021/acsphotonics.5b00528}
  {\bibfield  {journal} {\bibinfo  {journal} {ACS Photonics}\ }\textbf
  {\bibinfo {volume} {3}},\ \bibinfo {pages} {117} (\bibinfo {year}
  {2016})}\BibitemShut {NoStop}%
\bibitem [{\citenamefont {Weiner}(2000)}]{Weiner2000}%
  \BibitemOpen
  \bibfield  {author} {\bibinfo {author} {\bibfnamefont {A.~M.}\ \bibnamefont
  {Weiner}},\ }\bibfield  {title} {\bibinfo {title} {{Femtosecond pulse shaping
  using spatial light modulators}},\ }\href {https://doi.org/10.1063/1.1150614}
  {\bibfield  {journal} {\bibinfo  {journal} {Review of Scientific
  Instruments}\ }\textbf {\bibinfo {volume} {71}},\ \bibinfo {pages} {1929}
  (\bibinfo {year} {2000})}\BibitemShut {NoStop}%
\bibitem [{\citenamefont {Chattrapiban}\ \emph {et~al.}(2003)\citenamefont
  {Chattrapiban}, \citenamefont {Rogers}, \citenamefont {Cofield},
  \citenamefont {{Hill, III}},\ and\ \citenamefont {Roy}}]{Chattrapiban2003}%
  \BibitemOpen
  \bibfield  {author} {\bibinfo {author} {\bibfnamefont {N.}~\bibnamefont
  {Chattrapiban}}, \bibinfo {author} {\bibfnamefont {E.~A.}\ \bibnamefont
  {Rogers}}, \bibinfo {author} {\bibfnamefont {D.}~\bibnamefont {Cofield}},
  \bibinfo {author} {\bibfnamefont {W.~T.}\ \bibnamefont {{Hill, III}}},\ and\
  \bibinfo {author} {\bibfnamefont {R.}~\bibnamefont {Roy}},\ }\bibfield
  {title} {\bibinfo {title} {{Generation of nondiffracting Bessel beams by use
  of a spatial light modulator}},\ }\href
  {https://doi.org/10.1364/OL.28.002183} {\bibfield  {journal} {\bibinfo
  {journal} {Optics Letters}\ }\textbf {\bibinfo {volume} {28}},\ \bibinfo
  {pages} {2183} (\bibinfo {year} {2003})}\BibitemShut {NoStop}%
\bibitem [{\citenamefont {Guo}\ \emph {et~al.}(2007)\citenamefont {Guo},
  \citenamefont {Wang}, \citenamefont {Ni}, \citenamefont {Wang},\ and\
  \citenamefont {Ding}}]{Guo2007}%
  \BibitemOpen
  \bibfield  {author} {\bibinfo {author} {\bibfnamefont {C.-S.}\ \bibnamefont
  {Guo}}, \bibinfo {author} {\bibfnamefont {X.-L.}\ \bibnamefont {Wang}},
  \bibinfo {author} {\bibfnamefont {W.-J.}\ \bibnamefont {Ni}}, \bibinfo
  {author} {\bibfnamefont {H.-T.}\ \bibnamefont {Wang}},\ and\ \bibinfo
  {author} {\bibfnamefont {J.}~\bibnamefont {Ding}},\ }\bibfield  {title}
  {\bibinfo {title} {{Generation of arbitrary vector beams with a spatial light
  modulator and a common path interferometric arrangement}},\ }\href
  {https://doi.org/10.1364/ol.32.003549} {\bibfield  {journal} {\bibinfo
  {journal} {Optics Letters}\ }\textbf {\bibinfo {volume} {32}},\ \bibinfo
  {pages} {3549} (\bibinfo {year} {2007})}\BibitemShut {NoStop}%
\bibitem [{\citenamefont {Zhu}\ and\ \citenamefont {Wang}(2014)}]{Zhu2014}%
  \BibitemOpen
  \bibfield  {author} {\bibinfo {author} {\bibfnamefont {L.}~\bibnamefont
  {Zhu}}\ and\ \bibinfo {author} {\bibfnamefont {J.}~\bibnamefont {Wang}},\
  }\bibfield  {title} {\bibinfo {title} {{Arbitrary manipulation of spatial
  amplitude and phase using phase-only spatial light modulators}},\ }\href
  {https://doi.org/10.1038/srep07441} {\bibfield  {journal} {\bibinfo
  {journal} {Scientific Reports}\ }\textbf {\bibinfo {volume} {4}},\ \bibinfo
  {pages} {7441} (\bibinfo {year} {2014})}\BibitemShut {NoStop}%
\bibitem [{\citenamefont {Lodahl}\ \emph {et~al.}(2004)\citenamefont {Lodahl},
  \citenamefont {{Van Driel}}, \citenamefont {Nikolaev}, \citenamefont {Irman},
  \citenamefont {Overgaag}, \citenamefont {Vanmaekelbergh},\ and\ \citenamefont
  {Vos}}]{Lodahl2004}%
  \BibitemOpen
  \bibfield  {author} {\bibinfo {author} {\bibfnamefont {P.}~\bibnamefont
  {Lodahl}}, \bibinfo {author} {\bibfnamefont {A.~F.}\ \bibnamefont {{Van
  Driel}}}, \bibinfo {author} {\bibfnamefont {I.~S.}\ \bibnamefont {Nikolaev}},
  \bibinfo {author} {\bibfnamefont {A.}~\bibnamefont {Irman}}, \bibinfo
  {author} {\bibfnamefont {K.}~\bibnamefont {Overgaag}}, \bibinfo {author}
  {\bibfnamefont {D.}~\bibnamefont {Vanmaekelbergh}},\ and\ \bibinfo {author}
  {\bibfnamefont {W.~L.}\ \bibnamefont {Vos}},\ }\bibfield  {title} {\bibinfo
  {title} {{Controlling the dynamics of spontaneous emission from quantum dots
  by photonic crystals}},\ }\href {https://doi.org/10.1038/nature02772}
  {\bibfield  {journal} {\bibinfo  {journal} {Nature}\ }\textbf {\bibinfo
  {volume} {430}},\ \bibinfo {pages} {654} (\bibinfo {year}
  {2004})}\BibitemShut {NoStop}%
\bibitem [{\citenamefont {Ringler}\ \emph {et~al.}(2008)\citenamefont
  {Ringler}, \citenamefont {Schwemer}, \citenamefont {Wunderlich},
  \citenamefont {Nichtl}, \citenamefont {K{\"{u}}rzinger}, \citenamefont
  {Klar},\ and\ \citenamefont {Feldmann}}]{Ringler2008}%
  \BibitemOpen
  \bibfield  {author} {\bibinfo {author} {\bibfnamefont {M.}~\bibnamefont
  {Ringler}}, \bibinfo {author} {\bibfnamefont {A.}~\bibnamefont {Schwemer}},
  \bibinfo {author} {\bibfnamefont {M.}~\bibnamefont {Wunderlich}}, \bibinfo
  {author} {\bibfnamefont {A.}~\bibnamefont {Nichtl}}, \bibinfo {author}
  {\bibfnamefont {K.}~\bibnamefont {K{\"{u}}rzinger}}, \bibinfo {author}
  {\bibfnamefont {T.~A.}\ \bibnamefont {Klar}},\ and\ \bibinfo {author}
  {\bibfnamefont {J.}~\bibnamefont {Feldmann}},\ }\bibfield  {title} {\bibinfo
  {title} {{Shaping Emission Spectra of Fluorescent Molecules with Single
  Plasmonic Nanoresonators}},\ }\href
  {https://doi.org/10.1103/PhysRevLett.100.203002} {\bibfield  {journal}
  {\bibinfo  {journal} {Physical Review Letters}\ }\textbf {\bibinfo {volume}
  {100}},\ \bibinfo {pages} {203002} (\bibinfo {year} {2008})}\BibitemShut
  {NoStop}%
\bibitem [{\citenamefont {Bleuse}\ \emph {et~al.}(2011)\citenamefont {Bleuse},
  \citenamefont {Claudon}, \citenamefont {Creasey}, \citenamefont {Malik},
  \citenamefont {Gerard}, \citenamefont {Maksymov}, \citenamefont {Hugonin},\
  and\ \citenamefont {Lalanne}}]{Bleuse2011}%
  \BibitemOpen
  \bibfield  {author} {\bibinfo {author} {\bibfnamefont {J.}~\bibnamefont
  {Bleuse}}, \bibinfo {author} {\bibfnamefont {J.}~\bibnamefont {Claudon}},
  \bibinfo {author} {\bibfnamefont {M.}~\bibnamefont {Creasey}}, \bibinfo
  {author} {\bibfnamefont {N.~S.}\ \bibnamefont {Malik}}, \bibinfo {author}
  {\bibfnamefont {J.-M.}\ \bibnamefont {Gerard}}, \bibinfo {author}
  {\bibfnamefont {I.}~\bibnamefont {Maksymov}}, \bibinfo {author}
  {\bibfnamefont {J.-P.}\ \bibnamefont {Hugonin}},\ and\ \bibinfo {author}
  {\bibfnamefont {P.}~\bibnamefont {Lalanne}},\ }\bibfield  {title} {\bibinfo
  {title} {{Inhibition, Enhancement, and Control of Spontaneous Emission in
  Photonic Nanowires}},\ }\href
  {https://doi.org/10.1103/PhysRevLett.106.103601} {\bibfield  {journal}
  {\bibinfo  {journal} {Physical Review Letters}\ }\textbf {\bibinfo {volume}
  {106}},\ \bibinfo {pages} {103601} (\bibinfo {year} {2011})}\BibitemShut
  {NoStop}%
\bibitem [{\citenamefont {Novoselov}\ \emph {et~al.}(2005)\citenamefont
  {Novoselov}, \citenamefont {Jiang}, \citenamefont {Schedin}, \citenamefont
  {Booth}, \citenamefont {Khotkevich}, \citenamefont {Morozov},\ and\
  \citenamefont {Geim}}]{Novoselov2005}%
  \BibitemOpen
  \bibfield  {author} {\bibinfo {author} {\bibfnamefont {K.~S.}\ \bibnamefont
  {Novoselov}}, \bibinfo {author} {\bibfnamefont {D.}~\bibnamefont {Jiang}},
  \bibinfo {author} {\bibfnamefont {F.}~\bibnamefont {Schedin}}, \bibinfo
  {author} {\bibfnamefont {T.~J.}\ \bibnamefont {Booth}}, \bibinfo {author}
  {\bibfnamefont {V.~V.}\ \bibnamefont {Khotkevich}}, \bibinfo {author}
  {\bibfnamefont {S.~V.}\ \bibnamefont {Morozov}},\ and\ \bibinfo {author}
  {\bibfnamefont {A.~K.}\ \bibnamefont {Geim}},\ }\bibfield  {title} {\bibinfo
  {title} {{Two-dimensional atomic crystals}},\ }\href
  {https://doi.org/10.1073/pnas.0502848102} {\bibfield  {journal} {\bibinfo
  {journal} {Proc. Natl. Acad. Sci.}\ }\textbf {\bibinfo {volume} {102}},\
  \bibinfo {pages} {10451} (\bibinfo {year} {2005})}\BibitemShut {NoStop}%
\bibitem [{\citenamefont {Geim}\ and\ \citenamefont
  {Novoselov}(2007)}]{Geim2007}%
  \BibitemOpen
  \bibfield  {author} {\bibinfo {author} {\bibfnamefont {A.~K.}\ \bibnamefont
  {Geim}}\ and\ \bibinfo {author} {\bibfnamefont {K.}~\bibnamefont
  {Novoselov}},\ }\bibfield  {title} {\bibinfo {title} {{The rise of
  graphene}},\ }\href {https://doi.org/10.1038/nmat1849} {\bibfield  {journal}
  {\bibinfo  {journal} {Nat. Mater.}\ }\textbf {\bibinfo {volume} {6}},\
  \bibinfo {pages} {183} (\bibinfo {year} {2007})}\BibitemShut {NoStop}%
\bibitem [{\citenamefont {Koppens}\ \emph {et~al.}(2011)\citenamefont
  {Koppens}, \citenamefont {Chang},\ and\ \citenamefont {Abajo}}]{Koppens2011}%
  \BibitemOpen
  \bibfield  {author} {\bibinfo {author} {\bibfnamefont {F.~H.~L.}\
  \bibnamefont {Koppens}}, \bibinfo {author} {\bibfnamefont {D.~E.}\
  \bibnamefont {Chang}},\ and\ \bibinfo {author} {\bibfnamefont {F.~J. G.~D.}\
  \bibnamefont {Abajo}},\ }\bibfield  {title} {\bibinfo {title} {{Graphene
  Plasmonics: A Platform for Strong Light-Matter Interactions}},\ }\href
  {https://doi.org/10.1021/nl201771h} {\bibfield  {journal} {\bibinfo
  {journal} {Nano Lett.}\ }\textbf {\bibinfo {volume} {11}},\ \bibinfo {pages}
  {3370} (\bibinfo {year} {2011})}\BibitemShut {NoStop}%
\bibitem [{\citenamefont {Basov}\ \emph {et~al.}(2016)\citenamefont {Basov},
  \citenamefont {Fogler},\ and\ \citenamefont {{Garcia de Abajo}}}]{Basov2016}%
  \BibitemOpen
  \bibfield  {author} {\bibinfo {author} {\bibfnamefont {D.~N.}\ \bibnamefont
  {Basov}}, \bibinfo {author} {\bibfnamefont {M.~M.}\ \bibnamefont {Fogler}},\
  and\ \bibinfo {author} {\bibfnamefont {F.~J.}\ \bibnamefont {{Garcia de
  Abajo}}},\ }\bibfield  {title} {\bibinfo {title} {{Polaritons in van der
  Waals materials}},\ }\href {https://doi.org/10.1126/science.aag1992}
  {\bibfield  {journal} {\bibinfo  {journal} {Science}\ }\textbf {\bibinfo
  {volume} {354}},\ \bibinfo {pages} {aag1992} (\bibinfo {year}
  {2016})}\BibitemShut {NoStop}%
\bibitem [{\citenamefont {Low}\ \emph {et~al.}(2016)\citenamefont {Low},
  \citenamefont {Chaves}, \citenamefont {Caldwell}, \citenamefont {Kumar},
  \citenamefont {Fang}, \citenamefont {Avouris}, \citenamefont {Heinz},
  \citenamefont {Guinea}, \citenamefont {Martin-Moreno},\ and\ \citenamefont
  {Koppens}}]{Low2016}%
  \BibitemOpen
  \bibfield  {author} {\bibinfo {author} {\bibfnamefont {T.}~\bibnamefont
  {Low}}, \bibinfo {author} {\bibfnamefont {A.}~\bibnamefont {Chaves}},
  \bibinfo {author} {\bibfnamefont {J.~D.}\ \bibnamefont {Caldwell}}, \bibinfo
  {author} {\bibfnamefont {A.}~\bibnamefont {Kumar}}, \bibinfo {author}
  {\bibfnamefont {N.~X.}\ \bibnamefont {Fang}}, \bibinfo {author}
  {\bibfnamefont {P.}~\bibnamefont {Avouris}}, \bibinfo {author} {\bibfnamefont
  {T.~F.}\ \bibnamefont {Heinz}}, \bibinfo {author} {\bibfnamefont
  {F.}~\bibnamefont {Guinea}}, \bibinfo {author} {\bibfnamefont
  {L.}~\bibnamefont {Martin-Moreno}},\ and\ \bibinfo {author} {\bibfnamefont
  {F.}~\bibnamefont {Koppens}},\ }\bibfield  {title} {\bibinfo {title}
  {{Polaritons in layered two-dimensional materials}},\ }\href
  {https://doi.org/10.1038/nmat4792} {\bibfield  {journal} {\bibinfo  {journal}
  {Nat. Mater.}\ }\textbf {\bibinfo {volume} {16}},\ \bibinfo {pages} {182}
  (\bibinfo {year} {2016})}\BibitemShut {NoStop}%
\bibitem [{\citenamefont {Moskovits}(1985)}]{Moskovits1985}%
  \BibitemOpen
  \bibfield  {author} {\bibinfo {author} {\bibfnamefont {M.}~\bibnamefont
  {Moskovits}},\ }\bibfield  {title} {\bibinfo {title} {{Surface-enhanced
  spectroscopy}},\ }\href {https://doi.org/10.1103/RevModPhys.57.783}
  {\bibfield  {journal} {\bibinfo  {journal} {Rev. Mod. Phys.}\ }\textbf
  {\bibinfo {volume} {57}},\ \bibinfo {pages} {783} (\bibinfo {year}
  {1985})}\BibitemShut {NoStop}%
\bibitem [{\citenamefont {Nie}\ and\ \citenamefont {Emory}(1997)}]{Nie1997}%
  \BibitemOpen
  \bibfield  {author} {\bibinfo {author} {\bibfnamefont {S.}~\bibnamefont
  {Nie}}\ and\ \bibinfo {author} {\bibfnamefont {S.~R.}\ \bibnamefont
  {Emory}},\ }\bibfield  {title} {\bibinfo {title} {{Probing single molecules
  and single nanoparticles by surface-enhanced Raman scattering}},\ }\href
  {https://doi.org/10.1126/science.275.5303.1102} {\bibfield  {journal}
  {\bibinfo  {journal} {Science}\ }\textbf {\bibinfo {volume} {275}},\ \bibinfo
  {pages} {1102} (\bibinfo {year} {1997})}\BibitemShut {NoStop}%
\bibitem [{\citenamefont {Stiles}\ \emph {et~al.}(2008)\citenamefont {Stiles},
  \citenamefont {Dieringer}, \citenamefont {Shah},\ and\ \citenamefont {{Van
  Duyne}}}]{Stiles2008}%
  \BibitemOpen
  \bibfield  {author} {\bibinfo {author} {\bibfnamefont {P.~L.}\ \bibnamefont
  {Stiles}}, \bibinfo {author} {\bibfnamefont {J.~A.}\ \bibnamefont
  {Dieringer}}, \bibinfo {author} {\bibfnamefont {N.~C.}\ \bibnamefont
  {Shah}},\ and\ \bibinfo {author} {\bibfnamefont {R.~P.}\ \bibnamefont {{Van
  Duyne}}},\ }\bibfield  {title} {\bibinfo {title} {{Surface-Enhanced Raman
  Spectroscopy}},\ }\href
  {https://doi.org/10.1146/annurev.anchem.1.031207.112814} {\bibfield
  {journal} {\bibinfo  {journal} {Annu. Rev. Anal. Chem}\ }\textbf {\bibinfo
  {volume} {1}},\ \bibinfo {pages} {601} (\bibinfo {year} {2008})}\BibitemShut
  {NoStop}%
\bibitem [{\citenamefont {Caz{\'e}}\ \emph {et~al.}(2013)\citenamefont
  {Caz{\'e}}, \citenamefont {Pierrat},\ and\ \citenamefont
  {Carminati}}]{caze_pierrat_carminati_2013}%
  \BibitemOpen
  \bibfield  {author} {\bibinfo {author} {\bibfnamefont {A.}~\bibnamefont
  {Caz{\'e}}}, \bibinfo {author} {\bibfnamefont {R.}~\bibnamefont {Pierrat}},\
  and\ \bibinfo {author} {\bibfnamefont {R.}~\bibnamefont {Carminati}},\
  }\bibfield  {title} {\bibinfo {title} {Spatial coherence in complex photonic
  and plasmonic systems},\ }\href
  {https://doi.org/10.1103/physrevlett.110.063903} {\bibfield  {journal}
  {\bibinfo  {journal} {Physical Review Letters}\ }\textbf {\bibinfo {volume}
  {110}},\ \bibinfo {pages} {063903} (\bibinfo {year} {2013})}\BibitemShut
  {NoStop}%
\bibitem [{\citenamefont {Gonzaga-Galeana}\ and\ \citenamefont
  {Zurita-S{\'a}nchez}(2013)}]{gonzaga-galeana_zurita-sanchez_2013}%
  \BibitemOpen
  \bibfield  {author} {\bibinfo {author} {\bibfnamefont {J.~A.}\ \bibnamefont
  {Gonzaga-Galeana}}\ and\ \bibinfo {author} {\bibfnamefont {J.~R.}\
  \bibnamefont {Zurita-S{\'a}nchez}},\ }\bibfield  {title} {\bibinfo {title} {A
  revisitation of the {F{\"o}rster} energy transfer near a metallic spherical
  nanoparticle: (1) efficiency enhancement or reduction? (2) the control of the
  {F{\"o}rster} radius of the unbounded medium. (3) the impact of the local
  density of states},\ }\href {https://doi.org/10.1063/1.4847875} {\bibfield
  {journal} {\bibinfo  {journal} {The Journal of Chemical Physics}\ }\textbf
  {\bibinfo {volume} {139}},\ \bibinfo {pages} {244302} (\bibinfo {year}
  {2013})}\BibitemShut {NoStop}%
\bibitem [{\citenamefont {Gonzalez-Tudela}\ \emph {et~al.}(2011)\citenamefont
  {Gonzalez-Tudela}, \citenamefont {Martin-Cano}, \citenamefont {Moreno},
  \citenamefont {Martin-Moreno}, \citenamefont {Tejedor},\ and\ \citenamefont
  {Garcia-Vidal}}]{gonzalez-tudela_martin-cano_2011}%
  \BibitemOpen
  \bibfield  {author} {\bibinfo {author} {\bibfnamefont {A.}~\bibnamefont
  {Gonzalez-Tudela}}, \bibinfo {author} {\bibfnamefont {D.}~\bibnamefont
  {Martin-Cano}}, \bibinfo {author} {\bibfnamefont {E.}~\bibnamefont {Moreno}},
  \bibinfo {author} {\bibfnamefont {L.}~\bibnamefont {Martin-Moreno}}, \bibinfo
  {author} {\bibfnamefont {C.}~\bibnamefont {Tejedor}},\ and\ \bibinfo {author}
  {\bibfnamefont {F.~J.}\ \bibnamefont {Garcia-Vidal}},\ }\bibfield  {title}
  {\bibinfo {title} {Entanglement of two qubits mediated by one-dimensional
  plasmonic waveguides},\ }\href
  {https://doi.org/10.1103/physrevlett.106.020501} {\bibfield  {journal}
  {\bibinfo  {journal} {Physical Review Letters}\ }\textbf {\bibinfo {volume}
  {106}},\ \bibinfo {pages} {020501} (\bibinfo {year} {2011})}\BibitemShut
  {NoStop}%
\bibitem [{\citenamefont {Lassalle}\ \emph {et~al.}(2020)\citenamefont
  {Lassalle}, \citenamefont {Lalanne}, \citenamefont {Aljunid}, \citenamefont
  {Genevet}, \citenamefont {Stout}, \citenamefont {Durt},\ and\ \citenamefont
  {Wilkowski}}]{lassalle_long-lifetime_2020}%
  \BibitemOpen
  \bibfield  {author} {\bibinfo {author} {\bibfnamefont {E.}~\bibnamefont
  {Lassalle}}, \bibinfo {author} {\bibfnamefont {P.}~\bibnamefont {Lalanne}},
  \bibinfo {author} {\bibfnamefont {S.}~\bibnamefont {Aljunid}}, \bibinfo
  {author} {\bibfnamefont {P.}~\bibnamefont {Genevet}}, \bibinfo {author}
  {\bibfnamefont {B.}~\bibnamefont {Stout}}, \bibinfo {author} {\bibfnamefont
  {T.}~\bibnamefont {Durt}},\ and\ \bibinfo {author} {\bibfnamefont
  {D.}~\bibnamefont {Wilkowski}},\ }\bibfield  {title} {{\selectlanguage
  {en}\bibinfo {title} {Long-lifetime coherence in a quantum emitter induced by
  a metasurface}},\ }\bibfield  {journal} {\bibinfo  {journal} {Physical Review
  A}\ }\textbf {\bibinfo {volume} {101}},\ \href
  {https://doi.org/10.1103/PhysRevA.101.013837} {10.1103/PhysRevA.101.013837}
  (\bibinfo {year} {2020})\BibitemShut {NoStop}%
\bibitem [{\citenamefont {Gustafsson}\ \emph {et~al.}(2019)\citenamefont
  {Gustafsson}, \citenamefont {Schab}, \citenamefont {Jelinek},\ and\
  \citenamefont {Capek}}]{gustafsson2019upper}%
  \BibitemOpen
  \bibfield  {author} {\bibinfo {author} {\bibfnamefont {M.}~\bibnamefont
  {Gustafsson}}, \bibinfo {author} {\bibfnamefont {K.}~\bibnamefont {Schab}},
  \bibinfo {author} {\bibfnamefont {L.}~\bibnamefont {Jelinek}},\ and\ \bibinfo
  {author} {\bibfnamefont {M.}~\bibnamefont {Capek}},\ }\bibfield  {title}
  {\bibinfo {title} {Upper bounds on absorption and scattering},\ }\href
  {http://arxiv.org/abs/1912.06699} {\bibfield  {journal} {\bibinfo  {journal}
  {arXiv preprint arXiv:1912.06699}\ } (\bibinfo {year} {2019})}\BibitemShut
  {NoStop}%
\bibitem [{\citenamefont {Molesky}\ \emph
  {et~al.}(2020{\natexlab{a}})\citenamefont {Molesky}, \citenamefont {Chao},
  \citenamefont {Jin},\ and\ \citenamefont {Rodriguez}}]{Molesky2020a}%
  \BibitemOpen
  \bibfield  {author} {\bibinfo {author} {\bibfnamefont {S.}~\bibnamefont
  {Molesky}}, \bibinfo {author} {\bibfnamefont {P.}~\bibnamefont {Chao}},
  \bibinfo {author} {\bibfnamefont {W.}~\bibnamefont {Jin}},\ and\ \bibinfo
  {author} {\bibfnamefont {A.~W.}\ \bibnamefont {Rodriguez}},\ }\bibfield
  {title} {\bibinfo {title} {{Global T operator bounds on electromagnetic
  scattering: Upper bounds on far-field cross sections}},\ }\href
  {https://doi.org/10.1103/physrevresearch.2.033172} {\bibfield  {journal}
  {\bibinfo  {journal} {Phys. Rev. Res.}\ }\textbf {\bibinfo {volume} {2}},\
  \bibinfo {pages} {033172} (\bibinfo {year} {2020}{\natexlab{a}})}\BibitemShut
  {NoStop}%
\bibitem [{\citenamefont {Kuang}\ and\ \citenamefont
  {Miller}(2020)}]{Kuang2020}%
  \BibitemOpen
  \bibfield  {author} {\bibinfo {author} {\bibfnamefont {Z.}~\bibnamefont
  {Kuang}}\ and\ \bibinfo {author} {\bibfnamefont {O.~D.}\ \bibnamefont
  {Miller}},\ }\bibfield  {title} {\bibinfo {title} {Computational bounds to
  light-matter interactions via local conservation laws},\ }\href@noop {}
  {\bibfield  {journal} {\bibinfo  {journal} {arXiv:2008.13325}\ } (\bibinfo
  {year} {2020})},\ \Eprint {https://arxiv.org/abs/2008.13325}
  {arXiv:2008.13325} \BibitemShut {NoStop}%
\bibitem [{\citenamefont {Molesky}\ \emph
  {et~al.}(2020{\natexlab{b}})\citenamefont {Molesky}, \citenamefont {Chao},\
  and\ \citenamefont {Rodriguez}}]{Molesky2020c}%
  \BibitemOpen
  \bibfield  {author} {\bibinfo {author} {\bibfnamefont {S.}~\bibnamefont
  {Molesky}}, \bibinfo {author} {\bibfnamefont {P.}~\bibnamefont {Chao}},\ and\
  \bibinfo {author} {\bibfnamefont {A.~W.}\ \bibnamefont {Rodriguez}},\
  }\bibfield  {title} {\bibinfo {title} {Hierarchical mean-field t-operator
  bounds on electromagnetic scattering: Upper bounds on near-field radiative
  purcell enhancement},\ }\href {http://arxiv.org/abs/2008.08168} {\bibfield
  {journal} {\bibinfo  {journal} {arXiv:2008.08168}\ } (\bibinfo {year}
  {2020}{\natexlab{b}})},\ \Eprint {https://arxiv.org/abs/2008.08168}
  {arXiv:2008.08168} \BibitemShut {NoStop}%
\bibitem [{\citenamefont {Luo}\ \emph {et~al.}(2010)\citenamefont {Luo},
  \citenamefont {Ma}, \citenamefont {So}, \citenamefont {Ye},\ and\
  \citenamefont {Zhang}}]{Luo2010}%
  \BibitemOpen
  \bibfield  {author} {\bibinfo {author} {\bibfnamefont {Z.~Q.}\ \bibnamefont
  {Luo}}, \bibinfo {author} {\bibfnamefont {W.~K.}\ \bibnamefont {Ma}},
  \bibinfo {author} {\bibfnamefont {A.}~\bibnamefont {So}}, \bibinfo {author}
  {\bibfnamefont {Y.}~\bibnamefont {Ye}},\ and\ \bibinfo {author}
  {\bibfnamefont {S.}~\bibnamefont {Zhang}},\ }\bibfield  {title} {\bibinfo
  {title} {{Semidefinite relaxation of quadratic optimization problems}},\
  }\href {https://doi.org/10.1109/MSP.2010.936019} {\bibfield  {journal}
  {\bibinfo  {journal} {IEEE Signal Processing Magazine}\ }\textbf {\bibinfo
  {volume} {27}},\ \bibinfo {pages} {20} (\bibinfo {year} {2010})}\BibitemShut
  {NoStop}%
\bibitem [{\citenamefont {Shim}\ \emph
  {et~al.}(2019{\natexlab{b}})\citenamefont {Shim}, \citenamefont {Chung},\
  and\ \citenamefont {Miller}}]{Shim2019subm}%
  \BibitemOpen
  \bibfield  {author} {\bibinfo {author} {\bibfnamefont {H.}~\bibnamefont
  {Shim}}, \bibinfo {author} {\bibfnamefont {H.}~\bibnamefont {Chung}},\ and\
  \bibinfo {author} {\bibfnamefont {O.~D.}\ \bibnamefont {Miller}},\ }\bibfield
   {title} {\bibinfo {title} {Maximal free-space concentration of
  electromagnetic waves},\ }\href {http://arxiv.org/abs/1905.10500} {\bibfield
  {journal} {\bibinfo  {journal} {arXiv:1905.10500}\ } (\bibinfo {year}
  {2019}{\natexlab{b}})}\BibitemShut {NoStop}%
\bibitem [{\citenamefont {Yablonovitch}(1982)}]{Yablonovitch1982}%
  \BibitemOpen
  \bibfield  {author} {\bibinfo {author} {\bibfnamefont {E.}~\bibnamefont
  {Yablonovitch}},\ }\bibfield  {title} {\bibinfo {title} {{Statistical ray
  optics}},\ }\href {https://doi.org/10.1364/JOSA.72.000899} {\bibfield
  {journal} {\bibinfo  {journal} {J. Opt. Soc. Am.}\ }\textbf {\bibinfo
  {volume} {72}},\ \bibinfo {pages} {899} (\bibinfo {year} {1982})}\BibitemShut
  {NoStop}%
\bibitem [{\citenamefont {Buddhiraju}\ and\ \citenamefont
  {Fan}(2017)}]{buddhiraju_theory_2017}%
  \BibitemOpen
  \bibfield  {author} {\bibinfo {author} {\bibfnamefont {S.}~\bibnamefont
  {Buddhiraju}}\ and\ \bibinfo {author} {\bibfnamefont {S.}~\bibnamefont
  {Fan}},\ }\bibfield  {title} {{\selectlanguage {en}\bibinfo {title} {Theory
  of solar cell light trapping through a nonequilibrium {Green}'s function
  formulation of {Maxwell}'s equations}},\ }\href
  {https://doi.org/10.1103/PhysRevB.96.035304} {\bibfield  {journal} {\bibinfo
  {journal} {Physical Review B}\ }\textbf {\bibinfo {volume} {96}},\ \bibinfo
  {pages} {035304} (\bibinfo {year} {2017})}\BibitemShut {NoStop}%
\bibitem [{\citenamefont {Benzaouia}\ \emph {et~al.}(2019)\citenamefont
  {Benzaouia}, \citenamefont {Tokic}, \citenamefont {Miller}, \citenamefont
  {Yue},\ and\ \citenamefont {Johnson}}]{Benzaouia2019}%
  \BibitemOpen
  \bibfield  {author} {\bibinfo {author} {\bibfnamefont {M.}~\bibnamefont
  {Benzaouia}}, \bibinfo {author} {\bibfnamefont {G.}~\bibnamefont {Tokic}},
  \bibinfo {author} {\bibfnamefont {O.~D.}\ \bibnamefont {Miller}}, \bibinfo
  {author} {\bibfnamefont {D.~K.~P.}\ \bibnamefont {Yue}},\ and\ \bibinfo
  {author} {\bibfnamefont {S.~G.}\ \bibnamefont {Johnson}},\ }\bibfield
  {title} {\bibinfo {title} {From solar cells to ocean buoys: Wide-bandwidth
  limits to absorption by metaparticle arrays},\ }\href
  {https://doi.org/10.1103/PhysRevApplied.11.034033} {\bibfield  {journal}
  {\bibinfo  {journal} {Physical Review Applied}\ }\textbf {\bibinfo {volume}
  {11}},\ \bibinfo {pages} {034033} (\bibinfo {year} {2019})}\BibitemShut
  {NoStop}%
\bibitem [{\citenamefont {Angeris}\ \emph {et~al.}(2019)\citenamefont
  {Angeris}, \citenamefont {Vuckovic},\ and\ \citenamefont
  {Boyd}}]{Angeris2019}%
  \BibitemOpen
  \bibfield  {author} {\bibinfo {author} {\bibfnamefont {G.}~\bibnamefont
  {Angeris}}, \bibinfo {author} {\bibfnamefont {J.}~\bibnamefont {Vuckovic}},\
  and\ \bibinfo {author} {\bibfnamefont {S.~P.}\ \bibnamefont {Boyd}},\
  }\bibfield  {title} {\bibinfo {title} {{Computational Bounds for Photonic
  Design}},\ }\href {https://doi.org/10.1021/acsphotonics.9b00154} {\bibfield
  {journal} {\bibinfo  {journal} {ACS Photonics}\ }\textbf {\bibinfo {volume}
  {6}},\ \bibinfo {pages} {1232} (\bibinfo {year} {2019})}\BibitemShut
  {NoStop}%
\bibitem [{\citenamefont {Bergman}(1981)}]{Bergman1981}%
  \BibitemOpen
  \bibfield  {author} {\bibinfo {author} {\bibfnamefont {D.~J.}\ \bibnamefont
  {Bergman}},\ }\bibfield  {title} {\bibinfo {title} {{Bounds for the complex
  dielectric constant of a two-component composite material}},\ }\href
  {https://doi.org/10.1103/PhysRevB.23.3058} {\bibfield  {journal} {\bibinfo
  {journal} {Phys. Rev. B}\ }\textbf {\bibinfo {volume} {23}},\ \bibinfo
  {pages} {3058} (\bibinfo {year} {1981})}\BibitemShut {NoStop}%
\bibitem [{\citenamefont {Milton}(1981)}]{Milton1981}%
  \BibitemOpen
  \bibfield  {author} {\bibinfo {author} {\bibfnamefont {G.~W.}\ \bibnamefont
  {Milton}},\ }\bibfield  {title} {\bibinfo {title} {{Bounds on the complex
  permittivity of a two-component composite material}},\ }\href
  {https://doi.org/10.1063/1.329385} {\bibfield  {journal} {\bibinfo  {journal}
  {J. Appl. Phys.}\ }\textbf {\bibinfo {volume} {52}},\ \bibinfo {pages} {5286}
  (\bibinfo {year} {1981})}\BibitemShut {NoStop}%
\end{thebibliography}%

\end{document}


\title{Supplementary Material: Maximal single-frequency electromagnetic response}

\author{Zeyu Kuang}
\affiliation{Department of Applied Physics and Energy Sciences Institute, Yale University, New Haven, Connecticut 06511, USA}
\author{Lang Zhang}
\affiliation{Department of Applied Physics and Energy Sciences Institute, Yale University, New Haven, Connecticut 06511, USA}
\author{Owen D. Miller}
\affiliation{Department of Applied Physics and Energy Sciences Institute, Yale University, New Haven, Connecticut 06511, USA}

\date{\today}

\maketitle

\setcounter{tocdepth}{1}
\tableofcontents

\section{The optimization problem and its dual function}
The optimization problem is to maximize a response function $f(\phi) = \phi^\dagger \mathbb{A} \phi +  \Im\left[\beta^\dagger \phi\right]$ under the optical theorem constraint, where the variable $\phi$ is polarization current induced in the scatterer. Under a prespecified basis, parameter $\beta$ is a vector, and $\Abb$ is a Hermitian matrix. 
The same basis defines positive semidefinite matrix $\ImGO$ and $\Im\xi$, representing radiative and material loss in the system.
Following the standard optimization notation, we rewrite the original maximization problem as a minimization problem by adding a minus sign to the objective function:
\begin{equation}
\begin{aligned}
& \underset{\phi}{\text{minimize}}
& & -f(\phi) = -\phi^\dagger \mathbb{A} \phi -  \Im\left[\beta^\dagger \phi\right] \\
& \text{subject to}
& & \phi^\dagger \left\{\Im \xi + \ImGO \right\} \phi = \Im \left[\psi_{\rm inc}^\dagger \phi\right].
\end{aligned}
\label{eq:gen-form-min}
\end{equation}
The optimization problem stated in \eqref{gen-form-min} is known to have strong duality~\cite{boyd2004convex}, prompting us to find its dual function, which in turn is defined by its Lagrangian:
\begin{equation}
L(\phi,\nu) = \phi^\dagger\BB(\nu)\phi - \Im\left[(\beta + \nu \psiinc)^\dagger\phi\right],
\label{eq:def-L}
\end{equation}
where we introduce dual variable $\nu$ and simplify our notation by introducing matrix
\begin{equation}
\BB(\nu) = -\mathbb{A}+\nu(\Im\xi+\ImGO).
\label{eq:def-B}
\end{equation}

The dual function $g(\nu)$ is defined as the minimum of Lagrangian $L(\phi,\nu)$ over variable $\phi$. We denote $\nu_0$ as the value of $\nu$ when the minimum eigenvalue of $\BB(\nu)$ is zero, leaving $\BB(\nu_0)$ a positive semidefinite matrix with at least one zero eigenvalue. For $\nu<\nu_0$, the positivity of $\Im\xi+\ImGO$ implies that $\BB(\nu)=\BB(\nu_0)-(\nu_0-\nu)(\Im\xi+\ImGO)$ has negative eigenvalues and $L(\phi,\nu)$ is unbounded below. For $\nu>\nu_0$, $\BB(\nu)=\BB(\nu_0)+(\nu-\nu_0)(\Im\xi+\ImGO)$ is positive definite, and $L(\phi,\nu)$ is convex in $\phi$ with a finite minimal value. This minimum is obtained at
\begin{equation}
\phi(\nu) = \frac{i}{2}\BB^{-1}(\nu)(\beta+\nu\psi_{\rm inc}),
\label{eq:phi}
\end{equation}
with the resulting dual function:
\begin{align}
g(\nu)= \min_\phi L(\phi, \nu) =
\begin{cases}
-\frac{1}{4}(\beta + \nu \psiinc)^\dagger \BB^{-1}(\nu)(\beta + \nu \psiinc) & \nu > \nu_0 
\\
-\infty & \nu< \nu_0.
\end{cases}
\label{eq:gnu}
\end{align}
Lastly, at $\nu=\nu_0$, if $\beta + \nu_0\psiinc$ is in the range of $\BB(\nu_0)$, then $L(\phi,\nu)$ is still convex and $g(\nu_0)$ takes the value of the first case in \eqref{gnu} with the inverse operator replaced by the pseudo-inverse; if not, then \eqref{def-L} suggests that $L(\phi,\nu)$ is unbounded below and  $g(\nu_0)\rightarrow -\infty$.

Due to strong duality, the optimization problem, \eqref{gen-form-min}, is solved by finding the maximum of the dual function:
\begin{equation}
\begin{aligned}
& \underset{\nu}{\text{maximize}}
& & g(\nu),
\end{aligned}
\label{eq:1_1}
\end{equation}
According to \eqref{gnu}, dual function $g(\nu)$ is maximized at a value within range $[\nu_0,+\infty)$, which we denote as $\nu^*$. The maximum response function takes the (negative of the optimal dual) value:
\begin{align}
f_{\rm max} = \frac{1}{4}(\beta + \nu^* \psiinc)^\dagger \BB^{-1}(\nu^*)(\beta + \nu^* \psiinc),
\label{eq:Popt}
\end{align}
and the optimal polarization current $\phi$ is given by evaluating \eqref{phi} at $\nu^*$:
\begin{equation}
\phi^* = \frac{i}{2}\BB^{-1}(\nu^*)(\beta + \nu^* \psiinc) \label{eq:x},
\end{equation}
except when $\nu^*=\nu_0$, where the $\phi^*$ can not be uniquely determined due to the presence of zero eigenvalues in $\BB(\nu_0)$.

To solve for the maximum response function $f_{\rm max}$ in \eqref{Popt}, we need to find the optimal dual variable $\nu^*$, which can only occur either in the interior of the domain $[\nu_0,\infty)$ or its boundary. If $\nu^*$ is in the interior, it has to satisfy the condition:
\begin{align}
\frac{\partial g(\nu)}{\partial \nu}\biggr\rvert_{\nu=\nu^*} = 0.
\label{eq:first_div_zero}
\end{align}
This can be translated to a transcendental equation that determines the first possible optimum which we denote as $\nu_1$:

\begin{equation}
\begin{aligned}
2\Re\left\{\psiinc^\dagger\BB^{-1}(\nu_1)(\beta+\nu_1\psiinc)\right\}
- (\beta+\nu_1\psiinc)^\dagger\BB^{-1}(\nu_1)&\BB'(\nu_1)\BB^{-1}(\nu_1)(\beta+\nu_1\psiinc) = 0.
\label{eq:optlambda}
\end{aligned}
\end{equation}
The concavity of the dual function $g(\nu)$ guarantees the uniqueness of the solution $\nu_1$. The lefthand side of \eqref{optlambda} is proportional to $-\partial g(\lambda)/\partial\nu$. Its derivative, $-\partial^2 g(\lambda)/\partial\nu^2$, is always non-negative based on the second-order condition of a concave function~\cite{boyd2004convex}. Thus, if there is a $\nu_1$ satisfying \eqref{optlambda}, it can simply be solved by identifying where the sign of the lefthand side changes, using either bisection or Newtonâ€™s method.

Based on the concavity of the dual function, we can also argue that if $\nu_1$ exists in the domain $(\nu_0,\infty)$ then it must be the global optimizer of $g(\nu)$. If not, then there is no point in the domain at which the gradient is zero, and $\nu^*$ must be one of the boundary values of $[\nu_0,\infty)$; by the concavity of $g(\nu)$, the maximum must occur at $\nu_0$. Hence we have:
\begin{align}
\nu^* = 
\begin{cases}
\nu_1 & \textrm{if } \nu_1 \in (\nu_0,\infty) \\
\nu_0 & \textrm{else.}
\end{cases}
\label{eq:det_nu}
\end{align}
The self-consistency implicit in \eqref{optlambda} for $\nu_1$ can make it to difficult to ascertain whether $\nu_1$ or $\nu_0$ is optimal. Instead, if the derivative of $g(\nu)$ at $\nu_0$ is well-defined, we can check its value to determine whether $g(\nu)$ attains it extremum in the interior of its domain or on its boundary: if and only if it is positive, then $\nu_1$ will be in the interior of the domain $[\nu_0,\infty)$. Hence, if $\beta + \nu_0 \psiinc$ is in the range of  ${\BB^{-1}(\nu_0)}$, then we can also use the equivalent condition to determine $\nu^*$:
\begin{align}
\nu^* = 
\begin{cases}
\nu_1 & \textrm{if } 2\psiinc^\dagger\BB^{-1}(\nu_0)(\beta+\nu_0\psiinc) < (\beta+\nu_0\psiinc)^\dagger \BB^{-1}(\nu_0)\BB'(\nu_0)\BB^{-1}(\nu_0)(\beta+\nu_0\psiinc) \\
\nu_0 & \textrm{else.}
\end{cases}
\label{eq:det-nu-alt}
\end{align}

\section{Absorbed, scattered, and extinguished power expressions}
We start with extinguished power which is linear in polarization current $\phi$: $\Pext = \frac{\omega}{2} \Im \left[ \psiinc^\dagger \phi \right]$. For simplicity, we take the objective function as $\Im \left[ \psiinc^\dagger \phi \right]$, and set $\mathbb{A} = 0$ and $\beta = \psiinc$ in the optimization problem, \eqref{gen-form-min}. The matrix defined in \eqref{def-B} becomes $\BB(\nu) = \nu (\Im\xi+\ImGO)$. Its minimum eigenvalue reaches zero when $\nu=\nu_0=0$. Dual function in \eqref{gnu} takes the form:
\begin{align}
g(\nu) =
\begin{cases}
-\frac{(\nu+1)^2}{4}\psiinc^\dagger (\Im\xi+\ImGO)^{-1}\psiinc & \nu > 0
\\
-\infty & \nu \leq 0,
\end{cases}
\end{align}
where we identified $g(\nu_0)=-\infty$ since $\beta+\nu_0\psiinc$ is not in the range of $B(\nu_0)$. Since $g(\nu_0)=-\infty$, the optimal dual variable $\nu^*$ can only be chosen at $\nu_1$. Solving \eqref{optlambda} gives $\nu_1=1$ and the maximum extinction given by \eqref{Popt} is (after adding back the $\frac{\omega}{2}$ prefactor):
\begin{align}
\Pext^{\rm max} = \frac{\omega}{2} \psiinc^\dagger \left(\Im\xi+\ImGO\right)^{-1} \psiinc.
\label{eq:Pext_bound}
\end{align}
The optimum polarization current $\phi^*$ is given by \eqref{x}:
\begin{equation}
\phi^* = i(\Im\xi+\ImGO)^{-1}\psiinc.
\end{equation}

Absorption has the form $\Pabs = \frac{\omega}{2} \phi^\dagger (\Im\xi) \phi$. Taking the objective function as $\phi^\dagger (\Im\xi) \phi$, we have $\mathbb{A} = \Im \xi$ and $\beta = 0$ in the optimization problem, \eqref{gen-form-min}. The matrix defined in \eqref{def-B} becomes $\BB(\nu) = (\nu-1)\Im\xi+\nu\ImGO$. Dual function takes the form of \eqref{gnu}:
\begin{align}
g(\nu) =
\begin{cases}
-\frac{\nu^2}{4}\psiinc^\dagger [(\nu-1)\Im\xi+\nu\ImGO]^{-1}\psiinc & \nu > \nu_0
\\
-\infty & \nu < \nu_0,
\end{cases}
\label{eq:gnu_abs}
\end{align}
At $\nu=\nu_0$, the value of $g(\nu_0)\rightarrow-\infty$ if $\psiinc$ is not in the range of $\BB(\nu_0)$, otherwise $g(\nu_0)$ takes the form of the first case in \eqref{gnu_abs} with the inverse replaced by pseudo-inverse.
As in \eqref{det_nu}, the optimal dual variable $\nu^*$ is obtained either at the  interval $(\nu_0,\infty)$ or its boundary $\nu_0$. The value of $\nu_0$ depends on the nature of both $\Im\xi$ and $\ImGO$. The value of $\nu_1$ is given by \eqref{optlambda}:
\begin{equation}
\psiinc^\dagger\left[ 2\BB^{-1}(\nu_1) - \nu_1\BB^{-1}(\nu_1) (\Im\xi+\ImGO) \BB^{-1}(\nu_1) \right]\psiinc = 0.
\label{eq:nu1_abs}
\end{equation}
Using \eqref{Popt} and adding back the $\frac{\omega}{2}$ prefactor, we have maximum absorption:
\begin{align}
\Pabs^{\rm max} = \frac{\omega}{2}\frac{\nu^{*2}}{4} \psiinc^\dagger[(\nu^*-1)\Im\xi+\nu^*\ImGO]^{-1} \psiinc.
\label{eq:Pabs_bound}
\end{align}
The optimal current can be determined by \eqref{x} in the case of $\nu^*=\nu_1$:
\begin{equation}
\phi^* = i\frac{\nu^*}{2}[(\nu^*-1)\Im\xi+\nu^*\ImGO]^{-1}\psiinc.
\end{equation}

Scattering power has the form $\Pscat = \frac{\omega}{2} \phi^\dagger (\ImGO) \phi$, such that $\mathbb{A} = \ImGO$ and $\beta=0$ after suppressing the $\frac{\omega}{2}$ prefactor. Following a similar procedure as absorption, we have maximum scattering as:
\begin{align}
\Pscat^{\rm max} = \frac{\omega}{2}\frac{\nu^{*2}}{4} \psiinc^\dagger[\nu^*\Im\xi + (\nu^*-1)\ImGO]^{-1} \psiinc.
\label{eq:Psca_bound}
\end{align}
Again, $\nu^*$ takes two possible values: $\nu_1$ and $\nu_0$, as dictated by \eqref{det_nu}. The determinant equation for $\nu_1$ takes the same form as \eqref{nu1_abs} with $\BB(\nu) = \nu\Im\xi+(\nu-1)\ImGO$.

An equivalent formulation for all three power quantities is to write them as the difference (or sum) of the other two. For example, scattering power can be written as the difference between extinction and absorption: $\Pscat = \frac{\omega}{2} \Im \left[ \psiinc^\dagger \phi \right] - \frac{\omega}{2} \phi^\dagger (\Im\xi) \phi$. With $\Abb=-\Im\xi$ and $\beta=\psi_{\rm inc}$ after suppressing the $\frac{\omega}{2}$ prefactor, this gives the same optima as in \eqref{Psca_bound} but with a different form:
\begin{align}
\Pscat^{\rm max} = \frac{\omega}{2} \frac{(1+\nu^*)^2}{4} \psiinc^\dagger \left[(\nu^*+1)\Im\xi + \nu^*\ImGO \right]^{-1}\psiinc,
\label{eq:Psca_bound_2nd}
\end{align}
where the optimal dual variable $\nu^*$ is determined by \eqref{det_nu}.

\section{Bound for a nonmagnetic scalar material under plane wave incidence}
Let us consider a typical case where the incident field is a plane wave and the scatterer is composed of nonmagnetic scalar material. Here, we only need to consider the electric response in Eq. (\ref{eq:Pext_bound}, \ref{eq:Pabs_bound}, \ref{eq:Psca_bound}), so we can replace $\psiinc$ with $\ev_{\rm inc}$, $\Gamma_0$ with $\GG_0$, and $\xi=\Im\chi/|\chi|^2$ is now a scalar with $\chi$ being the electric susceptibility of the material. 
Because $\Im \GG_0$ is positive-semidefinite, we can simplify its eigendecomposition to write $\Im\GG_0 = \VV\VV^\dagger$, where the columns of $\VV$, which we denote $\vv_i$, form an orthogonal basis of polarization currents. They are normalized such that the set $\rho_i=\vv_i^\dagger \vv_i$ are the eigenvalues of $\Im \GG_0$ and represent the powers radiated by unit-normalization polarization currents. The expansion of incident plane wave, $\ev_{\rm inc}$, in these channels is assumed to be: $\ev_{\rm inc} = \frac{1}{k^{3/2}}\sum_i e_i \vv_i$, where the exact value of $|e_i|^2$ depends on the choice of $\vv_i$. Throughout the SM, we use uncapitalized symbol $\ev_{\rm inc}$ to denote the incident electric field to emphasis its vector nature.

We decompose general bounds given by Eq. (\ref{eq:Pext_bound}, \ref{eq:Pabs_bound}, \ref{eq:Psca_bound}) into contributions from these channels: 
\begin{align}
P_{\rm ext} &\leq \frac{\omega}{2}\frac{1}{k^3}\sum_i |e_i|^2\frac{\rho_i}{\Im{\xi} + \rho_i} \label{eq:pw_ext}\\
P_{\rm abs} &\leq \frac{\omega}{2} \frac{\nu^{*2}}{4}\frac{1}{k^3}\sum_i |e_i|^2\frac{\rho_i}{(\nu^*-1)\Im{\xi}+\nu^*\rho_i} \label{eq:pw_abs} \\
P_{\rm scat} &\leq \frac{\omega}{2} \frac{\nu^{*2}}{4}\frac{1}{k^3}\sum_i |e_i|^2\frac{\rho_i}{\nu^*\Im{\xi}+(\nu^*-1)\rho_i} \label{eq:pw_sca}.
\end{align}
Taking $\omega=k$ in our unitless convention and write $k=2\pi/\lambda$ gives the expressions presented in the main text.
Bounds for both absorption and scattering contain $\nu^*$, which is determined by \eqref{det_nu}. For absorption, $\nu_0 = 1$, and $\nu_1$ is computationally evaluated by solving the following equation:
\begin{equation}
\sum_i |e_i|^2 \frac{(\nu_1-2)\Im{\xi} + \nu_1\rho_i}{[(\nu_1-1)\Im{\xi}+\nu_1\rho_i]^2} = 0.
\label{eq:nu1}
\end{equation}
For scattering bound, $\nu_0=\rho_{\rm max}/(\rho_{\rm max}+\Im{\xi})$, where $\rho_{\rm max}$ is the largest $\rho_i$. The other potential optimum, $\nu_1$, is solved computationally through equation:
\begin{equation}
\sum_i |e_i|^2\rho_i \frac{(\nu_1-2)\rho_i + \nu_1\Im{\xi}}{[(\nu_1-1)\rho_i+\nu_1\Im{\xi}]^2} = 0.
\label{eq:Sca_nu1}
\end{equation}

Bounds on maximal cross sections for a finite-size scatterer is obtained by normalizing \eqrefrange{pw_ext}{pw_sca} by plane wave intensity $|E_0|^2/2$ (the vacuum resistance $Z_0=1$):
\begin{align}
\sigma_{\rm ext} &\leq \frac{\lambda^2}{4\pi^2|E_0|^2}\sum_i |e_i|^2\frac{\rho_i}{\Im{\xi} + \rho_i}  \\
\sigma_{\rm abs} &\leq \frac{\lambda^2}{4\pi^2|E_0|^2}\frac{\nu^{*2}}{4}\sum_i |e_i|^2\frac{\rho_i}{(\nu^*-1)\Im{\xi}+\nu^*\rho_i}  \\
\sigma_{\rm scat} &\leq \frac{\lambda^2}{4\pi^2|E_0|^2}\frac{\nu^{*2}}{4}\sum_i |e_i|^2\frac{\rho_i}{\nu^*\Im{\xi}+(\nu^*-1)\rho_i}. 
\end{align}

For a plane wave incidence with $|e_i|^2=\pi(2n+1)\delta_{m,\pm 1}|E_0|^2$, we can simplify the above expression by summing over index $m$ within $i=\{m,n,j\}$, leaving contributions indexed only by total angular momentum $n$ and polarization state $j$:
\begin{align}
\sigma_{\rm ext} &\leq \frac{\lambda^2}{2\pi}\sum_{n,j} (2n+1)\frac{\rho_{n,1,j}}{\Im{\xi} + \rho_{n,1,j}}  \label{eq:sigma_ext_n}\\
\sigma_{\rm abs} &\leq \frac{\lambda^2}{2\pi}\frac{\nu^{*2}}{4}\sum_{n,j} (2n+1)\frac{\rho_{n,1,j}}{(\nu^*-1)\Im{\xi}+\nu^*\rho_{n,1,j}}  \\
\sigma_{\rm scat} &\leq \frac{\lambda^2}{2\pi}\frac{\nu^{*2}}{4}\sum_{n,j} (2n+1)\frac{\rho_{n,1,j}}{\nu^*\Im{\xi}+(\nu^*-1)\rho_{n,1,j}}. 
\end{align}
In Fig. 2(c) of the main text, we use the notation $\sigma_{\rm ext,n}$ to denote the contribution from the n-th channel in the summation of \eqref{sigma_ext_n}.

\section{General bound for extended scatterers}
We assume the material is isotropic, nonmagnetic, and homogeneous so the extended scatterer only has electric response to the incident field. The most general far-field incidence has the expansion in its electric field:
\begin{equation}
\ev_{\rm inc} = \frac{1}{k^{3/2}} \sum_i \int_{k_\parallel\leq k} e_i(\kv_\parallel) \vv_i(\kv_\parallel)\frac{\text{d}\kv_\parallel}{(2\pi)^2},
\label{eq:cont_inc}
\end{equation}
where index $i=\{s,p\}$. Plugging the expansion of $\ev_{\rm inc}$ in Eq. (\ref{eq:Pext_bound}, \ref{eq:Pabs_bound}, \ref{eq:Psca_bound}) gives the integral form of cross-sections bounds after normalization by the z-directed plane wave intensity $|E_0|^2k_z/2k$:
\begin{align}
\sigma_{\rm ext} &\leq \frac{1}{kk_z|E_0|^2}\sum_i\int_{k_\parallel \leq k} |e_i(\kvp)|^2\frac{\rho_i(\kvp)}{\Im\xi + \rho_i(\kvp)}\frac{\text{d}\kv_\parallel}{(2\pi)^2} \label{eq:ext-cont} \\
\sigma_{\rm abs}\leq& \frac{1}{kk_z|E_0|^2}\frac{\nu^{*2}}{4} \sum_i\int_{k_\parallel \leq k} |e_i(\kvp)|^2\frac{\rho_i(\kvp)}{(\nu^*-1)\Im\xi + \nu^*\rho_i(\kvp)}\frac{\text{d}\kv_\parallel}{(2\pi)^2} \label{eq:abs-cont} \\
\sigma_{\rm scat} \leq& \frac{1}{kk_z|E_0|^2}\frac{\nu^{*2}}{4} \sum_i\int_{k_\parallel \leq k} |e_i(\kvp)|^2\frac{\rho_i(\kvp)}{\nu^*\Im\xi + (\nu^*-1)\rho_i(\kvp)}\frac{\text{d}\kv_\parallel}{(2\pi)^2}. \label{eq:sca-cont}
\end{align}

Now we restrict our scope to a plane wave incidence with total wave vector $\vect{k}=k_x\vect{\hat{e}}_x+k_y\vect{\hat{e}}_y+k_z\vect{\hat{e}}_z$  and polarization $p'$. We denote its parallel wave vector as $\kvp'=k_x\vect{\hat{e}}_x+k_y\vect{\hat{e}}_y$ where the $'$ symbol differentiates $\kvp'$ from $\kvp$ that is used to label different channels in \eqref{cont_inc}. The plane wave has the expression:
\begin{equation}
\ev_{\rm inc} = E_0\hat{\ev}e^{i\kvp'\cdot\vect{r}_\parallel}e^{ik_zz},
\label{eq:pw_in_pw}
\end{equation}
where $\hat{\ev}$ is a unit vector denotes incident polarization, taking the form $(k_y\vect{\hat{e}}_x-k_x\vect{\hat{e}}_y)/k_\parallel'$ for $p'=M$, and $(-k_z\vect{\hat{k}'_\parallel}+k_\parallel'\vect{\hat{e}}_z)/k$ for $p'=N$. Equating \eqref{pw_in_pw} with \eqref{cont_inc} gives the expansion coefficients, $e_i(\kvp)$. 
Plugging its absolute value $|e_i(\kvp)| = |E_0|\sqrt{2k_zk}(2\pi)^2\delta(\kvp'-\kvp)\delta_{p,p'}$ into \eqrefrange{ext-cont}{sca-cont} gives bounds for plane wave incidence:
\begin{align}
\sigma_{\rm ext} / A & \leq 2\sum_{s=\pm} \frac{\rho_{s,p'}(\kvp')}{\Im\xi+\rho_{s,p'}(\kvp')} \\
\sigma_{\rm abs} / A &\leq \frac{\nu^{*2}}{2}\sum_{s=\pm}\frac{\rho_{s, p'}(\kvp')}{(\nu^*-1)\Im{\xi}+\nu^*\rho_{s, p'}(\kvp')} \label{eq:abs-cha-extend} \\
\sigma_{\rm scat} / A &\leq \frac{\nu^{*2}}{2}\sum_{s=\pm}\frac{\rho_{s, p'}(\kvp')}{\nu^*\Im{\xi}+(\nu^*-1)\rho_{s, p'}(\kvp')}, \label{eq:sca-cha-extend}
\end{align}
where we identified factor $A=(2\pi)^2\delta^2(0)$ corresponding to total surface area.

Bounds for both absorption and scattering contain $\nu^*$, which is determined by \eqref{det_nu}. For absorption, $\nu_0 = 1$, and $\nu_1$ is computationally evaluated by solving the following equation:
\begin{equation}
\sum_{s=\pm} \frac{(\nu_1-2)\Im\xi + \nu_1\rho_{s,p'}(\kvp')}{\left[(\nu_1-1)\Im\xi+\nu_1\rho_{s,p'}(\kvp')\right]^2}=0.
\label{eq:ext-det-nu}
\end{equation}	
For scattering, $\nu_0=\rho_{\rm max}/(\rho_{\rm max}+\Im{\xi})$, where $\rho_{\rm max}$ is the largest $\rho_i$. The other potential optimum, $\nu_1$, is solved computationally through equation:
\begin{equation}
\sum_{s=\pm} \rho_{s,p'}(\kvp') \frac{(\nu_1-2)\rho_{s,p'}(\kvp') + \nu_1\Im\xi}{\left[(\nu_1-1)\rho_{s,p'}(\kvp')+\nu_1\Im\xi\right]^2}=0.
\label{eq:ext-det-nu-sca}
\end{equation}	

\begin{figure}[t!]
	\includegraphics[width=0.45\textwidth]{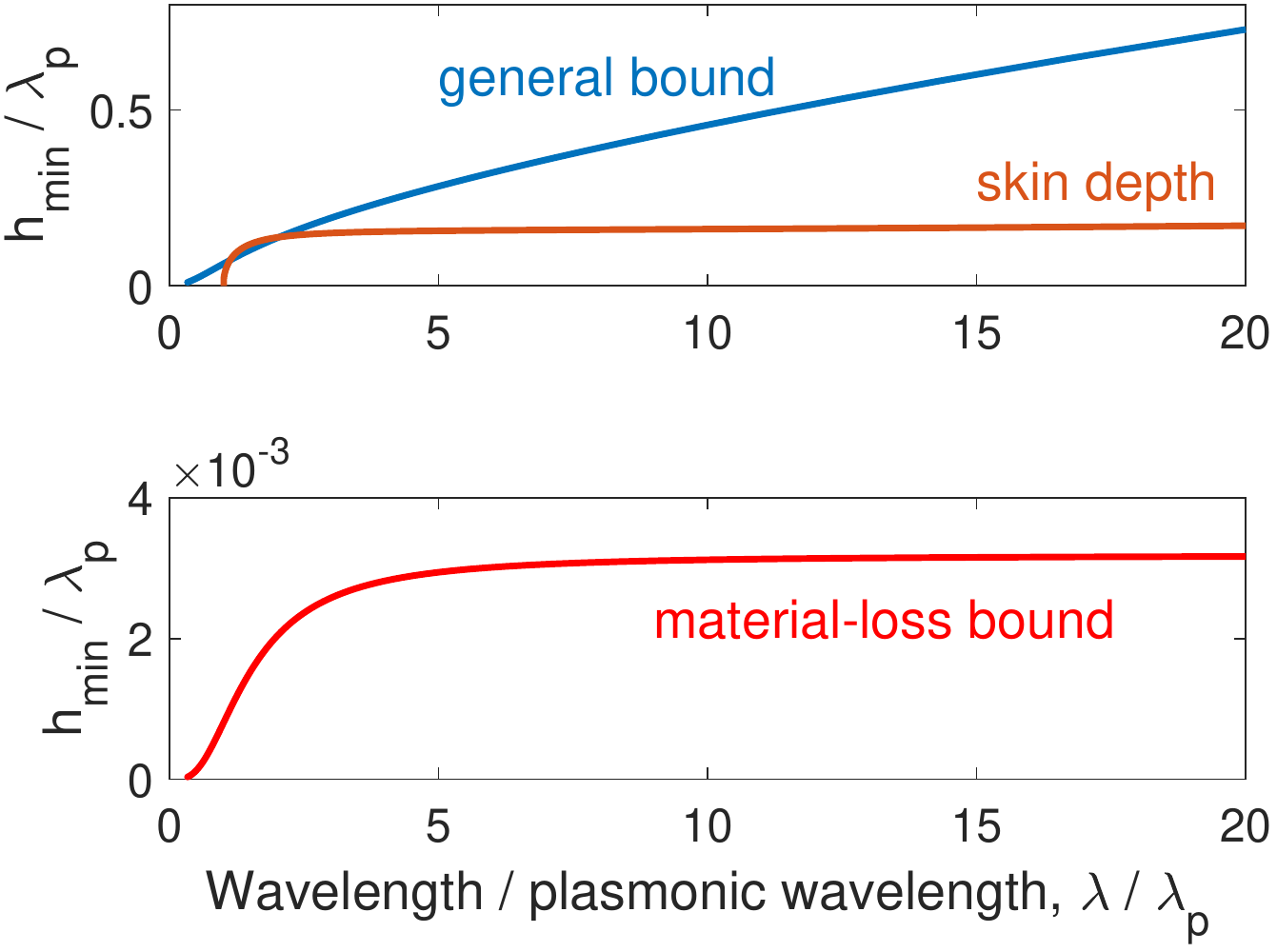}
	\centering
	\caption{Comparison between skin depth and minimum thickness, $h_{\rm min}$, for a perfect absorber as a function of wavelength $\lambda$ for a Drude metal. General bound and material-loss bound are shown seperately in two plots. The former one gives a much more modest prediction.
		All quantities are normalized by the plasmonic wavelength, $\lambda_p$, of the Drude metal.}
	\label{fig:figure7}
\end{figure}

\section{Minimum thickness for perfect absorbers}
Following Section IV in the SM, this section studies minimum thickness required for a perfect absorber that has 100\% absorption. Usually, one determines the optimal $\nu^*$ in \eqref{abs-cha-extend} by comparing the values of $\nu_0$ and $\nu_1$. Here, we take an alternative approach introduced through \eqref{det-nu-alt}, where the derivative of the dual function at $\nu_0$ is used as a threshold, giving a explicit expression for maximum absorption cross section:
\begin{align}
&\sigma_{\rm abs} / A  \leq \begin{cases}
\frac{\nu_1^2}{2}\sum_{s=\pm}\frac{\rho_{s, p'}(\kvp')}{(\nu_1-1)\Im{\xi}+\nu_1\rho_{s, p'}(\kvp')}, &\frac{1}{\Im{\xi}} > \frac{1}{2}\left[\frac{1}{\rho_{+,p'}(\vect{k}_\parallel)} + \frac{1}{\rho_{-,p'}(\vect{k}_\parallel)}\right] \\
1, &  \text{otherwise.}
\end{cases} 
\label{eq:film_abs}
\end{align}
Threshold for maximum absorption (at a given incident angle) corresponds to the condition:
\begin{equation}
\frac{1}{\Im{\xi}} = \frac{1}{2}\left[\frac{1}{\rho_{+,p'}(\vect{k}_\parallel)} + \frac{1}{\rho_{-,p'}(\vect{k}_\parallel)}\right],
\label{eq:threshold_pa}
\end{equation}
where we can solve for its required minimum thickness:
\begin{equation}
h_{\rm min} = \frac{k_z}{k^2}\frac{4\Im{\xi}}{1-\text{sinc}^2(k_zh_{\rm min})}.
\end{equation}
Under normal incidence ($k_z=k$), when the absorber is much thinner than the wavelength, $kh_{\rm min}\rightarrow 0$, it can be shown that:
\begin{equation}
kh_{\rm min} = (24\Im{\xi})^{1/3}.
\end{equation}
This prodicts a much more modest improvement over reduced material loss, compared with previous material-loss bound~\cite{miller_fundamental_2016} where the expression for minimum thickness under normal incidence is $kh_{\rm min} = \Im \xi$. \figref{figure7} shows that this contrast is on the order of $10^2$ for a Drude metal modeled by permittivity
\begin{equation}
\varepsilon(\omega) = - \frac{\omega_p^2}{\omega^2+i\gamma\omega},
\end{equation}
with loss rate $\gamma = 0.02\omega_p$. Plasmonic wavelength is  $\lambda_p = 2\pi c/\omega_p$, with $c=1$ being the speed of light in our unitless convention. 

It is also shown in \figref{figure7} that, minimum thickness predicted by the general bound is on the same length scale as skin depth in the metal~\cite{novotny2012principles} near plasmonic wavelength, $\lambda_p$. For $\lambda<\lambda_p$, there is no surface plasmonic mode inside a Drude metal and skin depth is ill-defined, though it is still possible to realize a perfect absorber according to the general bound. For $\lambda>\lambda_p$, while both skin depth and material-loss bound reach a plateau at large wavelength limit, general bound has $h_{\text{min}}$ increases proportionally to wavelength. This comes from the effectively thinner material under large wavelength incidence and explains the behavior of Al in Fig. 4(a) of the main text.

\section{Inverse design procedures for perfect absorbers}
In Fig. 3(b,c) and Fig. 4(c) of the paper, we showed examples of maximum absorption of topology-optimized metasurfaces with subwavelength periodicity, which are generally within $70\%$ of the bounds, and therefore confirming our bounds to be tight or nearly so. Here we present the details of the topology optimization procedures. Given the permittivity of the material $\epsm$, using a material density function $\alpha_i$, with the subscript $i$ standing for its spatial coordinate, $\alpha_i=1$ meaning material and $\alpha_i=0$ meaning air at pixel $i$, then the design problem of perfect absorbers is formulated as a maximization of the absorption cross section over all permissible choice of $\alpha_i$ at each pixel $i$:
\begin{equation}
\begin{aligned}
& \underset{\varepsilon_i}{\text{maximize}}
& &\sabs \\
& \text{subject to}
& &\varepsilon_i = 1 + \alpha_i(\epsm-1),\\
& & &\alpha_i \in [0,1],
\end{aligned}
\end{equation}
and the absorption cross section as function of the field $\psi$ is given by $\sabs(\psi)=\frac{1}{A}\int_{V}\frac{\omega}{2}\psi^{\dagger}\frac{\Im\chi}{|\chi|^2}\psi$, where $A$ is the unit cell area. The Maxwell constraint, i.e. that all solutions satisfy Maxwell equations is implied. 

Global optimization methods tend not to provide reasonable convergences with such large dimensionality of the problem. Hence local optimizations with random initial starting points were tested to approach the global bounds. Fast calculations of the gradients $\partial \sabs / \partial \alpha_i$ are facilitated with the adjoint method~\cite{miller2013}. Following the volume-integral formalism, one can take the variation of any generic figure of merit $f(\psi)$ due to changes in the susceptibility $\dchi$:
\begin{equation}
\begin{aligned}
\delta f & = 2\Re\int_{V}\left(\delta \psi\right)^{\rm T}\frac{\partial f}{\partial \psi}.
\end{aligned}
\end{equation}
Considering that the perturbed field:
\begin{equation}
\begin{aligned}
\delta \psi(x) = \int_{V}\Gamma_0(x,x')\dchi(x')\psi(x').
\end{aligned}
\end{equation}
The total variation can be written as:
\begin{equation}
\begin{aligned}
\delta f & = 2\Re\int_{V}\int_{V}\psi(x')^{\rm T}\dchi(x')\Gamma_0^{\rm T}(x,x')\frac{\partial f}{\partial \psi(x)}.
\end{aligned}
\end{equation}
Using reciprocity relations, $\Gamma_0^{\rm T}(x,x')=Q^{\rm T}\Gamma_0(x',x)Q$, where $Q=\begin{pmatrix}
1 & 0 \\
0 & -1 
\end{pmatrix}$ is the parity operator. Then by rearranging, the variation in the figure of merit is given by
\begin{equation}
\begin{aligned}
\delta f & = 2\Re\int_{V}\psi(x')^{\rm T}\dchi(x')Q\int_{V}\Gamma(x',x)Q\frac{\partial f}{\partial \psi(x)}.
\end{aligned}
\end{equation}
Now one can define the adjoint field $\psiadj(x')=\int_{V}\Gamma(x',x)Q\frac{\partial f}{\partial \psi(x)}$, which is essentially fields resulting from the current sources, the so-called adjoint sources, $\phi_{\rm adj}=Q\frac{\partial f}{\partial \psi}$.
In the case of absorption cross-section $\sabs$, the adjoint sources are given by
\begin{equation}
\phi_{\rm adj}=\frac{1}{A}\frac{\omega}{2}\frac{\Im\chi}{|\chi|^2}Q\psi^{\star},
\end{equation}
and so the variation in $\sabs$ is
\begin{equation}
\begin{aligned}
\delta \sabs = 2\Re\int_{V}\psi(x')^{\rm T}\dchi(x')Q\psiadj(x').
\end{aligned}
\end{equation}
Hence the fields $\psi$ from the prescribed structure with direct incidence plus the adjoint fields $\psi_{\rm adj}$ provide the gradients with respect to any number of design variables. Numerically, in each iteration of the topology optimizations, one direct simulation to compute $\psi$ and another simulation with $\phi_{\rm adj}$ as sources to compute $\psi_{\rm adj}$ are need.

The simulations are performed with a finite-difference time-domain~\cite{fdtd} open-source solver~\cite{MEEP}. In all design figures below, periodic conditions are imposed in the horizontal direction, and light is incident from below and propagating upward. For all sets of hyper-parameters, including material permittivities and thicknesses, we test at least 10 initial starting points, and run simulations with resolutions up to 110 grids per wavelength. Almost all optimizations converge within 700 iterations, and we show below the evolution of $\sabs$ in $\SI{1.2}{\micro\meter}$-thick SiC absorber optimization.

\begin{figure}[htb]
  \includegraphics[width=0.47\textwidth]{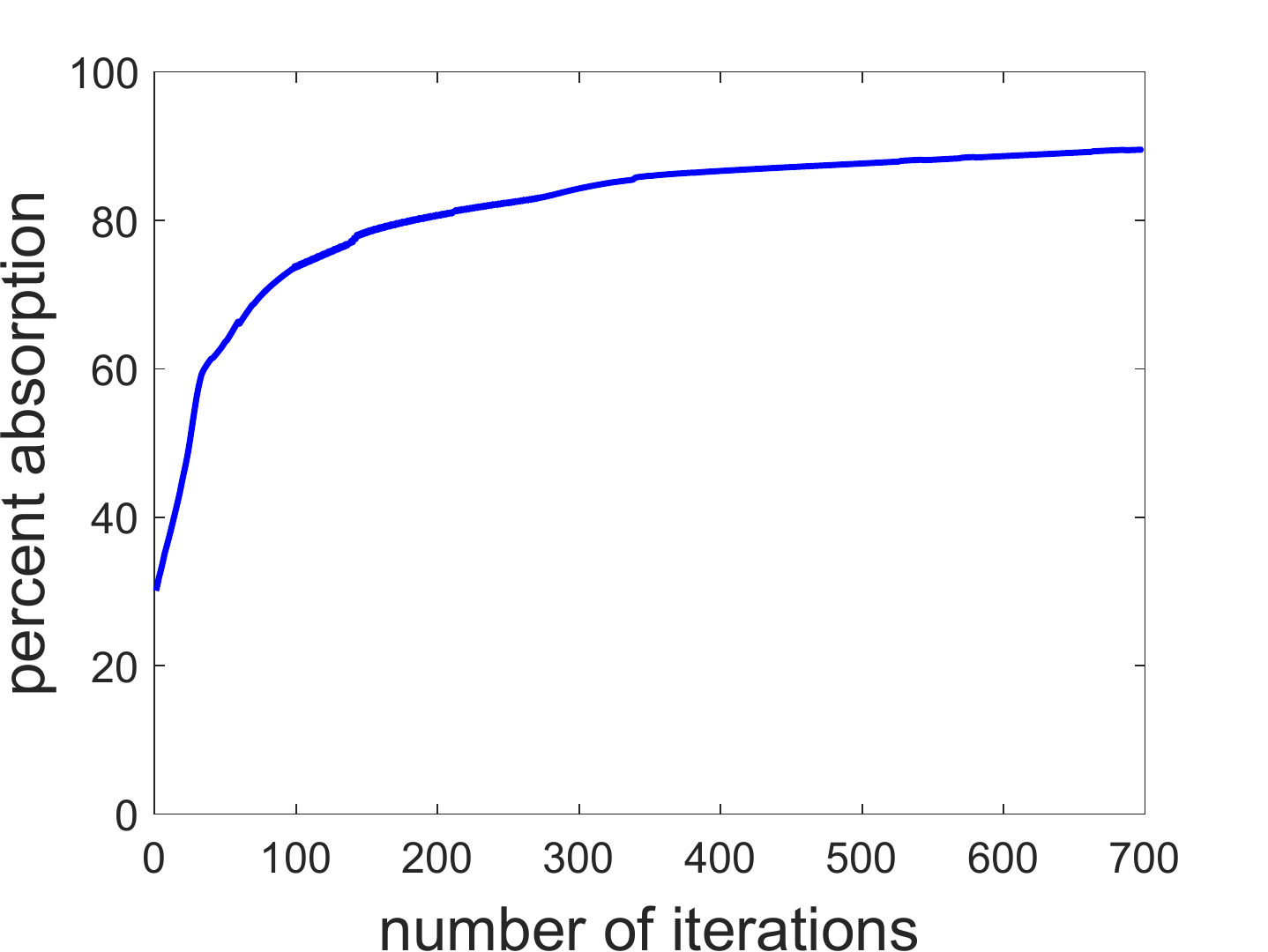}
  \caption{Percent absorption as a function the number of iterations for the best design optimization of $\SI{1.2}{\micro\meter}$-thick SiC absorber.}
\end{figure}

\section{Optimized designs}
\subsection{Different thicknesses of SiC absorbers at$\SI{11}{\micro\meter}$ wavelength}
As an example, we investigated absorber inverse designs with SiC at $\SI{11}{\micro\meter}$ wavelength and a range of thicknesses. The resolution is $\SI{0.1}{\micro\meter}$, and unit cell period is $\SI{1.1}{\micro\meter}$. Their percent absorption and designs are presented in Table \ref{tab:id1}.

\subsection{Minimum thicknesses of 70\% absorbers for different materials}
Thinnest perfect absorbers are designed for different types of materials, such as metals, doped semiconductors and polar dielectrics. In Table \ref{tab:id2}, we demonstrate designs of six representative materials at different wavelengths where 70\% absorption is achieved with minimum thicknesses of the metasurfaces.

\section{Deriving previous bounds from general bound formalism}
Different derivations of upper bounds can be formulated as optimization problems with same objective functions but different constraints. In this section, we showed that how the general bound, developed in this paper, can incoorperate previous bounds by either relaxing the energy equality constraint, or taking the result of the general bound in certain limit.

\begin{table*}[t]
	\centering
	\begin{tabular}{|l|c|c|c|}
		\hline
		& Extinction & Absorption & Scattering   \\ \hline
		General	bound & \makecell{max. $P_{\rm ext}$ \\ s.t. $P_{\rm scat}+P_{\rm abs}=P_{\rm ext}$}   & \makecell{max. $P_{\rm abs}$ \\ s.t. $P_{\rm scat}+P_{\rm abs}=P_{\rm ext}$} & \makecell{max. $P_{\rm scat}$ \\ s.t. $P_{\rm scat}+P_{\rm abs}=P_{\rm ext}$}   \\ \hline
		Material bound~\cite{miller_fundamental_2016} & \makecell{max. $P_{\rm ext}$ \\ s.t. $P_{\rm abs}\leq P_{\rm ext}$}   & \makecell{max. $P_{\rm abs}$ \\ s.t. $P_{\rm abs}\leq P_{\rm ext}$} & \makecell{max. $P_{\rm ext}-P_{\rm abs}$}    \\ \hline
		Channel	bound~\cite{hugonin_fundamental_2015} & \makecell{max. $P_{\rm ext}$ \\ s.t. $P_{\rm scat}\leq P_{\rm ext}$}   & \makecell{max. $P_{\rm ext}-P_{\rm scat}$}  & \makecell{max. $P_{\rm scat}$ \\ s.t. $P_{\rm scat}\leq P_{\rm ext}$}   \\ \hline
	\end{tabular}
	\caption{Upper bounds as optimization problems with the same objective function but different constraints. General bound uses an equality energy constraint, while the other two relax it to inequality constraints (or even unconstrained), resulting looser bounds.}
	\label{tab:opt-for}
\end{table*}

\begin{table*}[t]
	\centering
	\makebox[\textwidth]{%
		\begin{tabular}{|l|c|c|c|}
			\hline
			& Extinction & Absorption & Scattering   \\ \hline
			General	bound & $\psiinc^\dagger \left(\Im\xi+\ImGO\right)^{-1} \psiinc$  & $\frac{\nu^{*2}}{4} \psiinc^\dagger[(\nu^*-1)\Im\xi+\nu^*\ImGO]^{-1} \psiinc$ & $\frac{\nu^{*2}}{4} \psiinc^\dagger[(\nu^*-1)\Im\xi + \nu^*\ImGO]^{-1} \psiinc$ \\ \hline
			Material bound~\cite{miller_fundamental_2016} &  $\psiinc^\dagger (\Im\xi)^{-1} \psiinc$  & $\psiinc^\dagger (\Im\xi)^{-1} \psiinc$ & $\frac{1}{4}$ $\psiinc^\dagger (\Im\xi)^{-1} \psiinc$    \\ \hline
			Channel	bound~\cite{hugonin_fundamental_2015} & $\psiinc^\dagger (\ImGO)^{-1} \psiinc$  & $\frac{1}{4}\psiinc^\dagger (\ImGO)^{-1} \psiinc$ & $\psiinc^\dagger (\ImGO)^{-1} \psiinc$    \\ \hline
	\end{tabular}}
	\caption{Optimum of different upper-bound formulations presented in Table \ref{tab:opt-for}. Optimal dual variable of the general bound is determined by \eqref{det_nu}.}
	\label{tab:opt-sol}
\end{table*}

\paragraph{channel and material loss bounds}
Table \ref{tab:opt-for} compares general bound with material-loss bound (material bound) and channel bound.
General bound purposed in this paper utilizes the \textit{equality} energy conservation constraint: $P_{\rm scat}+P_{\rm abs}=P_{\rm ext}$. Throwing away either $P_{\rm scat}$ or $P_{\rm abs}$ gives the \textit{inequality} energy conservation constraint used in previous material bound~\cite{miller_fundamental_2016} or channel bound~\cite{hugonin_fundamental_2015}. In both formalisms, the disregarded term itself is treated by an unconstrained optimization.

All optimization problems in Table \ref{tab:opt-for} have strong duality, thus their optimums can be analytically determined by the optimal of their dual functions, given in Table \ref{tab:opt-sol} (with prefactor $\omega/2$ suppressed in every expression). Results for material bound appears in~\cite{miller_fundamental_2016}. Results for channel bound appears in~\cite{hugonin_fundamental_2015}. Moreover, expanding channel bound into VSWs for a spherical scatterer gives the expressions in~\cite{hamam_coupled-mode_2007,kwon2009optimal,ruan2011design,liberal2014least} (after adding back prefactor $\omega/2$): $P^{\rm max}_{\rm scat}=P^{\rm max}_{\rm ext}=4P^{\rm max}_{\rm abs}=\frac{|E_0|^2}{k^2}\sum_{n=1}^{+\infty}\pi(2n+1)$, where $E_0$ is the plane wave amplitude, $k$ is the amplitude of the wave vector, and $n$ is total angular momentum.

\paragraph{$\mathbb{T}$-operator bound}
As discussed in Section \ref{sec:thermal} in the SM, our bound is tighter than $\mathbb{T}$-operator bound~\cite{molesky2019t} for maximum absorption from a thermal incident field. Though using different approaches, the general bound can reproduce the same result as in $\mathbb{T}$-operator bound by relaxing the energy constraint to $\Pabs\leq\Pext$ and replacing objective function $\Pabs$ with $\Pext-\Pscat$:
\begin{equation}
\begin{aligned}
& \text{maximize} & \Pext-\Pscat \\
& \text{subject to} & \Pabs\leq\Pext.
\end{aligned}
\label{eq:dual-T}
\end{equation}
Similar to Section I in the SM, we solve \eqref{dual-T} by its dual function:
\begin{align}
g(\lambda)=\frac{(\lambda+1)^2}{4}\psiinc^\dagger\left[\ImGO+\lambda\Im\xi\right]^{-1}\psiinc,
\label{eq:dual-func-T}
\end{align}
where $\lambda$ is the notation used in~\cite{boyd2004convex} to denote dual variable for an inequality constraint. The range for $\lambda$ is $[0,+\infty)$. When $\lambda=0$, the inverse operator in \eqref{dual-func-T} is ill-defined and we replace it with pseudo inverse if $\psiinc\in\text{Range}\{\ImGO\}$, otherwise $g(0)\rightarrow -\infty$ 

Following assumptions made in T-operator bound, we assume far-field thermal incidence and nonmagnetic material, where $\psiinc$ and $\Gamma_0$ is replaced by $\ev_{\rm inc}$ and $\GG_0$. As discussed in Section IV in the main text, thermal incident field can be expanded by a set of uncorrelated orthogonal fields. We choose it to be $\{\vv_i\}$, the eigenvectors of $\Im\GG_0$, with expansion coefficients given by $|e_i|^2=\frac{4}{\pi\omega}\Theta(T)$ and $\Theta(T)=\hbar\omega/(e^{\hbar\omega/k_BT}-1)$ is the Planck energy of a harmonic oscillator at temperature $T$.

Maximizing $g(\lambda)$ gives the expression for optimal absorption of thermal incident fields in \cite{molesky2019t}:
\begin{equation}
P_\text{abs} \leq \frac{2}{\pi}\Theta(T) \sum_i
\begin{cases}
\frac{\rho_i}{\Im{\xi}} & \text{for } 2\rho_i \leq \Im{\xi} \\  
\frac{1}{4} & \text{for } 2\rho_i \geq \Im{\xi},
\end{cases}
\label{eq:thermal_bound}
\end{equation} 
where two cases correspond to optimal dual variable taking the value of either $\nu_1 \in (0,+\infty)$ or $\nu_0=0$. Such a bound is looser than the general bound presented in Section \ref{sec:thermal} of the SM, as a result of its inequality energy constraint in \eqref{dual-T}, rather than the equality energy constraint.

\paragraph{patterned thin film bound}
It is predicted that within a vacuum background, a patterned thin film with thickness much smaller than the incident wavelength has a maximum absorption of 50\%~\cite{thongrattanasiri_complete_2012}. To validate this, we take the limit $k_zh\rightarrow 0$ in \eqref{film_abs} and obtain:
\begin{equation}
\sigma_{\rm abs}/A \leq \frac{2(\Im{\xi})\rho_{+,p'}}{(\Im{\xi}+\rho_{+,p'})^2}.
\label{eq:thin_film_abs}
\end{equation}
Because a thin film only has dipole radiation that is symmetric respect to the $z=0$ plane, only mode with index $s=+$ survived in \eqref{thin_film_abs}. 

When $\Im\xi=\rho_{+,M}$, the absorption rate $\sigma_{\rm abs}/A$ in \eqref{thin_film_abs} reaches its maximum of 50\%, agreeing with the prediction made in~\cite{thongrattanasiri_complete_2012}. The advantage of our formalism is that we can also predict the minimum thickness for the patterned thin film to reach $50\%$ absorption: $h_{\rm min} = 2\Im\xi/k$, as solved from the optimal condition $\Im\xi=\rho_{+,M}$.


\section{Underestimation of the channel bounds from cutoff channels}
The channel bounds shown in Table \ref{tab:opt-sol} are in fact infinite for a plane wave incident. Physically, this is due to the negligible radiative loss in high-order VSW channels, corresponding to the eigenvectors of $\ImGO$ with near-zero eigenvalues. To regularize such divergence, one needs to truncate its radiation channels to a finite number based on certain threshold. Such an empirical truncation is certainly a disadvantage of the channel bound, moreover, as we will show below, it also introduces unwanted underestimation of the channel bound itself.

\begin{figure}[t!]
	\includegraphics[width=0.5\textwidth]{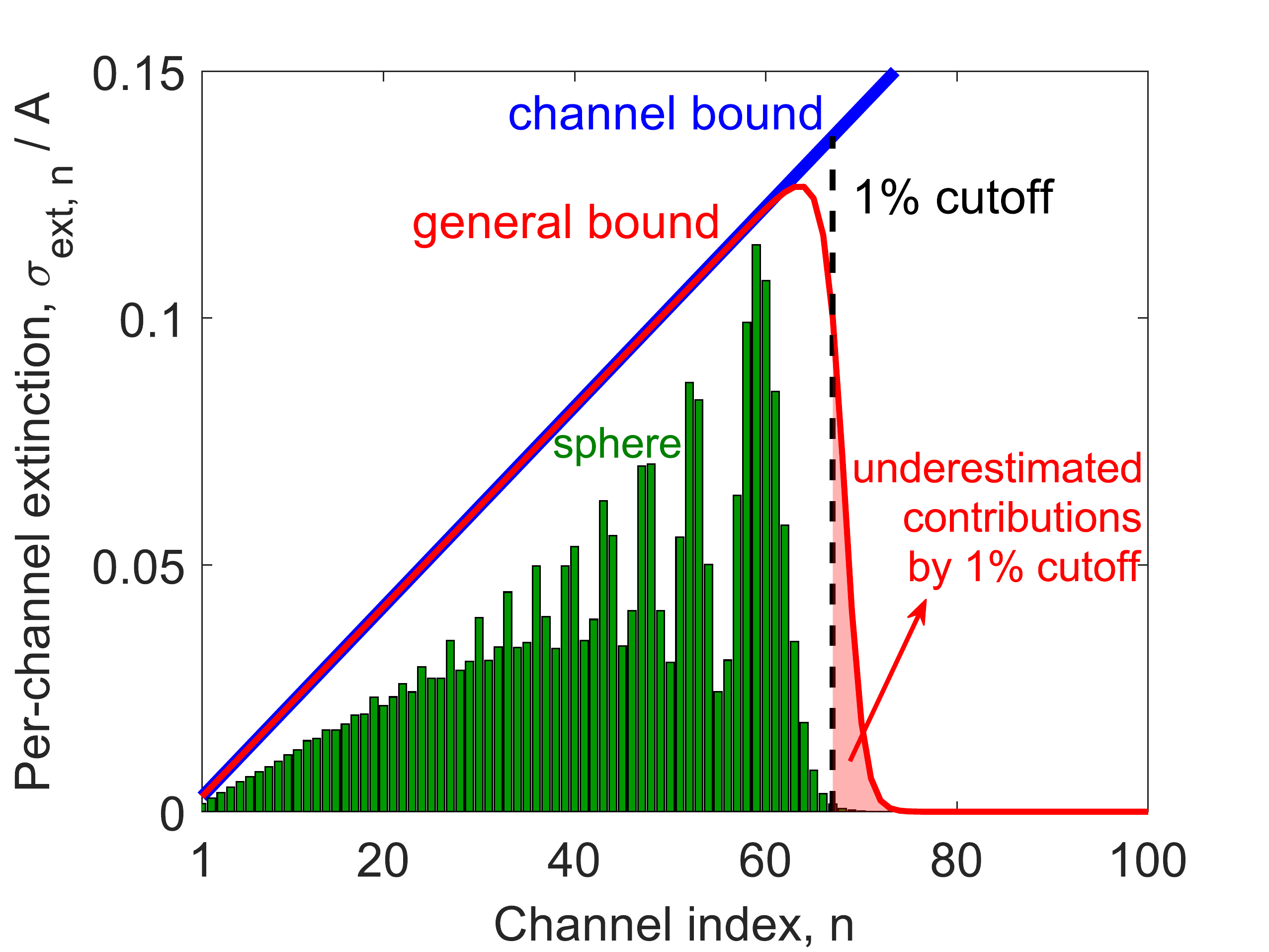}
	\centering
	\caption{Per-channel extinction $\sigma_{\text{ext},n}$ for material Ag in a spherical bounding volume with radius $R = 10\lambda$ at wavelength $\lambda = 360$ nm. Compared to the general bound, the 1\% cutoff threshold excludes channels whose potential contributions are marked by the red shaded region, resulting in an underestimated channel bound.}
	\label{fig:SM-fig3}
\end{figure}

As an example, Fig. \ref{fig:SM-fig3} shows channel bounds for per-channel extinction $\sigma_{\text{ext},n}$ within a bounding volume of radius $R=10\lambda$. The material is Ag and incident wavelength $\lambda=360$ nm. Also shown in the same figure are the general bound and spherical scattering. As expected, the channel bound diverges at high-order radiative channels, and is regularized by a 1\% cutoff line, which excludes channels for which the sphere scattering contributions are less than 1\% of the channel bound. 

Compared with the general bound, we see that the potential contribution of those excluded channels (red shaded region), are ignored by the 1\% threshold. Such an underestimation results in a seemly tighter bound in Fig. 2 of the paper at large radius limit. Of course, the 1\% threshold is empirical. One could reduce the threshold to eliminate the unwanted underestimation, but that usually results in an overall overestimation of the channel bound since more channels are now included without the inhibition of material loss. We found 1\% is a good empirical threshold for  estimating the channel bound.

\section{The imaginary part of the Green's function operator for a sphere}
The expressions of $\Im\GG_0$ is given in~\cite{tsang2004scattering}, whose imaginary part is Hermitian and can be decomposed as:
\begin{equation}
\Im{\GG_0}=\frac{1}{2i}(\GG_0 - \GGd_0) = \sum_{n,m,j} \vect{v}_{n,m,j}\vect{v}_{n,m,j}^\dagger,
\end{equation}
where $n=1,2, ...$, $m=-n,...,n$, $j=1,2$ represents two polarizations. $\vect{v}_{n,m,j}$ are reguarized VSWs whose definition can be found in~\cite{tsang2004scattering}: 
\begin{align}
\vect{v}_{n,m,1}(\vect{x}) &= k^{\frac{3}{2}} Rg \vect{M}_{n,m}(kr,\theta,\phi) \\
\vect{v}_{n,m,2}(\vect{x}) &= k^{\frac{3}{2}} Rg \vect{N}_{n,m}(kr,\theta,\phi).
\end{align}
The inner product of $\vect{v}_{n,m,j}$ with itself gives the eigenvalue of $\Im{\GG_0}$:
\begin{align}
\rho_{n,m,j} &= \vect{v}_{n,m,j}^\dagger \vect{v}_{n,m,j} = \int_V \vect{v}_{n,m,j}^*(\vect{x})\cdot\vect{v}_{n,m,j}(\vect{x}) \text{d}V. 
\end{align}
Integrating over angular coordinates gives the expression:
\begin{align}
\rho_{n,m,1} &= \int_0^{kR} x^2j_n^2(x)\text{d}x \\
\rho_{n,m,2} &= n(n+1) \int_0^{kR} j_n^2(x)\text{d}x + \int_0^{kR}[xj_n(x)]^{\prime 2}\text{d}x,
\end{align}
which can be computationally evaluated or even reduced to simpler analytical forms~\cite{bloomfield2017indefinite}.

\section{The imaginary part of the Green's function operator for a film}
As in~\cite{tsang2004scattering,kruger2012trace}, $\Im{\GG_0}$ in Cartesian coordinate can be decomposed into a complete set of plane waves:
\begin{equation}
\Im{\GG_0} = \sum_{s,p} \int_{k_{\parallel}\leq k} \vect{v}_{s,p}(\vect{k_{\parallel}})\vect{v}^\dagger_{s,p}(\vect{k_{\parallel}})\frac{\text{d}\vect{k_{\parallel}}}{(2\pi)^2}.
\end{equation}
Index $s=\{-1,+1\}$ represents odd and even parity, index $p=M,N$ represents different polarization, $\vect{k_{\parallel}}$ are in-plane wave vector whose integration only runs through propergating modes. 
Real-space expressions of $\vect{v}_{s,p}(\vect{k}_{\parallel})$ are:
\begin{align}
\vect{v}_{+, M}(\vect{k_{\parallel}},\vect{x}) &= ik\frac{e^{i\vect{k}_\parallel\cdot\vect{r}_\parallel}}{\sqrt{2k_z}k_\parallel}(k_y\vect{\hat{e}}_x-k_x\vect{\hat{e}}_y)\cos(k_zz)  \\
\vect{v}_{-, M}(\vect{k_{\parallel}},\vect{x}) &= -ik\frac{e^{i\vect{k}_\parallel\cdot\vect{r}_\parallel}}{\sqrt{2k_z}k_\parallel}(k_y\vect{\hat{e}}_x-k_x\vect{\hat{e}}_y)\sin(k_zz)\\
\vect{v}_{+, N}(\vect{k_{\parallel}},\vect{x}) &= \frac{e^{i\vect{k}_\parallel\cdot\vect{r}_\parallel}}{\sqrt{2k_z}}\left[k_\parallel\cos(k_zz)\vect{\hat{z}} - ik_z\sin(k_zz)\vect{\hat{k}_\parallel}\right]\\
\vect{v}_{-, N}(\vect{k_{\parallel}},\vect{x}) &= \frac{e^{i\vect{k}_\parallel\cdot\vect{r}_\parallel}}{\sqrt{2k_z}}\left[k_\parallel\sin(k_zz)\vect{\hat{z}} + ik_z\cos(k_zz)\vect{\hat{k}_\parallel}\right].
\end{align}

Inner products of $\vect{v}_{s, p}(\vect{k}_{\parallel})$ in a thin film (thickness $h$, centered at $z=0$) is~\cite{molesky2019t}:
\begin{align}
\vect{v}_{s, p}^\dagger(\vect{k}_{\parallel}) \vect{v}_{s', p'}(\vect{k}_{\parallel}') &= \int_V \vect{v}_{s, p}^*(\vect{k}_{\parallel},\vect{x}) \cdot \vect{v}_{s', p'}(\vect{k}_{\parallel}',\vect{x})\text{d}V \\
&= \rho_{s,p}(\vect{k}_\parallel)(2\pi)^2\delta(\vect{k}_\parallel-\vect{k}_\parallel')\delta_{s,s'}\delta_{p,p'},
\end{align}
where the eigenvalues are:
\begin{align}
\rho_{\pm, M}(\vect{k}_\parallel) &= \frac{k^2h}{4k_z}(1\pm\frac{\sin(k_zh)}{k_zh})\\
\rho_{\pm, N}(\vect{k}_\parallel) &= \frac{k^2h}{4k_z}(1\pm\frac{\sin(k_zh)}{k_zh})\mp\frac{\sin(k_zh)}{2}.
\end{align}

\section{Upper bounds at different generality}
\begin{enumerate}
	\item Most general form (include non-local, magnetic, inhomogeneous materials, any incident field, any geometry of the scatterer):
	\begin{align}
		\Pext &\leq \frac{\omega}{2} \psiinc^\dagger \left(\Im\xi+\ImGO\right)^{-1} \psiinc \label{eq:ext-gen}\\
		\Pabs &\leq \frac{\omega}{2}\frac{\nu^{*2}}{4} \psiinc^\dagger[(\nu^*-1)\Im\xi+\nu^*\ImGO]^{-1} \psiinc \label{eq:abs-gen} \\
		\Pscat &\leq \frac{\omega}{2}\frac{\nu^{*2}}{4} \psiinc^\dagger[\nu^*\Im\xi + (\nu^* - 1)\ImGO]^{-1} \psiinc, \label{eq:sca-gen}
	\end{align}
	where $\Im\xi$ and $\ImGO$ are matrices that depends on the exact shape and material compositions of the scatterer.
	\item Scalar material (electric or magnetic scalar material, any incident field, any geometry of the homogeneous scatterer):
	\begin{align}
	P_{\rm ext} &\leq \frac{\omega}{2}\frac{1}{\Im{\xi}}\Big[\psiinc^\dagger \psiinc - \psiinc^\dagger \VV (\Im{\xi}+\VV^\dagger \VV)^{-1} \VV^\dagger\psiinc\Big] \label{eq:scalar_bound1} \\
	P_{\rm abs} &\leq \frac{\omega}{2}\frac{\nu^{*2}}{4}\frac{\nu^*}{\nu^*-1}\frac{1}{\Im{\xi}}\Big\{ \frac{1}{\nu^*}\psiinc^\dagger\psiinc - \psiinc^\dagger \VV [(\nu^*-1)\Im{\xi}+\nu^*\VV^\dagger \VV]^{-1}\VV^\dagger \psiinc \Big\} \\
	P_{\rm scat} &\leq \frac{\omega}{2}\frac{\nu^{*2}}{4}\frac{\nu^*-1}{\nu^*}\frac{1}{\Im{\xi}}\Big\{ \frac{1}{\nu^*-1}\psiinc^\dagger\psiinc  - \psiinc^\dagger \VV [\nu^*\Im{\xi}+(\nu^*-1)\VV^\dagger \VV]^{-1}\VV^\dagger \psiinc \Big\}, \label{eq:scalar_bound3}
	\end{align}
	where $\Im\xi$ is a scalar represents either the isotropic electric or magnetic susceptibility and we write the eigendecomposition of $\ImGO$ as $\ImGO=\VV\VV^\dagger$.  
	\item Isotropic \emph{electric} material (electric scalar material, any incident field, any geometry): same form as \eqrefrange{scalar_bound1}{scalar_bound3} with $\psi_{\rm inc}$ replaced by $\vect{e}_{\rm inc}$, and $\Gamma_0$ replaced by $\GG_0$. Eigenbasis $\VV$ is now defined by the eigendecomposition: $\Im \GG_0 = \VV \VV^\dagger$, with $\vv_i$ being the i-th column of $\VV$.
	\begin{enumerate}
		\item For far field scattering, where the incident electric field $\vect{e}_{\rm inc}$ is characterized by the property $\vect{e}_{\rm inc} \in \text{Range}\{\Im \GG_0\}$, bounds in \eqrefrange{scalar_bound1}{scalar_bound3} can be dramatically simplified:
		\begin{align}
		P_{\rm ext} &\leq \frac{\omega}{2}\vect{e}_{\rm inc}^\dagger (\Im\xi+\Im \GG_0)^{-1}\vect{e}_{\rm inc} \\
		P_{\rm abs} &\leq \frac{\omega}{2}\frac{\nu^{*2}}{4}\vect{e}_{\rm inc}^\dagger [(\nu^*-1)\Im\xi+\nu^*\Im \GG_0]^{-1}\vect{e}_{\rm inc} \\
		P_{\rm scat} &\leq \frac{\omega}{2}\frac{\nu^{*2}}{4}\vect{e}_{\rm inc}^\dagger [\nu^*\Im\xi+(\nu^*-1)\Im \GG_0]^{-1}\vect{e}_{\rm inc}.
		\end{align}
		\begin{itemize}
			\item Plane wave incidence (applies to both finite and extended scatterers) with $\ev_{\rm inc} = \sum_i e_i\vv_i$. Explicitly written out contributions from different channels:
			\begin{align}
			P_{\rm ext} &\leq \frac{\omega}{2}\sum_i |e_i|^2\frac{\rho_i}{\Im{\xi} + \rho_i} \label{eq:ext-cha-bound} \\
			P_{\rm abs} &\leq \frac{\omega}{2} \frac{\nu^{*2}}{4}\sum_i |e_i|^2\frac{\rho_i}{(\nu^*-1)\Im{\xi}+\nu^*\rho_i} \label{eq:abs-cha-bound} \\
			P_{\rm scat} &\leq \frac{\omega}{2} \frac{\nu^{*2}}{4}\sum_i |e_i|^2\frac{\rho_i}{\nu^*\Im{\xi}+(\nu^*-1)\rho_i}, \label{eq:sca-cha-bound}
			\end{align}
			where $\rho_i = \vv_i^\dagger\vv_i$ is analytically known for highly symmetric bounding volumes.
			
			\item VSW incidence (applies to finite scatterers). Now the incident field is one specific VSW: $\ev_{\rm inc} = e_i \vv_i$, under which:
			\begin{align}
			P_{\rm ext} &\leq \frac{\omega}{2} \frac{|e_i|^2}{\Im\xi + \rho_i} \\
			P_{\rm abs} &\leq \frac{\omega}{2} \frac{\nu^{*2}}{4}\frac{|e_i|^2}{(\nu^*-1)\Im\xi + \nu^*\rho_i} 
			 = \begin{cases}
			\frac{\Im\xi}{(\Im\xi+\rho_i)^2}|e_i|^2 & \text{if\ \ } \rho_i\leq\Im\xi \\
			\frac{1}{4\rho_i}|e_i|^2 & \rm{else} 
			\end{cases} \\
			P_{\rm scat} &\leq \frac{\omega}{2} \frac{\nu^{*2}}{4}\frac{|e_i|^2}{\nu^*\Im\xi + (\nu^*-1)\rho_i} = 
			\begin{cases}
			\frac{\rho_i}{(\Im\xi+\rho_{\rm max})^2}|e_i|^2 & \text{if\ \ } \rho_i\geq\frac{\rho_{\rm max}\Im\xi}{2\Im\xi+\rho_{\rm max}} \\
			\frac{1}{4}\frac{\rho_{\rm max}^2}{\Im\xi(\rho_{\rm max}+\Im\xi)(\rho_{\rm max}-\rho_i)}|e_i|^2 & \rm{else},
			\end{cases}
			\end{align}
			where the choice of $\nu^*$ is simple enough that we can write out explicit two possible solutions of $P_{\rm abs}$ and $P_{\rm sca}$. We denote the maximum in $\{\rho_i\}$ as $\rho_{\rm max}$.
		\end{itemize}
		\item Incident field in near-field scattering is not necessarily in the range of $\Im \GG_0$ as evanescent waves may contribute (for an extended scatter). Expression for its bound takes the most general form as \eqrefrange{scalar_bound1}{scalar_bound3} with $\psi_{\rm inc}$ replaced by $\vect{e}_{\rm inc}$, and $\Gamma_0$ replaced by $\GG_0$.
		For arbitrary dipole sources $\pv$, the incident field can be written as $\ev_{\rm inc}=\GG_{0, \pv\rightarrow V}\pv$ where $\GG_{0, \pv\rightarrow V}$ is an integral Green's function mapped from the region of dipole source $\pv$ to the scatterer $V$. Taking the singular vector decomposition of $\GG_{0, \pv\rightarrow V}=\UU\WW^\dagger$, bounds for near field scattering can be written as:
	\end{enumerate}
\end{enumerate}
	\begin{align}
	P_{\rm ext} &\leq \frac{\omega}{2}\frac{1}{\Im\xi}\pv^\dagger\WW[\UU^\dagger\UU - \UU^\dagger\VV(\Im\xi + \VV^\dagger\VV)^{-1}\VV^\dagger\UU]\WW^\dagger\pv \nonumber \\
	&= \frac{\omega}{2}\frac{1}{\Im\xi} \sum_i |\pv^\dagger\wv_i|^2 \left( \uv_i^\dagger\uv_i - \frac{|\uv_i^\dagger\vv_i|^2}{\Im\xi+\vv_i^\dagger\vv_i} \right) \label{eq:nf_ext_bnd} \\
	P_{\rm abs} &\leq \frac{\omega}{2}\frac{\nu^{*2}}{4(\nu^*-1)}\frac{1}{\Im\xi}\pv^\dagger\WW\{\UU^\dagger\UU  - \UU^\dagger\VV[(\nu^*-1)\Im\xi/\nu^* + \VV^\dagger\VV]^{-1}\VV^\dagger\UU\}\WW^\dagger\pv \nonumber \\
	&= \frac{\omega}{2}\frac{\nu^{*2}}{4(\nu^*-1)}\frac{1}{\Im\xi}  \sum_i |\pv^\dagger\wv_i|^2 \left( \uv_i^\dagger\uv_i - \frac{|\uv_i^\dagger\vv_i|^2}{(\nu^*-1)\Im\xi/\nu^*+\vv_i^\dagger\vv_i} \right) \label{eq:nf_abs_bnd} \\
	P_{\rm scat} &\leq \frac{\omega}{2}\frac{\nu^*}{4}\frac{1}{\Im\xi}\pv^\dagger\WW\{\UU^\dagger\UU - \UU^\dagger\VV[\nu^*\Im\xi/(\nu^*-1) + \VV^\dagger\VV]^{-1}\VV^\dagger\UU\}\WW^\dagger\pv \nonumber \\
	&= \frac{\omega}{2}\frac{\nu^*}{4}\frac{1}{\Im\xi} \sum_i |\pv^\dagger\wv_i|^2 \left( \uv_i^\dagger\uv_i - \frac{|\uv_i^\dagger\vv_i|^2}{\nu^*\Im\xi/(\nu^*-1)+\vv_i^\dagger\vv_i} \right). \label{eq:nf_sca_bnd}
	\end{align}

\section{Bound for local density of states (LDOS)}
\paragraph{General bounds for LDOS}
We start with the expressions of total, non-radiative, radiative electric LDOS in their volume integral form~\cite{miller_fundamental_2016}: 	
\begin{align}
\rho_{\rm tot} &= \rho_0 + \frac{1}{\pi\omega}\sum_j\Im \left(\tilde{\psi}_{\rm inc, j}^\dagger\phi_j\right) \label{eq:tot_LDOS_gen_bound} \\
\rho_{\rm nr} &= \frac{1}{\pi\omega}\sum_j \phi_j^\dagger(\Im\xi)\phi_j \label{eq:nr_LDOS_gen_bound} \\
\rho_{\rm rad} &= \rho_0 + \frac{1}{\pi\omega}\sum_j \left[\Im \left(\tilde{\psi}_{\rm inc, j}^\dagger\phi_j\right) - \phi_j^\dagger(\Im\xi)\phi_j \right], \label{eq:rad_LDOS_gen_bound}
\end{align}
where $\rho_0$ is the electric LDOS of the background material, and takes the value of $\frac{\omega^2}{2\pi^2c^3}$ for a scatterer in the vacuum~\cite{joulain2005surface}.
Summation $j=1,2,3$ denotes power quantities from three orthogonally polarized unit dipoles. Incident field from dipole $j$ is denoted by $\psi_{\rm inc, j}=(\vect{e}_{\rm inc, j}, \vect{h}_{\rm inc, j})$. Here we use lowercase notations for both electric and magnetic fields to emphasis their vector nature, as opposed to capitalized characters that are usually reserved for operators and matrices. Such incident field excites polarization current $\phi_j$ in the scatterer. Complex conjugate of $\psi_{\rm inc, j}$ (with a minus sign in front of magnetic fields) is denoted by $\tilde{\psi}_{\rm inc, j} = (\vect{e}^*_{\rm inc, j}, -\vect{h}^*_{\rm inc, j})$.

Because three dipoles are uncorrelated, we can first solve the bound for one unit dipole. For simplicity, we omit its index $j$ and write its incident field as $\psi_{\rm inc}$, which excites polarization current $\phi$ in the body. For this dipole, its non-radiative LDOS can be bounded by maximum absorption in \eqref{Pabs_bound} by identifying the objective function as $\phi^\dagger(\Im\xi)\phi$. Bounds on total and radiative LDOS are less straightforward and are discussed below.

Objective function for total LDOS is $\Im \left(\tilde{\psi}_{\rm inc}^\dagger\phi\right)$ with energy conservation constraint $\phi^\dagger \left(\Im \xi + \ImGO \right) \phi = \Im \left(\psi_{\rm inc}^\dagger \phi\right)$. This echos with \eqref{gen-form-min} with $\Abb=0$ and $\beta=\tpsiinc$. Its maximum is given by \eqref{Popt}:
\begin{equation}
\begin{aligned}
\max_{\phi}&
\left\{
\Im \left(\tilde{\psi}_{\rm inc}^\dagger\phi\right)\right\} = \frac{1}{4\nu^*}(\tpsiinc+\nu^*\psiinc)^\dagger \left(\Im \xi + \ImGO \right)^{-1} (\tpsiinc+\nu^*\psiinc),
\end{aligned}
\label{eq:tot_LDOS_gen}
\end{equation}
where the optimal dual variable $\nu^*$ is always chosen at $\nu_1>\nu_0=0$ given by \eqref{optlambda}:
\begin{equation}
\nu^* =\nu_1= \left[\frac{\tpsiinc^\dagger\left(\Im \xi + \ImGO \right)^{-1}\tpsiinc}{\psiinc^\dagger\left(\Im \xi + \ImGO \right)^{-1}\psiinc}\right]^{\frac{1}{2}}.
\label{eq:nu_tot_LDOS}
\end{equation}
For non-magnetic scatterer, the above expression can be significantly simplified. No magnetic current can be excited in the non-magnetic scatterer such that $\vect{M}=0$. Examining the object function $\Im \left(\tilde{\psi}_{\rm inc}^\dagger\phi\right)$, we can find that it is equivalent to set $\vect{h_{\rm inc}}=0$. \Eqref{nu_tot_LDOS} gives $\nu^*=1$ and the maximum objective function for non-magnetic scatterer can be simplified to:
\begin{align}
\max_{\phi}
\left\{
\Im \left(\tilde{\psi}_{\rm inc}^\dagger\phi\right)\right\} &= \left[\Re\vect{e}_{\rm inc}\right]^\dagger \left(\Im \xi + \Im\GG_0 \right)^{-1} \left[\Re\vect{e}_{\rm inc}\right] \\
&\leq \vect{e}_{\rm inc}^\dagger \left(\Im \xi + \Im\GG_0 \right)^{-1} \vect{e}_{\rm inc} \\
&= \frac{2}{\omega}P^{\rm max}_{\rm ext} \label{eq:tot_LDOS_rlx}.
\end{align}
where in the last two lines, we relax the bound to the maximum-extinction bound given in \eqref{Pext_bound} with the same assumption of non-magnetic scatterer. 

Objective function for radiative LDOS defined in \eqref{rad_LDOS_gen_bound} can be chosen as $\Im \left(\tilde{\psi}_{\rm inc}^\dagger\phi\right) - \phi^\dagger(\Im\xi)\phi$. Thus, $\Abb=-\Im\xi$, $\beta=\tpsiinc$. Maximal objective function given be \eqref{Popt} can be written as:
\begin{equation}
\begin{aligned}
\max_\phi & \left\{\Im \left(\tilde{\psi}_{\rm inc}^\dagger\phi\right) - \phi^\dagger(\Im\xi)\phi\right\}
= \frac{1}{4} (\tpsiinc+\nu^*\psiinc)^\dagger \left[(\nu^*+1)\Im\xi + \nu^*\ImGO \right]^{-1} (\tpsiinc+\nu^*\psiinc)
\end{aligned}
\label{eq:fmax_radiative_LDOS}
\end{equation}
with optimal dual variable $\nu^*$ given by \eqref{det_nu}. For non-magnetic scatterer (effectively $\vect{h_{\rm inc}}=0$ in \eqref{fmax_radiative_LDOS}), bound in \eqref{fmax_radiative_LDOS} reduces to:
\begin{equation}
\begin{aligned}
\max_\phi& \left\{\Im \left(\tilde{\psi}_{\rm inc}^\dagger\phi\right) - \phi^\dagger(\Im\xi)\phi\right\} = \frac{1}{4} (\tilde{\ev}_{\rm inc}+\nu^*\einc)^\dagger \left[(\nu^*+1)\Im\xi + \nu^*\Im\GG_0 \right]^{-1} (\tilde{\ev}_{\rm inc}+\nu^*\einc).
\end{aligned}
\label{eq:fmax_radiative_LDOS_2}
\end{equation}
Radiative LDOS bound in \eqref{fmax_radiative_LDOS_2} can be relaxed to scattering bound in \eqref{Psca_bound_2nd} by observing that the dual function of the former, $g_1(\nu)$, is always greater than or equal to the latter (after suppressing its $\frac{\omega}{2}$ factor), $g_2(\nu)$, for any $\nu\geq\nu_0$:
\begin{align}
g_1(\nu) &= -\frac{1}{4} (\tilde{\ev}_{\rm inc}+\nu\einc)^\dagger \left[(\nu+1)\Im\xi + \nu\Im\GG_0 \right]^{-1} (\tilde{\ev}_{\rm inc}+\nu\einc) \\
&\geq -\frac{(1+\nu)^2}{4}\einc^\dagger \left[(\nu+1)\Im\xi + \nu\Im\GG_0 \right]^{-1} \einc = g_2(\nu). \label{eq:g1_g2}
\end{align}
The last inequality can be proved by performing Cholesky decomposition on the Hermitian matrix $\left[(\nu+1)\Im\xi + \nu\Im\GG_0 \right]^{-1} = \Lbb^\dagger \Lbb$ and using Cauchyâ€“Schwarz inequality to relax the cross term:
\begin{equation}
\begin{aligned}
\Re\{\ev_{\rm inc}^\dagger \Lbb^\dagger \Lbb  \ev^*_{\rm inc}\} &\leq \left|\ev_{\rm inc}^\dagger \Lbb^\dagger \Lbb  \ev^*_{\rm inc}\right| \leq  \left\Vert \Lbb\ev_{\rm inc} \right\Vert\cdot \left\Vert \Lbb\ev^*_{\rm inc} \right\Vert = \left\Vert\Lbb \ev_{\rm inc}\right\Vert^2 = \ev_{\rm inc}^\dagger \Lbb^\dagger \Lbb  \ev_{\rm inc}.
\end{aligned}
\end{equation}
It follows from \eqref{g1_g2} that the maximum of $g_1(\nu)$ is greater than the maximum of $g_2(\nu)$. The optimum of a primal function is given by the negative of the maximum of a dual function, so the optimal objective function considered here is smaller than the optimal scattering bound in \eqref{Psca_bound_2nd}, and equivalently \eqref{Psca_bound}:
\begin{equation}
\max_\phi \left\{\Im \left(\tilde{\psi}_{\rm inc}^\dagger\phi\right) - \phi^\dagger(\Im\xi)\phi\right\} \leq \frac{2}{\omega}P_{\rm sca}^{\rm max}.
\label{eq:rad_LDOS_rlx}
\end{equation}
To summarize, we derive general LDOS bounds for any material. For non-magnetic material specifically, LDOS can be directly bounded by maximum power response in Eqs. (\ref{eq:Pext_bound}), (\ref{eq:Pabs_bound}), and (\ref{eq:Psca_bound}):
\begin{align}
\rho_{\rm tot} &\leq \frac{2}{\pi\omega^2}\sum_j P^{\rm max}_{\text{ext}, j} + \rho_0 \label{eq:rhotot_bnd} \\ 
\rho_{\rm nr} &\leq \frac{2}{\pi\omega^2}\sum_j P^{\rm max}_{\text{abs}, j} \\
\rho_{\rm rad} &\leq \frac{2}{\pi\omega^2}\sum_j P^{\rm max}_{\text{sca}, j} + \rho_0,
\label{eq:rhorad_bnd}
\end{align}
where $j=1,2,3$ denotes the summation of maximum power quantities from three orthogonally polarized unit dipoles.

\begin{figure}[t!]
	\includegraphics[width=0.5\textwidth]{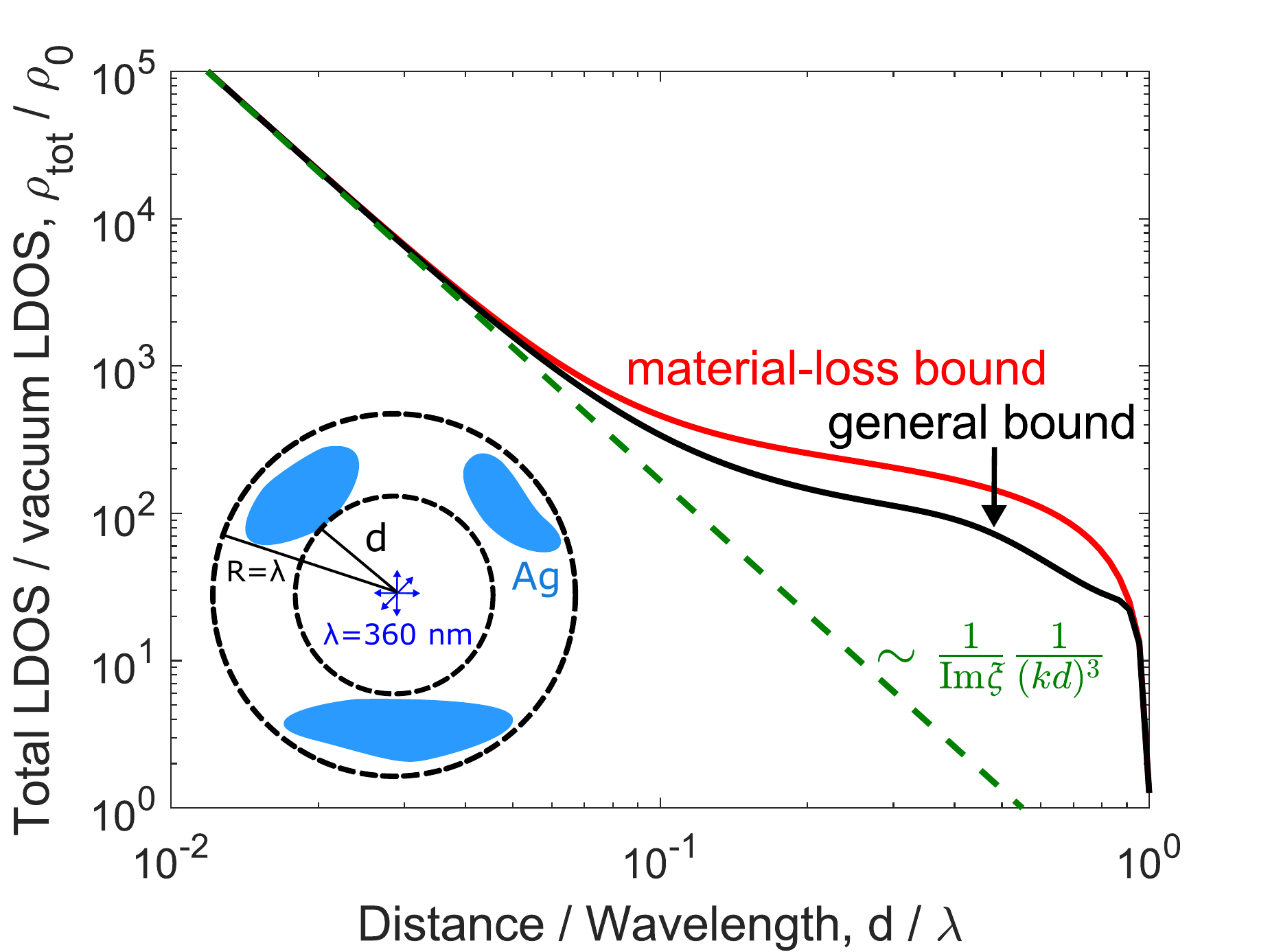}
	\centering
	\caption{Bounding volume for the LDOS problem is chosen to be a spherical shell with three randomly oriented dipoles in the center, radiating at wavelength $\lambda=360$ nm. Inner radius of the shell is determined by the minimum distance $d$ to the scatterer comprising only Ag. Outer radius $R$ of the shell covers the far end of the scatterer and is assumed to be one wavelength in the figure. In the far field, general bound is tighter than material-loss bound. In the near field, general bound converge to material-loss bound, and both follow the same divergence as $\frac{1}{\Im\xi}\frac{1}{k^3d^3}$.}
	\label{fig:figure5}
\end{figure}

\paragraph{LDOS bounds for a finite non-magnetic scatterer}
In the following, we assume the scatterer is non-magnetic and finite, embedded in the vacuum. The non-magnetic nature of the scatterer allows us to use \eqrefrange{rhotot_bnd}{rhorad_bnd} to decompose LDOS bounds to previous power bounds for three orthogonally polarized unit dipoles. In \eqrefrange{nf_ext_bnd}{nf_sca_bnd}, we presented power bounds for arbitrary dipole distributions $\pv(\xv)$. Here, we start with a point dipole oriented along $\ehv_j$ at origin $\pv(\xv) = \pv_j(\xv) = p_0 \delta(\xv)\ehv_j$ with $p_0=1$, and later sum up the contributions from three orthogonal polarizations.
We also assume the scatterer is finite, thus can be enclosed by a spherical shell (see Fig. \ref{fig:figure5} inset).
A shell-like bounding volume has spherical symmetry, so $\vv_i$ and $\wv_i$ in \eqrefrange{nf_ext_bnd}{nf_sca_bnd} are regular VSWs:
\begin{align}
\vv_{mn1}(\xv) &= \wv_{mn1}(\xv) = k^{\frac{3}{2}}Rg\Mv_{mn}(kr,\theta,\phi) \\
\vv_{mn2}(\xv) &= \wv_{mn2}(\xv) = k^{\frac{3}{2}}Rg\Nv_{mn}(kr,\theta,\phi),
\end{align}
$\uv_i$ are outgoing VSWs:
\begin{align}
\uv_{mn1}(\xv) &= k^{\frac{3}{2}}\Mv_{mn}(kr,\theta,\phi) \\
\uv_{mn2}(\xv) &= k^{\frac{3}{2}}\Nv_{mn}(kr,\theta,\phi).
\end{align} 

Power bounds in \eqrefrange{nf_ext_bnd}{nf_sca_bnd} require us to evaluate four overlap integrals: $\pv_j^\dagger\wv_i, \uv_i^\dagger\uv_i, \uv_i^\dagger\vv_i, \vv_i^\dagger\vv_i$.
We first evaluate overlap integral between the point dipole and regular VSWs in the source volume $V_s$:
\begin{align}
\pv_j^\dagger\wv_i &= \int_{V_s}\pv_j^*(\xv)\cdot\wv_i(\xv)\text{d}\xv \\
&= p_0 \ehv_j\cdot\wv_i(\xv=0) \\
&= k^{\frac{3}{2}}p_0
\begin{cases}
\ev_j\cdot Rg\Nv_{m,1}(0,\theta,\phi) & \text{if  } j=2, n=1 \\
0 & \text{else,}
\end{cases}
\end{align}
where we used the fact that only $Rg\Nv_{m,1}$ has nonzero value at the origin. Exact value of the dot product $\ev_j\cdot Rg\Nv_{m,1}(0,\theta,\phi)$ depends on the orientation of the dipole:
\begin{equation}
\ev_j\cdot Rg\Nv_{m,1}(0,\theta,\phi) = 
\begin{cases}
\pm \frac{1}{2\sqrt{3\pi}}\delta_{m,\pm 1} & \text{if\ } \ehv_j = \vect{\hat{x}} \\
\frac{1}{2i\sqrt{3\pi}}\delta_{m,\pm 1} & \text{if\ } \ehv_j = \vect{\hat{y}} \\
-\frac{1}{\sqrt{6\pi}}\delta_{m,0} & \text{if\ } \ehv_j = \vect{\hat{z}}. 
\end{cases}
\end{equation}
Later for LDOS, we will need to evaluate averaged power from three randomly oriented dipoles, which is related to the quantity:
\begin{equation}
\frac{1}{3}\sum_j |\pv_j^\dagger\wv_i|^2 = k^3 \frac{p_0^3}{18\pi}\delta_{n,1}\delta_{j,2},
\label{eq:random_dipole}
\end{equation}
where $\ehv_j$ runs through directions $\vect{\hat{x}}$, $\vect{\hat{y}}$, and $\vect{\hat{z}}$.
We now evaluate overlap integrals between different VSWs within the bounding volume $V$:
\begin{align}
\vv_i^\dagger\vv_i &= \int_V \vv_{mnj}^*(\xv)\cdot\vv_{mnj}(\xv)\text{d}\xv = I_j\left(j_n^{(1)}(x), j_n^{(1)}(x)\right) \\
\uv_i^\dagger\uv_i &= \int_V \uv_{mnj}^*(\xv)\cdot\uv_{mnj}(\xv)\text{d}\xv = I_j\left(h_n^{(1)*}(x), h_n^{(1)}(x)\right) \\
\uv_i^\dagger\vv_i &= \int_V \uv_{mnj}^*(\xv)\cdot\vv_{mnj}(\xv)\text{d}\xv = I_j\left(h_n^{(1)*}(x), j_n^{(1)}(x)\right), 
\end{align}
where we defined function:
\begin{equation}
I_j\left(z_n^{(1)}(x), z_n^{(2)}(x)\right) =  
\begin{cases}
\int_{kR_1}^{kR_2}x^2z_n^{(1)}(x)z_n^{(2)}(x)\text{d}x &\text{if } j=1 \\
n(n+1)\int_{kR_1}^{kR_2}z_n^{(1)}(x)z_n^{(2)}(x)\text{d}x \\
\quad + \int_{kR_1}^{kR_2}[xz_n^{(1)}(x)]'[xz_n^{(2)}(x)]'\text{d}x &\text{if } j=2.
\end{cases}
\end{equation}

Bound for total extinction from three randomly oriented dipoles is bounded by:
\begin{align}
\frac{1}{3}\sum_j P_{\text{ext}, j} \leq \frac{1}{3}\sum_j \sum_i \frac{\omega}{2} |\pv_j^\dagger\wv_i|^2 \underbrace{\frac{1}{\Im\xi}\left( \uv_i^\dagger\uv_i - \frac{|\uv_i^\dagger\vv_i|^2}{\Im\xi+\vv_i^\dagger\vv_i} \right)}_{f_i},
\label{eq:def_f}
\end{align}
where we defined enhancement factor $f_i$ (depends only on $n$ and $j$). Using \eqref{random_dipole}, we can show that:
\begin{equation}
\frac{\frac{1}{3}\sum_j P_{\text{ext}, j}}{P_0} \leq f_{n=1,j=2},
\end{equation}
where $P_0=\omega k^3p_0^3/12\pi$ is the power radiated by a dipole with amplitude $p_0$ in vacuum. Similarly, one can show that:
\begin{equation}
\frac{\rho_{\rm tot}}{\rho_0} \leq 1+f_{n=1,j=2}.
\end{equation}
The enhancement factor $f_{n=1,j=2}$ shows how large the light extinction of three uncorrelated dipoles can be, compared to the vacuum. While the first term in \eqref{def_f} appears in previous material-loss bound~\cite{miller_fundamental_2016}, the second term comes from radiation coupling between the bounding volume and the vacuum.
In near field when material loss dominates, $f_{n=1,j=2}$ can be simplified to the material-loss bound:
\begin{align}
f_{n=1,j=2} &= \frac{1}{\Im\xi}\uv_i^\dagger\uv_i \\
&= \frac{1}{\Im\xi}\left(x-\frac{1}{x}-\frac{1}{x^3}\right)\bigg|_{kR_1}^{kR_2} \\
&\rightarrow \frac{1}{\Im\xi}\frac{1}{k^3R_1^3} \label{eq:nf_exp},
\end{align}
where, in the last line, we take the limit of extreme near field where $kR_1\ll 1, kR_2$. 
In Fig. \ref{fig:figure5}, we showed the general bound and material-loss bound for LDOS enhancement at wavelength $360$ nm by Ag surroundings. It is clear that both bounds follow \eqref{nf_exp} in near field limit. In far field, general bound is slightly tighter than the material-loss bound due to the consideration of additional radiative loss.

Absorption and scattering bounds can also be written through an enhancement factor over the vacuum radiation:
\begin{align}
\frac{\frac{1}{3}\sum_j P_{\text{abs}, \pv_j}}{P_0} \leq f^{\rm abs}_{n=1,j=2}(\nu^*) \\
\frac{\frac{1}{3}\sum_j P_{\text{sca}, \pv_j}}{P_0} \leq f^{\rm sca}_{n=1,j=2}(\nu^*).
\end{align}
Though they are more complicated in the sense that both enhancement factors (defined below) are functions of $\nu^*$, the optimal dual variable. Similarly, for non-radiative and radiative LDOS we can write:
\begin{align}
\frac{\rho_{\rm nr}}{\rho_0} &\leq f^{\rm abs}_{n=1,j=2}(\nu^*) \\
\frac{\rho_{\rm rad}}{\rho_0} &\leq 1+f^{\rm sca}_{n=1,j=2}(\nu^*).
\end{align}

Lastly, we present the explicit expressions of absorptive and scattering enhancement factors.
For absorption, the enhancement factor is:
\begin{equation*}
f^{\rm abs}_{n=1,j=2}(\nu^*) = \frac{\nu^{*2}}{4(\nu^*-1)}\frac{1}{\Im\xi}\left(\uv_i^\dagger\uv_i-\frac{|\uv_i^\dagger\vv_i|^2}{(\nu^*-1)\Im\xi/\nu^* + \vv_i^\dagger\vv_i}\right),
\end{equation*}
where $\nu^*$ is determined by solving $a = (\nu^*-1)\Im\xi/\nu^*$ in the following equation:
\begin{align*}
& 2a\left(\uv_i^\dagger\uv_i-\frac{|\uv_i^\dagger\vv_i|^2}{a+\vv_i^\dagger\vv_i}\right) = \left\{\uv_i^\dagger\uv_i\Im\xi + |\uv_i^\dagger\vv_i|^2\left[1 - \frac{(\Im\xi+\vv_i^\dagger\vv_i)(2a+\vv_i^\dagger\vv_i)}{(a+\vv_i^\dagger\vv_i)^2}\right]\right\}.
\end{align*}
For scattering, the enhancement factor is:
\begin{equation*}
f^{\rm sca}_{n=1,j=2}(\nu^*) = \frac{\nu^*}{4}\frac{1}{\Im\xi}\left(\uv_i^\dagger\uv_i-\frac{|\uv_i^\dagger\vv_i|^2}{\nu^*\Im\xi/(\nu^*-1) + \vv_i^\dagger\vv_i}\right),
\end{equation*}
where $\nu^*$ is determined by solving $a = \nu^*\Im\xi/(\nu^*-1)$ in the following equation:
\begin{align*}
& 2\Im\xi\left(\uv_i^\dagger\uv_i-\frac{|\uv_i^\dagger\vv_i|^2}{a+\vv_i^\dagger\vv_i}\right) = \left\{\uv_i^\dagger\uv_i\Im\xi + |\uv_i^\dagger\vv_i|^2\left[1 - \frac{(\Im\xi+\vv_i^\dagger\vv_i)(2a+\vv_i^\dagger\vv_i)}{(a+\vv_i^\dagger\vv_i)^2}\right]\right\}.
\end{align*}

\begin{figure}[t!]
	\includegraphics[width=0.5\textwidth]{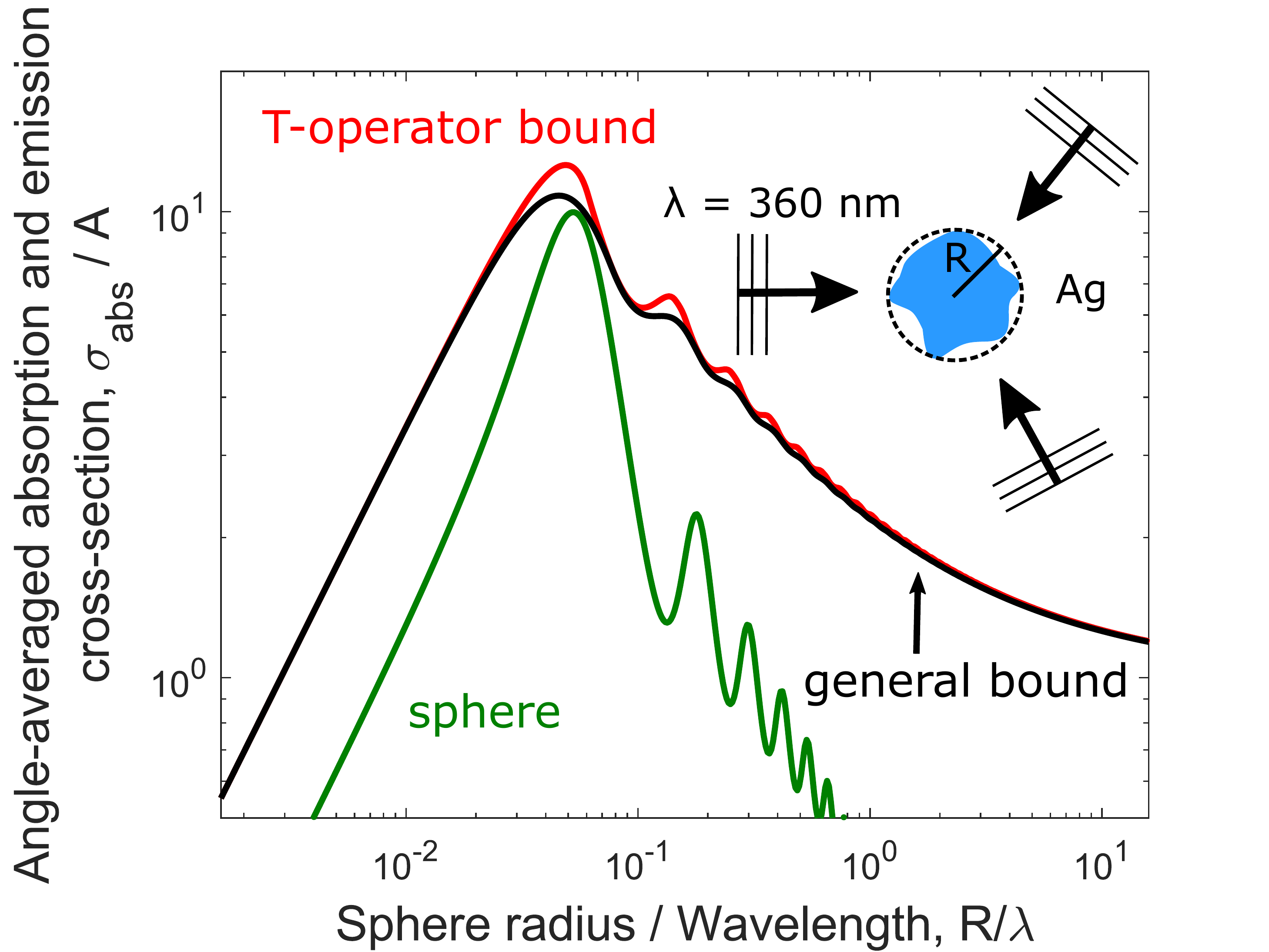}
	\centering
	\caption{General bound for maximum thermal absorption and emission, compared with $\mathbb{T}$-operator bound and thermal absorption of an actual Ag \cite{johnson1972optical} sphere with radius $R$.}
	\label{fig:figure6}
\end{figure}

\section{Thermal absorption and emission} 
\label{sec:thermal}
Our formalism applies equally to thermal absorption and emission. By Kirchhoff's Law (reciprocity), or its nonreciprocal generalization~\cite{Miller2017a}, total thermal absorption and emission are equivalent and can be found by considering a weighted average of incoherent, orthogonal incoming fields $\vect{E}_{\textrm{inc},i}$:
\begin{equation}
\langle \left| \vect{E}_{\rm inc} \right|^2 \rangle = \sum_i w_i \left| \vect{E}_{\text{inc}, i} \right|^2,
\end{equation}
where $w_i$ is a weighting factor. For a continuum of incoming fields the sum is instead an integral with a differential weight. A direct consequence of the incoherent averaging is that an upper bound to the average absorptivity/emissivity is given by the average of the bounds for each independent incident field. Surprisingly, the bounds computed by this averaging procedure varies depending on which basis is used for the incoming fields. If the incident field is treated as an incoherent sum of plane waves, over all propagation angles, for example, then the absorptivity/emissivity cross-section bounds would simply be a scalar multiple of \eqref{abs-cha-bound}. However, the bound can be tightened (decreased) if the incident fields are instead decomposed in vector spherical waves, for which the weight function $w_i$ is determined by the fluctuation-dissipation theorem~\cite{kruger2012trace}: $w_i = \frac{4}{\pi\omega}\Theta(T)$, where $\Theta(T)=\hbar\omega/(e^{\hbar\omega/k_BT}-1)$ is the Planck energy of a harmonic oscillator at temperature $T$ without the zero-point energy. The resulting bound is a sum over all VSW channels $i$:
\begin{equation}
P_\text{abs} \leq \frac{2}{\pi}\Theta(T) \sum_i
\begin{cases}
\frac{\rho_i \Im\xi}{(\Im{\xi}+\rho_i)^2} & \text{for }\rho_i \leq \Im{\xi} \\  
\frac{1}{4} & \text{for } \rho_i \geq \Im{\xi}
\end{cases}
\label{eq:thermal_bound2}
\end{equation}  
where $i=\{n,m,j\}$ includes all VSW channels: $n=1,2,...$, $m = -n,...,n$, $j=1,2$, and the sum converges for any nonzero $\Im \xi$. \eqref{thermal_bound} shows a distinct threshold behavior within each VSW channel. In the asymptotic limits of radiation-dominant ($\rho_i \gg \Im\xi$) or material-loss-dominant ($\rho_i \ll \Im \xi$) scenarios, \eqref{thermal_bound} simplifies to the known channel-~\cite{pendry1999radiative} and material-loss bounds~\cite{miller_fundamental_2016}. In tandem, accounting for both mechanisms yields a significantly tighter bound than any previous approach.

Taking the same approach as in Sec. 3 in the paper, the bound for any arbitrary shape is no larger than the bound for any bounding volume of that shape, and thus we can compute analytical bounds for finite-sized thermal absorbers with spherical bounding volumes. \Figref{figure6} shows the thermal absorption/emission cross-section as a function of the size of a spherical silver~\cite{johnson1972optical} nanoparticle at wavelength $\lambda = \SI{360}{nm}$. Included is the bound of \eqref{thermal_bound}, which is nearly achieved by the sphere at its ideal resonant size. We also include the recently published $\mathbb{T}$-operator bound of \citeasnoun{molesky2019t}, which considered the effect of radiation and material losses separately for thermal sources. As shown in \figref{figure6}, by incorporating both losses in one optical theorem constraint, even for thermal fields the new bounds are slightly tighter.

\providecommand{\noopsort}[1]{}\providecommand{\singleletter}[1]{#1}%

\clearpage
%

\begin{table*}[t]
\centering
\begin{tabular}{ | m{1.5cm} | m{1.4cm}| m{5cm} | m{1.5cm} | m{1.4cm}| m{5cm} | }
\hline
    thickness ($\SI{}{\micro\meter}$)  & absorption (\%) &  \hspace{19mm} design & thickness ($\SI{}{\micro\meter}$)  & absorption (\%)   &  \hspace{19mm} design\\
    \hline
    \hspace{15mm} 0.4 &   \hspace{15mm} 52 & \vspace{1mm}\hspace{6.5mm} \phototm{SM-fig5-SiC_h04um} & \hspace{15mm} 0.6 &   \hspace{15mm} 57 & \vspace{1mm}\hspace{6.5mm} \phototm{SM-fig5-SiC_h06um}\\
    \hspace{15mm} 0.8 &   \hspace{15mm} 70 & \vspace{1mm}\hspace{6.5mm} \phototm{SM-fig5-SiC_h08um} &
    \hspace{15mm} 1.0 &   \hspace{15mm} 76 & \vspace{1mm}\hspace{6.5mm} \phototm{SM-fig5-SiC_h10um} \\
    \hspace{15mm} 1.2 &   \hspace{15mm} 90 & \vspace{1mm}\hspace{6.5mm} \phototm{SM-fig5-SiC_h12um} &
    \hspace{15mm} 1.4 &   \hspace{15mm} 94 & \vspace{1mm}\hspace{6.5mm} \phototm{SM-fig5-SiC_h14um} \\
    \hspace{15mm} 1.6 &   \hspace{15mm} 95 & \vspace{1mm}\hspace{6.5mm} \phototm{SM-fig5-SiC_h16um} & & & \\
\hline
\end{tabular}
\caption{Inverse-designed SiC ultra-thin absorbers at $\SI{11}{\micro\meter}$. These are grey-scale designs with material ranges from pure air (purely white) to pure SiC~\cite{francoeur2010spectral} (dark blue).}
\label{tab:id1}
\end{table*}

\clearpage

\begin{table*}[t]
	\centering
	\begin{tabular}{ | m{3cm} | m{3cm} | m{2cm} | m{2cm} | m{3cm}| m{4cm} | }
\hline
   \hspace{8mm} material  &\hspace{5.5mm} wavelength  &\hspace{4.5mm} period &\hspace{2.5mm} thickness &\hspace{13.5mm} $\varepsilon$ &\hspace{14mm} design\\
    \hline
    \hspace{11.5mm} Au~\cite{palik1998handbook}  &\hspace{9mm} 500 nm  &\hspace{5.5mm} 55 nm &\hspace{5mm} 80 nm  &\hspace{5mm} -2.99+2.93i &\vspace{1mm}\hspace{1.5mm} \phototm{SM-fig5-Au_h80nm}\\
    \hspace{11.5mm} Ag~\cite{palik1998handbook}  &\hspace{9mm} 500 nm  &\hspace{5.5mm} 55 nm &\hspace{5mm} 40 nm &\hspace{5mm} -7.63+0.73i &\vspace{1mm}\hspace{1.5mm} \phototm{SM-fig5-Ag_h40nm}\\
    \hspace{11.5mm} Al~\cite{palik1998handbook}  &\hspace{9mm} 500 nm  &\hspace{5.5mm} 55 nm &\hspace{5mm} 40 nm &\hspace{5mm} -34.23+8.98i &\vspace{1mm}\hspace{1.5mm} \phototm{SM-fig5-Al_h40nm}\\
    
    \hspace{11.5mm} SiO$_2$~\cite{popova1972optical}  &\hspace{10mm} $\SI{9}{\micro\meter}$  &\hspace{4mm} $\SI{1.1}{\micro\meter}$ &\hspace{4.5mm} $\SI{1.4}{\micro\meter}$ &\hspace{5mm} -4.71+3.20i &\vspace{1mm}\hspace{1.5mm} \phototm{SM-fig5-SiO2_h14um}\\
    \hspace{6.5mm} doped InAs~\cite{law2013all}  &\hspace{10mm} $\SI{7.5}{\micro\meter}$  &\hspace{4mm} $\SI{1.1}{\micro\meter}$ &\hspace{4.5mm} $\SI{0.6}{\micro\meter}$ &\hspace{5mm} -10.39+1.80i &\vspace{1mm}\hspace{1.5mm} \phototm{SM-fig5-InAs_h06um}\\
    \hspace{11.5mm} SiC~\cite{francoeur2010spectral}  &\hspace{10mm} $\SI{11}{\micro\meter}$  &\hspace{4mm} $\SI{1.1}{\micro\meter}$ &\hspace{4.5mm} $\SI{0.8}{\micro\meter}$ &\hspace{5mm} -3.81+0.23i &\vspace{1mm}\hspace{1.5mm} \phototm{SM-fig5-SiC_h08um}\\
\hline
\end{tabular}
\caption{Inverse-designed ultra-thin absorbers with 70\% absorption rate. These are grey-scale designs with material ranges from pure air (purely white) to pure material (dark blue).}
\label{tab:id2}
\end{table*}